\documentclass[draft,tightenlines,nofootinbib,preprint,aps,eqsecnum,amsmath,amssymb]{revtex4}

\begin{document}

\newcommand{\sgn}{\mbox{\boldmath $\epsilon$}}

\newcommand{\beq}{\begin{equation}}
\newcommand{\eeq}{\end{equation}}
\newcommand{\bea}{\begin{eqnarray}}
\newcommand{\eea}{\end{eqnarray}}
\newcommand{\cir}{{\buildrel \circ \over =}}

\newcommand{\on}{\stackrel{\circ}{=}}
\newcommand{\byd}{\stackrel{def}{=}}
\baselineskip 20pt

\title{The Chrono-Geometrical Structure of Special and General
Relativity: towards a Background-Independent Description of the
Gravitational Field and Elementary Particles.}

\author{Luca Lusanna}

\affiliation
{Sezione INFN di Firenze\\
Via G. Sansone 1\\
50019 Sesto Fiorentino (FI), Italy\\
E-mail: LUSANNA@FI.INFN.IT}

\bigskip

Invited paper for the book {\it Progress in General Relativity and
Quantum Cosmology Research} (Nova Science).

\bigskip

\begin{abstract}

Since the main open problem of contemporary physics is to find a
unified description of the four interactions, we present a
possible scenario which, till now only at the classical level, is
able to englobe experiments ranging from experimental space
gravitation to atomic and particle physics. After a reformulation
of special relativistic physics in a form taking into account the
non-dynamical chrono-geometrical structure of Minkowski space-time
(parametrized Minkowski theories and rest-frame instant form) and
in particular the conventionality of simultaneity (re-phrased as a
gauge freedom), a model of canonical metric and tetrad gravity is
proposed in a class of space-times where the deparametrization to
Minkowski space-time is possible. In them it is possible to give a
post-Minkowskian background-independent description of the
gravitational field and of matter. The study of the dynamical
chrono-geometrical structure of these space-times allows to face
interpretational problems like the physical identification of
point-events (the Hole Argument), the distinction between inertial
(gauge) and tidal (Dirac observables) effects, the dynamical
nature of simultaneity in general relativity and to find
background-independent gravitational waves. These developments are
possible at the Hamiltonian level due to a systematic use  of
Dirac-Bergmann theory of constraints. Finally there is a proposal
for a new coordinate- and background-independent quantization
scheme for gravity.

\end{abstract}

\maketitle

\section{Introduction.}

After having reached a reasonable understanding of the
electro-magnetic, weak and strong interactions with the $SU(3)
\times SU(2) \times U(1)$ standard model of elementary particles
\cite{1} and with its subsequent tentative extensions, the main
open theoretical problem is the incorporation of gravity in the
framework. This is a highly non-trivial task already at the
classical level, where the standard formulation of field theory in
curved space-times denies any role to the concept of particle
\cite{2}. On the other hand, no one has found an acceptable blend
of general relativity and quantum mechanics \cite{2a}, so that it
is possible that we must modify  one or both the theories. The
foundational problems of quantum mechanics like the {\it
entanglement of macroscopic bodies} and the connected problem of
giving a meaning to the words {\it measuring apparatus}, which are
undefined in every approach, are still formulated at the
non-relativistic level without an accepted extension to either
special or general relativity due to the absence of an absolute
notion of simultaneity \cite{3}. The foundational problems of
general relativity like the physical individuation of the
point-events of space-time due to Einstein's Hole Argument
\cite{4} and the double role of the metric tensor, which is not
only the potential of the gravitational field but also implies a
{\it dynamical definition of the chrono-geometrical structure} of
Einstein's space-times, show how far we are from a control upon
relativistic causality and, in absence of realistic solutions of
Einstein's equations for macroscopic bodies, from a theory of
measurement involving {\it dynamical} and not {\it test} matter
\cite{5}. A naive quantization of the metric tensor destroys basic
notions like {\it being time-, light- or space-like} which are
essential for relativistic causality \cite{6}. Theories on a
background space-time like effective quantum field theory and
string theory \cite{7} have a background non-dynamical
chrono-geometrical structure: as a consequence the gravitational
field, instead of teaching causality to the other fields
determining the null curves to be followed by massless particles,
is reduced, after linearization, to a spin-2 graviton moving on a
fixed light-cone like photons and gluons. On the other hand the
best developed background-independent theory, loop quantum gravity
\cite{8}, does not lead to a Fock space, so that no one knows how
to incorporate in it the standard model of elementary
particles.\medskip

This host of big unsolved problems is accompanied by a lot of
other either technical or conceptual secondary problems, sometimes
remnants of older ones, which, in absence of a solution, have been
hidden under the rug. Here is a partial list:\medskip

i) All the physically relevant theories are formulated in terms of
singular Lagrangians. Therefore  for their Hamiltonian description
and for the definition of a well posed Cauchy problem a good
understanding of Dirac-Bergmann theory of constraints
\cite{9,10,11,12} is needed. Moreover, a technique for the
separation of the arbitrary {\it gauge variables} from the {\it
gauge invariant predictable Dirac observables} has to be defined.

ii) The need of a different interpretation of the {\it gauge
variables} in theories invariant under local inner Lie groups,
like gauge theories, and in theories invariant under time or
space-time diffeomorphisms, like relativistic particles, strings
and every formulation of gravity.

iii) The theory of relativistic bound states \cite{1}, which
requires the understanding of the instantaneous approximations to
quantum field theory so to arrive at an effective relativistic
wave equation and to an acceptable scalar product. In turn, the
wave equation must also result from the quantization of a
relativistic action-at-a-distance two-body problem in relativistic
mechanics (but with the particles interpreted as asymptotic states
of quantum field theory), since only in this way we can get a
solution of the interpretational problems connected to the gauge
nature of the {\it relative times}, a problem going back to
Tetrode and Fokker \cite{13}, not to speak of Droste \cite{14} for
the beginning of the still unsolved general relativistic two-body
problem, and lying at the basis of the interpretational problems
with the Bethe-Salpeter equation \cite{15}.

iv) The {\it problem of time} in general relativity \cite{16,17}:
Mach's influence on Einstein resulted in a preference for compact
space-times without boundary and, then, in the Wheeler-DeWitt
interpretation implying local evolution in an internal either
extrinsic (York) or intrinsic (Misner) time but also a frozen
formalism with no evolution in the reduced phase space \cite{18}.
On the other hand, in non-compact space-times the many-fingered
time notion of evolution \cite{17} was introduced as the general
relativistic counterpart of the Tomonaga-Schwinger special
relativistic approach to quantum field theory \cite{19}. However,
already in special relativity, the Torre-Varadarajan no-go theorem
\cite{16} shows that there is a ultraviolet obstruction to its
implementation at the quantum level.

v) The claimed superiority of the configurational manifestly
covariant approach over the Hamiltonian approach requiring an
explicit definition of {\it time} through a 3+1 decomposition of
space-time. However, this illusion is broken by the necessity of a
well posed formulation of the Cauchy problem to identify which are
the {\it predictable} quantities.

vi) The necessity of the Hamiltoniabn approach to study: a) the
notions of relativistic center  of mass and relative variables for
the relativistic N-body problem and for relativistic fluids and
fields; b) the impossibility, in the rest frame, to make a unique
separation of global rotations from vibrations due to the
non-Abelian nature of the angular momentum Noether constants of
motion both at the non-relativistic and relativistic levels (as a
consequence Machian concepts may be consistently introduced only
for zero angular momentum); c) how to recover notions like the
tensor of inertia, which, together with the Jacobi coordinates,
exist only in Galilei space-time, by means of relativistic
multipolar expansions.

vii) Due to the absence of absolute simultaneity, the 3+1
decomposition of space-time is fundamental both in special and
general relativity to introduce well defined {\it equal time
Cauchy 3-spaces} and notions of spatial distance and one-way
velocity of light. The lack of a good notion of simultaneity,
generalizing Einstein's convention for clock synchronization in
inertial frames, is at the basis of pathologies like the
coordinate singularities appearing in rotating frames (the
rotating disk, the Sagnac effect; see the bibliography of
Refs.\cite{20,21}) and in the Fermi coordinates of a non-inertial
observer.

viii) The lack of an accepted theory of the measurements made by
non-inertial observers and of which are the {\it relativistic
inertial forces} seen by them Ref.\cite{21}. The problem of the
description of gyroscopes in space experiments as pole-dipole
systems and the need to replace metric gravity with tetrad gravity
(a theory of non-inertial observers) when fermions are present.
\bigskip

Therefore, we felt the exigence of a revisitation of both special
and general relativity at the classical level to see whether it is
possible to arrive at a scenario with a unified description of
particle physics and gravitational field in some suitable class of
non-compact space-times allowing, when we switch off the Newton
constant, a deparametrization to a description of particle physics
in Minkowski space-time in all those general non-inertial frames
where it possible to define a good notion of simultaneity. It is
hoped that this scenario will allow to arrive to a new
quantization scheme of the four interactions in a way compatible
with relativistic causality and the dynamical chrono-geometrical
structure of space-time.

Here we delineate the general framework of this research program
and its status after various recent achievements. After a review
of constraint theory, we discuss the problem of the admissible
notions of simultaneity in special relativity, where only the
chrono-geometrical structure of Minkowski space-time is absolute,
and its treatment according to the 3+1 point of view, leading to
parametrized Minkowski theories for the description of every
isolated system. The associated special relativistic general
covariance of this theory shows that all such notions of
simultaneity are {\it gauge equivalent} and that there is a
preferred intrinsic notion leading to the {\it rest-frame instant
form of dynamics}. In Ref.\cite{22} there is a complete review of
all the special relativistic systems which have been treated in
this way. Then we identify a model for metric and tetrad gravity
in a suitable class of non-compact space-times, in which the
admissible 3+1  splittings tend to space-like hyper-planes at
spatial infinity in such a way that the deparametrization of
general relativity with matter leads to the rest-frame instant
form description of the  same matter in Minkowski space-time.
Since in general relativity the chorono-geometrical structure of
the space-time is dynamical, we find that the admissible notions
of simultaneity, besides being gauge equivalent like in special
relativity, are now {\it dynamically} determined by the solutions
of the Hamilton-Dirac equations, equivalent to Einstein's
equations.

\vfill\eject

\section{Constraint's Theory}

Most of the physically relevant systems are described by means of
singular Lagrangians. This  means that the Hessian matrix, whose
elements are the second derivatives of the Lagrangian with respect
to the velocities, has zero eigenvalues. This implies that\medskip

i) the action is (quasi-) invariant under gauge transformations
depending on arbitrary functions of time (finite degrees of
freedom) or space and time (field theory) and the second Noether
theorems applies;

ii) the Euler-Lagrange equations cannot be put in normal form,
i.e. solved in the accelerations;

iii) the Noether identities implied by the second Noether theorem
determine the full set of extra equations of motion to be added to
the original Euler-Lagrange equations for consistency;

iv) the solutions of the Euler-Lagrange equations depend on
arbitrary functions of time or space and time and, especially in
field theory, it may difficult to arrive at a well posed Cauchy
problem;

v) there is a number of arbitrary functions of the configuration
variables and velocities (primary generalized velocity functions)
less or equal to the number of null eigenvalues of the Hessian
matrix (assuming that it has constant rank as it happens in many
physically relevant systems) determined by the Noether identities;

vi) as a consequence the original configuration variables do not
have a predictable evolution in time;

vii) the main task is to identify an equal number of functions of
the configuration variables and velocities such that a subset of
them has a predictable evolution ({\it gauge-invariant
observables}), while the elements of the complementary subset are
left completely arbitrary ({\it gauge variables}; in general their
number is higher than that of the primary generalized velocity
functions due to the propagation of their indeterminateness
implied by the Noether identities).

\medskip

Since the Lagrangian formalism does not have a natural technology
to face all these problems, we must go to the Hamiltonian
formalism, where the theory of canonical transformations allows to
find their solution. The degeneracy of the Hessian matrix and the
second Noether theorem \cite{12} imply the modification of the
Hamiltonian theory known as Dirac-Bergmann theory of constraints
\cite{9,10,11,12}.

The degeneracy of the Lagrangian implies that the original phase
space is restricted to a sub-manifold by as many primary
constraints $\phi_A \approx 0$ as null eigenvalues of the Hessian
matrix. If $H_c$ is the canonical Hamiltonian, the existence of
the constraints leads to the introduction of the Dirac Hamiltonian
$H_D = H_c + \sum_A\, \lambda_A\, \phi_A$, where the $\lambda_A$'s
(the {\it Dirac multipliers}) are arbitrary functions of time or
space and time corresponding to generalized velocity functions
non-projectable to phase space and left arbitrary by the
Euler-Lagrange equations. Then, the so called Dirac algorithm,
equivalent to the Noether identities, restricts the number of the
arbitrary $\lambda_A$'s to that of the primary generalized
velocity functions and determines the secondary, tertiary, ...
constraints, if any. The full set of constraints determines the
final constraint sub-manifold to which the dynamics of the system
is restricted. Then the constraints are divided in two groups: i)
the {\it first class} constraints, having weakly zero Poisson
brackets with all the constraints and being the generators of the
Hamiltonian gauge transformations; ii) the {\it second class}
constraints, which are even in number since they correspond to
pairs of redundant canonical variables present, for instance, for
reasons of manifest covariance. The final Dirac Hamiltonian is
$H_D = H^{'}_c + \sum_a\, \lambda_a\, \phi_a$, where the
$\phi_a$'s are the {\it primary first class} constraints and the
$\lambda_a$'s are the Dirac multipliers left arbitrary. Even if
there is no general demonstration, in the physically relevant
cases $H^{'}_c$ contains all the non-primary first class
constraints, each one multiplied by some function of the canonical
variables left indetermined by the Hamilton equations. In general
$H^{'}_c$ contains also all the second class constraints either
linearly or quadratically. When there is reparametrization or
diffeomorphism invariance of the action, the canonical Hamiltonian
vanishes, $H_c \equiv 0$, and $H^{'}_c \approx 0$.

\medskip

When there are no second class constraints, as in all the physical
system we consider, the constraint manifold is a {\it
presymplectic} manifold and we have the following Poisson bracket
algebra: $\{ \phi_A, \phi_B \} = C_{ABC}\, \phi_C \approx 0$, $\{
H_c, \phi_A \} = C_{AB}\, \phi_B \approx 0$. The Hamilton-Dirac
equations generated by the final Dirac Hamiltonian $H_D = H^{'}_c
+ \sum_a\, \lambda_a\,  \phi_a = H_c^{"} + \sum_{\alpha}\,
f_{\alpha}\, \phi_{\alpha} + \sum_a\, \lambda_a\, \phi_a$
($\phi_{\alpha}$ are the non-primary first class constraints,
$f_{\alpha}$ functions on phase space and $H^{"}_c$ is the final
canonical Hamiltonian) are equivalent to the Euler-Lagrange
equations and to all the extra equations of motion implied by the
Noether identities. The {\it predictable} quantities are the
gauge-invariant {\it Dirac observables} $O_r$ satisfying $\{ O_r,
\phi_a \} \approx \{ O_r, \phi_{\alpha} \} \approx 0$ and with the
deterministic evolution in time ruled by $H_c$, which is a Dirac
observable. The presymplectic constraint sub-manifold is foliated
with {\it gauge orbits}. Each gauge orbit has its points connected
by Hamiltonian gauge transformations generated by the first class
constraints. Therefore the gauge orbits are coordinatized by the
gauge variables, whose number is equal to the number of first
class constraints and which are left arbitrary by the
Hamilton-Dirac equations due to a dependence on the arbitrary
Dirac multipliers. The reduced phase space, a symplectic manifold
coordinatized by the Dirac observables, is the quotient of the
presymplectic sub-manifold by the foliation with gauge orbits.
Usually it is not a manifold, since rarely we have a nice
foliation with all the gauge orbits diffeomorphic. When $H_c
\equiv 0$, we get a frozen reduced phase space like in the
Hamilton-Jacobi description: the Dirac observables are Jacobi
data, an evolution can be reintroduced by using the energy
generator of the associated realization of the Poincare' group and
it can be shown to be consistent with the gauge motion induced by
$H_D$ in the gauge orbits.
\medskip

The {\it gauge fixing procedure} is a method to build copies of
the reduced phase space by adding as many gauge fixing constraints
$\chi_A \approx 0$ as first class constraints $\phi_A \approx 0$.
The surface determined by the equations $\chi_A \approx 0$ must
intersect each gauge orbit of the presymplectic sub-manifold in
one and only one point, namely the gauge variables must be
uniquely determined. Locally this is equivalent to the orbit
condition $det\, | \{ \chi_A, \phi_B \} | \not= 0$, but globally
it can be a very difficult topological problem to define a good
gauge fixing. The orbit condition means that the set of
constraints $\phi_A \approx 0$ and $\chi_A \approx 0$ is second
class and a symplectic structure isomorphic to that of the reduced
phase space is introduced by evaluating the {\it Dirac brackets}
associated to the gauge fixing. The detailed study of the Dirac
algorithm and of the Noether identities shows that all the
non-primary first class constraints $\phi_{\alpha} \approx 0$ lie
in {\it chains} whose progenitors are the primary first class
constraints $\phi_a \approx 0$ and with the $\lambda_a$'s as
arbitrary primary generalized velocity functions. This implies the
following cascade method as the natural way to introduce gauge
fixings. Given for instance a 3-chain $\phi_a \approx 0$
(primary), $\phi_{\alpha_1} \approx 0$ (secondary),
$\phi_{\alpha_2}\approx 0$ (tertiary), we first introduce a gauge
fixing $\chi_{\alpha_2} \approx 0$ to the tertiary constraint
$\phi_{\alpha_2} \approx 0$. Its preservation in time, $\{
\chi_{\alpha_2}, H_D \} \approx 0$, generates the gauge fixing
$\chi_{\alpha_1} \approx 0$ to the secondary constraint
$\phi_{\alpha_1} \approx 0$. Again $\{ \chi_{\alpha_1}, H_D \}
\approx 0$ generates the gauge fixing $\chi_a \approx 0$ to the
primary constraint $\phi_a \approx 0$ and $\{ \chi_a, H_D \}
\approx 0$ determines the Dirac multiplier $\lambda_a$.

\bigskip

See Section II of Ref.\cite{22} for a detailed description, based
on Refs. \cite{12,23}, of the theory of singular Lagrangians, of
Hamiltonian constraints and of the equivalence of the two
approaches.

\bigskip

Let us now consider the main problem, namely the determination of
the Dirac observables and of the gauge variables, or at least of
how the original canonical variables depend on them.\medskip

A method to find how the original variables depend on the gauge
variables makes use of the {\it multitemporal equations}
\cite{24}, also called the generalized Lie equations in
Ref.\cite{25} for the case in which the constraints satisfy a
Poisson bracket Lie algebra. Since the phase space functions
$f_{\alpha}$ in front of the non-primary first class constraints
in $H_D$ have an arbitrariness induced by the Dirac multipliers
$\lambda_a$, we can define an extended Dirac Hamiltonian $H^{'}_D
= H_c + \sum_A\, \lambda_A\, \phi_A$ where they are replaced by
new Dirac multipliers. Let us consider a finite dimensional case
with canonical variables $q^i(t)$, $p_i(t)$ and with $\{ H_c,
\phi_A(q,p) \} = 0$, $\{ \phi_A(q,p), \phi_B(q,p) \} = C_{ABC}\,
\phi_C(q,p)$ with $C_{ABC}$ the structure constants of a Lie
algebra for the sake of simplicity. Let us define {\it generalized
times} by means of the equations $d\tau_A = \lambda_A(t)\, dt$ and
vector fields $Y_A = A_{AB}(\tau_C)\, {{\partial}\over {\partial
\tau_B}}$ satisfying $[Y_A, Y_B] = - C_{ABC}\, Y_C$. Let us
rewrite $q^i(t)$, $p_i(t)$ as $q^i(t, \tau_A)$, $p_i(t, \tau_A)$
and replace the Hamilton equations generated by $H^{'}_D$ with the
following coupled multitemporal equations [$F = F(q,p)$]: i)
${{\partial F(t, \tau_A)}\over {\partial t}} = \{ F, H_c \}$
(namely $H_c$ generates the deterministic $t$-evolution); ii)
$Y_A\, F(t, \tau_C) = \{ F, \phi_A \}$ (these equations generate
the gauge orbit through each point of the constraint
sub-manifold). The whole set of equations is integrable due to the
first-class property of the constraints. In the ideal case in
which the gauge foliation is nice, all the gauge orbits are
diffeomorphic and in the simplest case all of them are
diffeomorphic to the group manifold of a Lie group. In this ideal
case to rebuild a gauge orbit from one of its points one needs the
Lie equations associated with the given Lie group: the Hamiltonian
multitemporal equations are generalized Lie equations describing
all the gauge orbits simultaneously. In a generic case this
description holds only locally for a set of diffeomorphic orbits,
also in the case of systems invariant under diffeomorphisms (in
general, when the $C_{ABC}$ are phase space dependent structure
functions, the vector fields $Y_A$ have also a dependence upon the
canonical variables).
\bigskip

Even if we do not succeed to solve the multitemporal equations,
their importance is the connection with the remarkable class of
the  Shanmugadhasan canonical transformations \cite{23}, which
allow to separate the gauge variables from the Dirac observables
(the tool lacking to the configurational Lagrangian approach). In
the finite dimensional case general theorems \cite{26} connected
with the Lie theory of function groups \cite{27} ensure the
existence of local canonical transformations from the original
canonical variables $q^i$, $p_i$ restricted by the first class
constraints (assumed globally defined) $\phi_A(q,p)\approx 0$, to
canonical bases $P_A$, $Q_A$, $P_r$, $Q_r$, such that the
equations $P_A \approx 0$ locally define the same original
constraint sub-manifold (the $P_A$ are an Abelianization of the
first class constraints); the $Q_A$ are the adapted Abelian gauge
variables describing the gauge orbits (they are a realization of
the generalized times, $\tau_A = Q_A$ of the multitemporal
equations); the $Q_r$, $P_r$ are an adapted canonical basis of
Dirac observables. These canonical transformations are the basis
of the Hamiltonian definition of the Faddeev-Popov measure of the
path integral \cite{28} and give a trivialization of the BRS
construction of observables (the BRS method works when the first
class constraints may be Abelianized \cite{11}). Second class
constraints, when present, are also taken into account by the
Shanmugadhasan canonical transformation \cite{22}: they correspond
to pairs of weakly vanishing irrelevant canonical
variables.\medskip

Therefore the problem of the search of the Dirac observables
becomes the problem of finding Shanmugadhasan canonical
transformations. Often, especially in field theories, it is not
known how to express the gauge variables and the Dirac observables
in terms of the original canonical variables. But the study of the
finite Hamiltonian gauge transformations, namely of the
multitemporal equations, usually allows to find the inverse
canonical transformation (or at least its restriction to the
constraint sub-manifold) $q^i = q^i[Q_A, Q_r, P_r]$, $p_i =
p_i[Q_A, Q_r, P_r]$. This is enough to define the preferred gauge
fixings $Q_A \approx 0$, in which all the physical quantities can
be explicitly expressed in terms of the predictable Dirac
observables. If a system with constraints admits one (or more)
global Shanmugadhasan canonical transformations, one obtains one
(or more) privileged global gauges in which the physical Dirac
observables are globally defined and globally separated from the
gauge degrees of freedom. These privileged gauges (when they
exist) can be called {\it generalized Coulomb gauges}. When the
system under investigation has some global symmetry group, the
associated theory of the momentum map is a source of
globality.\bigskip

Let us add some remarks \cite{22} on the properties of the
physical gauge systems defined on flat Minkowski space-time
deriving from the existence of the {\it global Poincare'
symmetry}. In this case we must study the structure of the
constraint sub-manifold  from the point of view of the orbits of
the Poincare' group. If $p^{\mu}$ is the total momentum of the
system, the constraint manifold has to be divided in four strata
(some of them may be absent for certain systems) according to
whether $\sgn\, p^2 > 0$, $p^2=0$, $\sgn\, p^2 < 0$ or $p^{\mu}=0$
[the metric is $\sgn\, (+ - - -)$ with $\sgn = \pm 1$ according to
the particle physics or general relativity convention]. Due to the
different little groups of the various Poincare' orbits, the gauge
orbits of different sectors will not be diffeomorphic. Therefore
the constraint sub-manifold is a stratified manifold and the gauge
foliations of relativistic systems are nearly never nice, but
rather one has to do with singular foliations. For an acceptable
relativistic system the stratum $\sgn\, p^2 < 0$ has to be absent
to avoid tachyons. To study the strata $p^2=0$ and $p^{\mu}=0$ one
has to add these relations as extra constraints. For all the
strata the next step is to do a canonical transformation from the
original variables to a new set consisting of center-of-mass
variables $x^{\mu}$, $p^{\mu}$ and of variables relative to the
center of mass. Let us now consider the stratum $\sgn\, p^2 > 0$.
By using the standard Wigner boost $L^{\mu}_{\nu}(p, {\buildrel
\circ \over p})$ ($p^{\mu}=L^{\mu}_{\nu}(p,{\buildrel \circ \over
p}){\buildrel \circ \over p}^{\nu}$, ${\buildrel \circ \over
p}^{\mu}=\eta \sqrt {\sgn\, p^2} (1;\vec 0 )$, $\eta = sign\,
p^o$), one boosts the relative variables at rest. The new
variables are still canonical and the base is completed by
$p^{\mu}$ and by a new center-of-mass coordinate ${\tilde
x}^{\mu}$, differing from $x^{\mu}$ for spin terms. The variable
${\tilde x}^{\mu}$ has complicated covariance properties; instead
the new relative variables are either Poincare' scalars or Wigner
spin-1 vectors, transforming under the group O(3)(p) of the Wigner
rotations induced by the Lorentz transformations. A final
canonical transformation \cite{29}, leaving fixed the relative
variables, sends the center-of-mass coordinates ${\tilde
x}^{\mu}$, $p^{\mu}$ in the new set $p\cdot {\tilde x}/\eta \sqrt
{\sgn\, p^2}=p\cdot x/\eta \sqrt {\sgn\, p^2}$ (the time in the
rest frame), $\eta \sqrt {\sgn\, p^2}$ (the total mass), $\vec k
=\vec p /\eta \sqrt {\sgn\, p^2}$ (the spatial components of the
4-velocity $k^{\mu}= p^{\mu}/\eta \sqrt {\sgn\, p^2}$, $k^2=1$),
$\vec z=\eta \sqrt {\sgn\, p^2}( {\vec {\tilde x}}-{\tilde
x}^o\vec p/p^o)$. $\vec z$ is a non-covariant center-of-mass
canonical 3-coordinate multiplied by the total mass: it is the
classical analog of the Newton-Wigner position operator (like it,
$\vec z$ is covariant only under the little group O(3)(p) of the
time-like Poincar\'e orbits). This techniques are useful to find
Lorentz scalar Abelianizations of the first class constraints and
shows that the breaking of manifest Lorentz covariance is
restricted to the decoupled, physically irrelevant, center-of-mass
motion.
\bigskip

 In gauge field theories the situation is more complicated,
because the theorems ensuring the existence of the Shanmugadhasan
canonical transformation have not been extended to the
infinite-dimensional case. One of the reasons is that some of the
constraints can now be interpreted as elliptic equations and they
can have {\it zero modes}. Let us consider the stratum $\sgn\, p^2
> 0$ of free Yang-Mills theory as a prototype and its first class
constraints, given by the Gauss laws and by the vanishing of the
time components of the canonical momenta. The problem of the zero
modes will appear as a singularity structure of the gauge
foliation of the allowed strata, in particular of the stratum
$\sgn\, p^2 > 0$. This phenomenon was discovered in Ref.\cite{30}
by studying the space of solutions of Yang-Mills and Einstein
equations, which can be mapped onto the constraint manifold of
these theories in their Hamiltonian description. It turns out that
the space of solutions has a {\it cone over cone} structure of
singularities: if we have a line of solutions with a certain
number of gauge symmetries, in each point of this line there is a
cone of solutions with one less symmetry. In the Yang-Mills case
the {\it gauge symmetries} of a gauge potential are connected with
the generators of its stability group, i.e. with the subgroup of
those special gauge transformations which leave invariant that
gauge potential (this is the {\it Gribov ambiguity} for gauge
potentials; there is also a more general Gribov ambiguity for
field strengths, the {\it gauge copies} problem; see
Refs.\cite{31} for a review). The analog of gauge symmetries in
general relativity is the existence of {\it Killing vectors}
implying that the space-time has symmetries.

Since the Gauss laws are  generators of Hamiltonian gauge
transformations (and depend on the chosen gauge potential through
the covariant derivative), this means that for a gauge potential
with non trivial stability group those combinations of the Gauss
laws corresponding to the generators of the stability group cannot
be any more first class constraints, since they do not generate
effective gauge transformations but special symmetry
transformations. This problematic has still to be clarified, but
it seems that in this case these components of the Gauss laws
become {\it third class} constraints, which are not generators of
true gauge transformations. This new kind of constraints was
introduced in Refs.\cite{12,23} in the finite dimensional case as
a result of the study of some examples, in which the Jacobi
equations (the linearization of the Euler-Lagrange equations) are
singular, i.e. some of their solutions are not infinitesimal
deviations between two neighboring extremals of the Euler-Lagrange
equations. This interpretation seems to be confirmed by the fact
that the singularity structure discovered in Ref.\cite{30} follows
from the existence of singularities of the linearized Yang-Mills
and Einstein equations. These problems are part of the Gribov
ambiguity, which, as a consequence, induces an extremely
complicated stratification and also singularities in each
Poincar\'e stratum of the constraint sub-manifold.

Other possible sources of (not yet exploredd) singularities of the
gauge foliation of Yang-Mills theory in the stratum $\sgn\, p^2 >
0$ may be: i) different classes of gauge potentials identified by
different values of the field invariants; ii) the orbit structure
of the rest frame (or Thomas) spin $\vec S$, identified by the
Pauli-Lubanski Casimir $W^2=- \sgn\, p^2\, {\vec S}^2$ of the
Poincare' group.

The final outcome of this structure of singularities is that the
reduced phase-space, i.e. the space of the gauge orbits, is in
general a stratified manifold with singularities \cite{32}. In the
stratum $\sgn\, p^2 > 0$ of the Yang-Mills theory these
singularities survive the Wick rotation to the Euclidean
formulation and it is not clear how the ordinary path integral
approach and the associated BRS method can take them into account.
The search of a global canonical basis of Dirac observables for
each stratum of the space of the gauge orbits can give a
definition of the measure of the phase space path integral, but at
the price of a non polynomial Hamiltonian. Therefore, if it is not
possible to eliminate the Gribov ambiguity (assuming that it is
only a mathematical obstruction without any hidden physics), the
existence of global Dirac observables for Yang-Mills theory is
very problematic.\bigskip

See Ref.\cite{22} for the list of special relativistic systems,
from relativistic particle mechanics to the Nambu string and the
$SU(3) \times SU(2) \times U(1)$ standard model of elementary
particles \cite{33}, whose Dirac observables and Abelianized gauge
variables have been determined by means of Shanmugadhasan
canonical transformations.

\vfill\eject

\section{ Simultaneity and
Parametrized Minkowski Theories.}

In this Section we explore those aspects of special relativistic
systems which suitably modified are present also in general
relativity, whose formulation must be made in a way allowing a
deparametrization to special relativity.

\subsection{The Lesson of Relativistic Mechanics.}

Relativistic particle mechanics in presence of interactions with a
finite time delay  goes back \cite{13} to the Tetrode-Fokker
action principle , to the Feynman-Wheeler electrodynamics  and to
its generalization by Van Dam and Wigner. The particle world-lines
$q^{\mu}_i(\tau^i)$, $i=1,..,N,$ are parametrized with independent
affine parameters $\tau^i$ and these action principles are
invariant under separate reparametrizations of each world-line,
since this is geometrically possible even in presence of
interactions. Since the dynamical correlation among the points on
the particle's world-lines is not in general one-to-one in these
approaches, it was impossible to develop a Hamiltonian formulation
starting from the Euler-Lagrange integro-differential equations of
motion implied by the delay. The natural development of these
approaches was field theory, for instance the study of the coupled
system of relativistic charged particles plus the electro-magnetic
field.\medskip

These difficulties and the need of a description of relativistic
bound states led to the development of relativistic mechanics with
action-at-a-distance interactions described by suitable potentials
implying a one-to-one correlation among the world-lines. As
already said, geometrically each particle has its world-line
described by a four-vector (the {\it four-position}) $q_i^{\mu} =
q_i^{\mu}(\tau^i)$, $i=1,..,N$, parametrized with an independent
arbitrary affine scalar parameter $\tau^i$ \footnote{The standard
choice in the manifestly covariant approach with a
$4N$-dimensional configuration space is $\tau^1 = .. = \tau^N =
\tau$. Another possibility is the choice of proper times $\tau^i =
\tau^i_{PT}$.}.  By inverting $q_i^{o}(\tau^i)$ to get $\tau^i =
\tau^i(q_i^{o})$, we can identify the world-line in a
non-manifestly covariant way with ${\vec q}_i = {\vec
q}_i(q_i^{o})$: in this form they are named {\it predictive
coordinates}. The instant form amounts to put $q_1^{o} = .. =
q_N^{o} = x^o$ and to describe the world-lines with the functions
${\vec q}_i(x^o)$. Each one of these configuration variables has a
different associated notion of {\it velocity}: ${{d
q_i^{\mu}(\tau^i)}\over {d\tau^i}}$ (or ${{d q_i^{\mu}(\tau
)}\over {d \tau}}$) , ${{d {\vec q}_i(q_i^{o})}\over {d q_i^{o}}}$
({\it predictive velocities}), ${{d {\vec q}_i(x^o)}\over {d
x^o}}$, and of {\it acceleration}: ${{d^2 q_i^{\mu}(\tau^i)}\over
{(d\tau^i)^2}}$ (or ${{d^2 q_i^{\mu}(\tau )}\over {d \tau^2}}$) ,
${{d^2 {\vec q}_i(q_i^{o})}\over {(d q_i^{o})^2}}$ ({\it
predictive accelerations}), ${{d^2 {\vec q}_i(x^o)}\over {(d
x^o)^2}}$.\medskip

Bel's  {\it non-manifestly covariant  predictive mechanics}
\cite{34} is the attempt to describe relativistic mechanics with
$N$-time predictive equations of motion for the predictive
coordinates ${\vec q}_i(q_i^{o})$ in Newtonian form: $m_i\, {{d^2
{\vec q}_i(q_i^{o})}\over {(d q_i^{o})^2}}\, \cir \,$\break $
{\vec {\cal F}}_i(q_k^{o}, {\vec q}_k(q_k^{o}), {{d {\vec
q}_k(q_k^{o})}\over {d q_k^{o}}}) $. Since the left hand side of
these equations depends only on $q_i^{o}$, the {\it predictive
forces} must satisfy the {\it predictive conditions} ${{d {\vec
{\cal F}}_i}\over { d q_k^{o}}} = 0$ for $k \not= i$. Moreover
they must be invariant under space translations and behave like
space three-vectors under spatial rotations. Finally they must
satisfy the Currie-Hill equations \cite{35} (or {\it Currie-Hill
world-line conditions}), whose satisfaction implies that the
predictive positions ${\vec q}_i(x^o)$ behave under Lorentz boosts
like the spatial components of four-vectors. Bel \cite{36} proved
that these equations constitute the necessary and sufficient
conditions, which guarantee that the dynamics is Lorentz invariant
with respect to finite Lorentz transformations. However the
Currie-Hill equations are so non-linear that it is practically
impossible to find consistent predictive forces and develop this
point of view.

The first well posed Hamiltonian formulation of relativistic
mechanics was given by Dirac \cite{37} with the {\it instant,
front (or light) and point forms} of relativistic Hamiltonian
dynamics and the associated canonical realizations of the
Poincare' algebra. In the instant form the simultaneity
hyper-surfaces defining a parameter for the time evolution are
space-like hyper-planes $x^o = const.$, in the front form
hyper-planes $x^{-} = {1\over 2}\, (x^o - x^3) = const.$ tangent
to future light-cones, while in the point form the future branch
of a two-sheeted hyperboloid $x^2 > 0$. In a $6N$-dimensional
phase space for $N$ scalar particles the ten generators of the
Poincare' algebra are classified into {\it kinematical} generators
(the generators of the stability group of the simultaneity
hyper-surface) and {\it dynamical} generators (the only ones to be
modified with respect to the free case in presence of
interactions) according to the chosen concept of simultaneity.
While in the instant and point forms there are four dynamical
generators (in the former energy and boosts, in the latter the
four-momentum), the front form has only three of them. After the
pioneering work of Thomas, Bakamjian and Foldy \cite{38} on the
non-manifestly covariant Hamiltonian instant form, much work has
been done in elucidating the classical and quantum aspects of this
approach, which has a well defined non-relativistic limit and
$1/c$ expansions containing the deviations (potentials) from the
free case.

A big obstacle for the development of Hamiltonian models was the
{\it no-interaction theorem} of Currie, Jordan and Sudarshan
\cite{39} (see Refs.\cite{40} for reviews). Its original form was
formulated in the Hamiltonian Dirac instant form in the
$6N$-dimensional phase space $\Big( {\vec q}_i(x^o), {\vec
p}_i(x^o) \Big)$ of $N$ particles. The no-interaction theorem
states that the hypotheses i) the configuration variables ${\vec
q}_i(x^o)$ are canonical, i.e. $\{ {\vec q}_i(x^o), {\vec
q}_j(x^o) \} =0$; ii) the Lorentz boosts can be implemented as
canonical transformations (existence of a canonical realization of
the Poincare' group) and the ${\vec q}_i(x^o)$ are the space
components of four-vectors; iii) the system is non-singular (the
transformation from positions and velocity to canonical
coordinates is non singular; it is not assumed the existence of a
Lagrangian); imply {\it only free motion}. \medskip

As a consequence of the theorem, if we denote $x_i^{\mu}(\tau )$,
$p_{i\mu}(\tau )$ the canonical coordinates of the manifestly
covariant approach and ${\vec x}_i(x^o)$, ${\vec p}_i(x^o)$ their
equal time restriction in the instant form, we have ${\vec
x}_i(x^o) \not= {\vec q}_i(x^o)$ except for free motion. Let us
remark that, {\it since the manifestly covariant approach gives
the classical basis for the theory of covariant wave equations,
the four-coordinates $x_i^{\mu}(\tau )$ (and not the geometrical
four-positions $q_i^{\mu}(\tau )$) are the coordinates locally
minimally coupled to external fields}.

Many attempts were made to avoid this theorem by relaxing one of
its hypotheses or by renouncing to the concept of world-line. One
was the already quoted {\it non-manifestly covariant manifestly
predictive approach}. Then the {\it manifestly covariant,
non-manifestly predictive approach} was developed \cite{41}: in it
the world-lines are described by $N$ four-vectors
$q_i^{\mu}(\tau^i_{PT})$ parametrized with $N$ proper times
$\tau^i_{PT}$: $\Big( {{d q_i^{\mu}(\tau^i_{PT})}\over {d
\tau^i_{PT}}} \Big)^2\, =\, m^2_i$. The covariant equations of
motion are ${{d^2 q_i^{\mu}(\tau^i_{PT})}\over {(d
\tau^i_{PT})^2}}\, \cir \, \theta_i^{\mu}(q_k^{\nu}(\tau^k_{PT}),
{{d q_k^{\mu}(\tau^k_{PT})}\over {d \tau^k_{PT}}})$ with
$\theta_i^{\mu}\, {{d q_i^{\mu}}\over {d \tau^i_{PT}}}\, = 0$. In
this many-time formalism the predictive conditions are equivalent
to the existence of $N$ Abelian, Poincare\ invariant vector fields
(identified for the first time by Droz Vincent) ${\cal H}_i$, $[
{\cal H}_i, {\cal H}_j ] = 0$, such that ${\cal H}_i\,
q_j^{\mu}(\tau^j_{PT}) = \delta_{ij}\, {{d
q_i^{\mu}(\tau^i_{PT})}\over {d \tau^i_{PT}}}$, ${\cal H}_i\, {{d
q_j^{\mu}(\tau^j_{PT})}\over {d \tau^j_{PT}}}\, = \delta_{ij}\,
\theta_i^{\mu}$. The connection with the non-manifestly covariant
manifestly predictive approach is obtained by imposing
$q_1^{o}(\tau^1_{PT}) =...= q_N^{o}(\tau^N_{PT}) = x^o$ and by
means of the identification ${\vec q}_i(\tau^i_{PT}) = {\vec
q}_i(x^o)$. For the predictive forces one gets ${\cal F}_i^{ h}\,
= {1\over {m^2_i}}\, \Big( 1 - ({{d {\vec q}_i(x^o)}\over {d
x^o}})^2\Big)\, \Big( \delta^{hk} - {{d q_i^{h}(x^o)}\over { d
x^o}}\, {{d q_i^{k}(x^o)}\over {d x^o}}\Big)\, \theta_i^{k}$ with
the Currie-Hill conditions satisfied. \medskip

Finally independently Droz Vincent's many-time Hamiltonian
formalism \cite{42} (a refinement of the manifestly covariant
non-manifestly predictive approach; it is the origin of the
multi-temporal equations quoted in the previous  Section),
Todorov's quasi-potential approach to bound states \cite{43} and
Komar's study of toy models for general relativity \cite{44} {\it
converged towards manifestly covariant models based on singular
Lagrangians and/or Dirac-Bergmann theory of constraints}. Since
quite often the Lagrangian formulation is not known, a system of
$N$ relativistic scalar particles is usually described in a
manifestly covariant $8N$-dimensional phase space with coordinates
$\Big( x_i^{\mu}(\tau ), p_{i\mu}(\tau ) \Big)$ [$\{
x_i^{\mu}(\tau ), p_{j\nu}(\tau ) \} = - \delta_{ij} \,
\delta^{\mu}_{\nu}$, $\{ x_i^{\mu}(\tau ), x_j^{\nu}(\tau ) \} =
\{ p_{i\mu}(\tau ), p_{j\nu}(\tau ) \} = 0$], where $\tau$ is a
scalar evolution parameter. The description is independent from
the choice of $\tau$: the Lagrangian (even if usually not
explicitly known) is $\tau$-reparametrization invariant, while at
the Hamiltonian level the canonical Hamiltonian vanishes
identically, ${\bar H}_c \equiv 0$. Since the physical degrees of
freedom for $N$ scalar particles are $6N$, there are constraints,
which, in the case of $N$ free scalar particles of mass $m_i$, are
just the mass-shell conditions

\bea
 {\bar \phi}_i(q, p) &=& p^2_i - m^2_i \approx 0,\quad
i=1,..,N,\qquad \Rightarrow\,\,\,
 p^o_i \approx \pm \sqrt{m^2_i + {\vec p}_i^2},\nonumber \\
 && \{ {\bar \phi}_i(q, p), {\bar \phi}_j(q, p) \} = 0.
 \label{3.1.1}
 \eea

These constraints say that the {\it time} variables $x_i^{o}(\tau
)$ are the {\it gauge variables of a $\tau$-reparametrization
invariant theory with ${\bar H}_c \equiv 0$}. The Dirac
Hamiltonian is ${\bar H}_D = \sum_{i=1}^N\, \lambda^i(\tau )\,
{\bar \phi}_i$ if all the first class constraints are primary. In
the free case this is true because these constraints are implied
by the action principle \footnote{An alternative N-time
Hamiltonian description is the multi-temporal one of of the
previous Section, in which the N scalar time parameters $\tau^i$
are defined by $d\tau^i=\lambda^i(\tau ) d\tau$.}

\bea
 S &=& \int\, d\tau\, L,\qquad L = - \sum_{i=1}^N\,\, m_i\,
 \sqrt{{\dot x}_i^{2}(\tau )},\nonumber \\
 &&\Rightarrow\,\,\, p_{i\mu} = - {{\partial L}\over {\partial
 {\dot x}_i^{\mu}}} = m_i\, {{{\dot x}_i^{\mu}(\tau )}\over
 {\sqrt{{\dot x}_i^{ 2}(\tau )}}},\quad p^2_i - m^2_i \approx 0.
 \label{3.1.2}
 \eea

The Euler-Lagrange equations are $L_{i\mu} = {d\over {d\tau}}\,
{{m_i\, {\dot x}_i^{\mu}(\tau )}\over {\sqrt{{\dot x}_i^{ 2}(\tau
)}}}\, \cir 0$, $i=1,..,N$, while the Hamilton-Dirac equations are
${\dot x}_i^{\mu}(\tau ) \cir -2\, \lambda^i(\tau )\, p_i^{\mu}$,
${\dot p}_{i\mu} \cir 0$, $p^2_i - m^2_i \approx 0$. The final
constraint sub-manifold is the union of $2^N$ (for generic masses
$m_i$) disjoint sub-manifolds corresponding the choice of the
either positive- or negative-energy branch of each two-sheeted
mass-shell hyperboloid. Each branch is a non-compact sub-manifold
of phase space on which each particle has a well defined sign of
the energy and $2^N$ is a topological number (the zeroth homotopy
class of the constraint sub-manifold) \footnote{Let us remark that
when the particles are coupled to weak external fields the $2^N$
sub-manifolds are deformed but remain disjoint. But when the
strength of the external fields increases the various
sub-manifolds may intersect each other and this topological
discontinuity is the signal that we are entering in a
non-classical regime where quantum pair production becomes
relevant due to the disappearance of mass gaps.}.

\bigskip

Only in the case of two-body systems it is known how to introduce
interactions due to the DrozVincent-Todorov-Komar model
\cite{42,43,44} with an {\it arbitrary action-at-a-distance
interaction instantaneous in the rest frame}  described by the two
first class constraints ${\bar \phi}_i = p_i^2-m_i^2+
V(r^2_{\perp})\approx 0$, i=1,2, with
$r^{\mu}_{\perp}=(\eta^{\mu\nu}-p^{\mu}p^{\nu} /\sgn\,
p^2)r_{\nu}$, $r^{\mu}=x_1^{\mu}-x^{\mu}_2$,
$p_{\mu}=p_{1\mu}+p_{2\mu}$. For $N > 2$ a closed form of the $N$
first class constraints is not known explicitly (there is only an
existence proof): only versions of the model with explicit gauge
fixings, so that all the constraints except one are second class,
are known.

This model has been completely understood both at the classical
and quantum level \cite{29}. Its study led to the identification
of a class of canonical transformations (utilizing the standard
Wigner boost for time-like Poincar\'e orbits) which allowed to
understand how to define suitable center-of-mass and relative
variables (in particular a suitable relative energy is determined
by a combination of the two first class constraints, so that the
relative time variable is a gauge variable), how to find a
quasi-Shanmugadhasan canonical transformation adapted to the
constraint determining the relative energy, how to separate the
four, topologically disjoined, branches of the mass spectrum (it
is determined by the other independent combination of the
constraints; therefore, there is a distinct Shanmugadhasan
canonical transformation for each branch). At the quantum level it
was possible to find four physical scalar products, compatible
with both the resulting coupled wave equations (i.e. independent
from the relative and the absolute rest-frame times): they have
been found as generalization of the two existing scalar products
of the Klein-Gordon equation: all of them are non-local even in
the limiting free case and differ among themselves for the sign of
the norm of states on different mass-branches. This example shows
that the physical scalar product knows the functional form of the
constraints.
\medskip

The no-interaction theorem is initially avoided due to the
singular nature of the Lagrangian: there is a canonical
realization of the Poincare' group and the canonical coordinates
$x^{\mu}_i$ are four-vectors. However, when we restrict ourselves
to the constraint sub-manifold and look for canonical coordinates
adapted to it and to the Poincare' group, it  turns out that among
the final canonical coordinates will always appear the canonical
{\it non-covariant} center of mass ${\tilde x}^{\mu}$ of the
particle system. Therefore,  all these models have the following
properties: i) the canonical and predictive four-positions do not
coincide (except in the free case); ii) the decoupled canonical
center of mass is not covariant.
\medskip

These models with $N$ first class constraints have the following
interpretation. Since there are $N$ arbitrary Dirac multipliers
$\lambda^i(\tau )$ [the Dirac Hamiltonian is $H_D = \sum_{i=1}^N\,
\lambda^i(\tau )\, \phi_i(q,p)$], the solutions of the
Euler-Lagrange equations are $x_i^{\mu}(\tau )\, \cir\,
x_i^{\mu}[\lambda^1(\tau ),.., \lambda^N(\tau )] \, {\buildrel
{def}\over =}\, x_i^{\mu}(\tau^1, .., \tau^N)\, \not=\,
q_i^{\mu}(\tau^i)$. Only for free particles we get $x_i^{\mu}(\tau
)\, \cir\, x_i^{\mu}[\lambda^i(\tau )] \, {\buildrel {def}\over
=}$\break $ x_i^{\mu}(\tau^i)\, = \, q_i^{\mu}(\tau^i)$. Therefore
{\it in the interacting case the canonical coordinates cannot
coincide with the predictive ones except in the free case} as
required by the no-interaction theorem. Each {\it gauge-dependent}
canonical coordinate $x_i^{\mu}$ spans a $N$-dimensional
hyper-surface (parametrized by the multi-times $\tau^1$, .. ,
$\tau^N$), the Hamiltonian world-sheet, instead of a world-line
(parametrized by $\tau^i$). Only if we make $N-1$ pre-gauge
fixings $\lambda^i(\tau ) = \Lambda^i( \lambda (\tau ))$ with
given functions $\Lambda^i$ of only one arbitrary function $\tilde
\tau = \lambda (\tau )$, we can select {\it a well defined
world-line} ${\hat x}_i^{\mu}(\tilde \tau )$ (parametrized by
$\tilde \tau = \lambda (\tau )$) for each particle: {\it it will
be interpreted as the predictive world-line ${\hat
q}^{\mu}_i(\tilde \tau )$ of that gauge}. The $N-1$ pre-gauge
fixings can be replaced by $N-1$ real gauge fixing constraints
(implying them) ${\bar \chi}_a(q, p) \approx 0$, $a=1,..,N-1$,
which are interpreted as a statement on the $N-1$ gauge variables
{\it relative times} ($N-1$ independent combinations of the
variables $x_i^o(\tau ) - x_j^o(\tau )$). But this is {\it
equivalent to a choice of which one-to-one space-like correlation
among the particles we are going to use to describe the system of
$N$ interacting particles} and, therefore, of which kind of
triggering of the particles the inertial observer in the
laboratory is going to use to detect them. Since a change of
Hamiltonian gauge is equivalent to a (in general) {\it non-point}
canonical transformation in phase space, corresponding to a
Lie-Backlund transformation on the velocity space, to {\it each
gauge is associated a different configuration space} (to be
reached by inverse Legendre transfornation). All these
configuration spaces can be identified with different copies of
Minkowski space-time containing different set of world-lines
connected by the velocity-dependent Lie-Backlund transformations.
In this way we reproduce all the possible world-lines spanning the
world-sheet in the Hamiltonian description. We need a {\it
semantic} statement \cite{44} about which is the configuration
space which contains those specific world-lines which are {more
natural from a physical interpretational viewpoint}. This choice
is equivalent to select a {\it natural} (i.e. preferred for the
physical interpretation) set of gauge fixings and this, in turn,
selects a certain one-to-one space-like correlation to be
associated to the given type of interaction. The chosen
world-lines in Minkowski space-time  will be the {\it natural
predictive world-lines} according to the chosen interpretation.
The natural set of gauge fixings for nearly all the models till
now proposed is to choose the one-to-one space-like correlation
corresponding to {\it simultaneity in the (inertial) rest frame of
the isolated $N$-body system}: the $N-1$ natural gauge fixing
constraints are $N-1$ independent combinations of $p_{\mu}\, [
x^{\mu}_i(\tau ) - x^{\mu}_j(\tau )] \approx 0$, where $p^{\mu}$
is the conserved total four-momentum of the isolated $N$-body
system.\bigskip

To clarify the interpretation we need a quasi-Shanmugadhasan
canonical transformation adapted to those $N-1$ combinations  of
the $N$ first class constraints, whose $N-1$ gauge fixings select
the natural world-lines for the given interaction (usually those
whose points are in one-to-one correlation in the rest frame). In
the $N=2$ case (DVTK model) it is given by \cite{29}

\bea
\begin{minipage}[t]{1cm}
\begin{tabular}{|l|} \hline
$x^{\mu}_i$ \\ $p^{\mu}_i$ \\ \hline
\end{tabular}
\end{minipage} \ {\longrightarrow \hspace{.2cm}} \
\begin{minipage}[t]{2 cm}
\begin{tabular}{|ll|l|l|} \hline
T & $T_R$ & $\vec z$ & $\vec \rho$    \\  $\epsilon$ & ${\hat
\epsilon}_R\, (\approx 0)$ & $\vec k$ & $\vec \pi$
\\ \hline
\end{tabular}
\end{minipage},
 \label{3.1.21}
 \eea

\noindent where $T_R = p \cdot r/\epsilon$, $\epsilon =
\sqrt{p^2}$, $T = p \cdot x/\epsilon$ [$x^{\mu} = {1\over 2}\,
(x_1^{\mu} + x^{\mu}_2)$], $\epsilon_R = {1\over {2 \epsilon}}\,
({\bar \phi}_1 - {\bar \phi}_2) = {1\over 2}\, [p \cdot q -
{1\over 2}\, (m^2_1 - m^2_2)] \approx 0$ [$q^{\mu} = {1\over 2}\,
(p_1^{\mu} - p_2^{\mu})$]. While the 3-center-of-mass quantities
$\vec z$ and $\vec k$ have been defined at the end of the previous
Section, the relative 3-vectors $\rho^r = \epsilon^r_{\mu}(p)\,
r^{\mu}$, $\pi^r = \epsilon^r_{\mu}(p)\, q^{\mu}$ require the
columns of the standard Wigner boost for time-like Poincare'
orbits for their definition. The gauge variable conjugate to the
constraint $\epsilon_R \approx 0$ is the Lorentz-scalar relative
time $T_R$. As a consequence the natural gauge fixing, identifying
the natural world-lines of this model inside the Hamiltonian
world-sheet, is ${\bar \chi}_{-} = T_R \approx 0$, i.e.
instantaneous interaction in the rest frame. A different gauge
fixing would identify a different pair of world-lines in the
world-sheet: its being non-natural is also shown by the
Shanmugadhasan canonical transformation, which takes into account
the fact that the potential depends upon $r^{\mu}_{\perp}$, the
mutual separation in the rest frame.

The other combination ${\bar \phi}_{+} = {1\over 2}({\bar \phi}_1
+ {\bar \phi}_2) \approx 0$ of the two first class constraints is
an equation for the mass spectrum $\epsilon$ of the system. The
natural gauge fixing to it, ${\bar \chi}_{+} = T - \tau \approx
0$, is a choice of the Lorentz-scalar rest-frame time $T$ as the
parameter to be used for the overall evolution.

For $N$ particles there is a combination ${\bar \phi}_{+} \approx
0$ of the $N$ first class constraints determining the mass
spectrum and generating $\tau$-repametrizations of the overall
isolated system, and $N-1$ combinations ${\bar \phi}_a \approx 0$,
$a=1,..,N-1$, which do not generate reparametrizations but
Hamiltonian gauge transformations implying the gauge nature of
$N-1$ independent relative times, to be fixed with $N-1$ natural
gauge fixings ${\bar \chi}_a \approx 0$.\bigskip

Let us remark, that, as shown in Ref. \cite{45}, once we have done
the semantic choice natural for the given interactions, a set of
$N$ gauge fixing ${\bar \chi}_i \approx 0$ is {\it admissible} if:
i) they imply that the $N-1$ natural gauge fixings ${\bar \chi}_a
\approx 0$, identifying the natural world-lines in the Hamiltonian
world-sheet, allow to express all the Dirac multipliers
$\lambda_i(\tau )$ as Poincare-invariant functions of a unique
arbitrary multiplier $\lambda (\tau )$; ii) they imply a final
gauge fixing ${\bar \chi}_{+} \approx 0$ which is Lorentz
invariant (otherwise, like it happens with $x^o(\tau ) \approx ...
\approx x^o_N(\tau ) \approx \tau$, the selected world-lines are
not stable under Poincare' transformations and the whole
Hamiltonian world-sheet reappears).
\medskip

Since we have $\{ x^{\mu}_i, {\bar \phi}_j \} \not= 0$ for $i\not=
j$ (the constraints do not generate reparametrizations of a single
world-line) except in the free case, in general, as already said,
the canonical coordinates $x^{\mu}_i(\tau )$ do not coincide with
any set of predictive ones $q^{\mu}_i(\tau^i)$ (associated to the
various world-lines existing in the Hamiltonian world-sheet),
which, if expressed in terms of $x^{\mu}_i$, $p^{\mu}_i$, should
satisfy $\{ q^{\mu}_i, {\bar \phi}_j \} = 0$ for $i\not= j$.
However, if we add a set of admissible gauge fixings ${\bar
\chi}_i \approx 0$, we select well defined world-lines for which
we must have geometrically $x^{\mu}_i{|}_{{\bar \chi}_j=0}\,
\approx q^{\mu}_i$. However, if we go to the Dirac brackets
implied by ${\bar \phi}_i \approx 0$, ${\bar \chi}_i \approx 0$,
we find that in the $6N$-dimensional reduced phase space we get
$x^{\mu}_i(\tau ) \equiv q^{\mu}_i (\tau^i(\tau ))$ but with these
quantities not being any more four-vectors in accord with the
no-interaction theorem. Actually the Dirac brackets force us to
make Poincare' transformations with a fixed parameter $\tau \equiv
T$ and this violates the Hamiltonian world-line condition and also
the predictive Currie-Hill one: i) in the Hamiltonian approach to
implement manifest Poincare' covariance we need to add to each
Poincare' transformations (essentially to the boosts) a
compensating $\tau$-reparametrization which is forbidden in the
reduced phase space; ii) to implement the individual
$\tau^i$-reparametrization of each world-line we have to use the
predictive accelerations which correlate world-line points
simultaneous in a generic (in general non inertial) frame and not
only in the inertial rest frame as implied by the Dirac brackets.

\bigskip

The lesson of relativistic mechanics is the need of i) a good
choice of simultaneity adapted to the type of action-at-a-distance
interaction under investigation; ii) a set of relativistic
center-of-mass and relative canonical variables compatible with
this choice ($p \cdot r_a \approx 0$). Given a set of first class
constraints the main tool for solving these problems are the
Shanmugadhasan canonical transformations adapted to as many
constraints as possible.

\bigskip

This state of affairs becomes a necessity when we consider a
charged scalar particle interacting with the electro-magnetic
field. Starting from the traditional action principle $S = \int
d\tau\, \Big[-m\, \sqrt{\sgn\, {\dot x}^2(\tau )} + e\, {\dot
x}^{\mu}(\tau )\, A_{\mu}(x(\tau ))\Big] + {1\over 2} \int d^4z\,
F^{\mu\nu}(z^o, \vec z)\, F_{\mu\nu}(z^o, \vec z)$, we arrive at
the primary constraints $[p_{\mu} - e\, A_{\mu}(x(\tau ))]^2 - m^2
\approx 0$, $\Pi^o(z^o, \vec z) \approx 0$ in a phase space
spanned by $x^{\mu}(\tau )$, $p_{\mu}(\tau )$, $A_{\mu}(z^o, \vec
z)$, $\Pi^{\mu}(z^o, \vec z)$. Since there is no concept of {\it
equal time} valid both for the particle and the field, we do not
know how to define the Poisson bracket of these two constraints.

To solve this problem without explicitly breaking manifest Lorentz
covariance [for instance by imposing by hand $x^o(\tau ) = z^o$ as
a restriction on the affine parameter $\tau$], we need to revisit
the notion of simultaneity in special relativity and to introduce
the 3+1 point of view which is implied by the Tomonaga-Schwinger
approach to quantum field theory \cite{46} and is a prerequisite
to move to globally hyperbolic pseudo-Riemannian manifolds as it
is required by the Hamiltonian formulation of metric and tetrad
gravity.

\subsection{Simultaneity Notions in Special Relativity.}

Let us review some recent results \cite{21} induced by a
revisitation of the problem of simultaneity in special relativity.
\medskip

In absence of gravity the special relativistic description of
physical systems is done in Minkowski space-time $M^4$, a flat
pseudo-Riemannian 4-manifold with Lorentz signature. The {\it
relativity principle} states that the laws of physics are the same
in a special family of {\it rigid systems of reference}, the {\it
inertial systems}, in uniform translational motion one with
respect to the other, endowed with pseudo-Cartesian (Lorentzian)
4-coordinates where the metric has the form $\sgn\, (+ - - -)$. In
them the laws of physics are manifestly covariant under the
kinematical group of Poincare' transformations (constant
translations and Lorentz transformations) and velocity has no
absolute meaning.

\medskip

An {\it observer} is a time-like future oriented world-line
$\gamma$ in $M^4$; in Cartesian coordinates we have $\gamma : R
\mapsto M^4$, $\tau \mapsto x^{\mu}(\tau )$, $\sgn\, {\dot
x}^2(\tau ) > 0$ [${\dot x}^{\mu}(\tau ) = dx^{\mu}(\tau
)/d\tau$]. An {\it inertial observer} has a constant 4-velocity
${\dot x}^{\mu}(\tau ) = const.$, i.e. the world-line is a
straight line. An {\it instantaneous inertial observer} is any
point $P$ on $\gamma$ together with the unit time-like vector
$e^{\mu}_{(o)} = {\dot x}^{\mu}/ \sqrt{\sgn\, {\dot x}^2}$ tangent
to $\gamma$ at $P$. An {\it inertial system} $I_P$ with origin at
$P$ has $\gamma$ as time axis and three orthogonal space-like
straight lines orthogonal to $\gamma$ in $P$, with unit tangent
vectors $e^{\mu}_{(r)}$, $r=1,2,3,$ as space axes. It corresponds
to a congruence of inertial observers defined by the constant unit
vector field $e^{\mu}_{(o)}$.  The four orthonormal vectors
$e^{\mu}_{(\alpha )}$ are a tetrad at $P$ and each inertial
observer of the congruence is endowed with a standard atomic clock
\footnote{Usually, in the case of an isolated inertial observer,
the clock is assumed to measure the proper time of the observer.
However, let us remark that a priori the notion of proper time
requires the solution of the equations of motion generating the
observer world-line, something which is beyond the duties of an
experimentalist. As a consequence, the standard unit of time is a
{\it coordinate second} \cite{47} and not a proper time both in
special and general relativity. In the case of a congruence of
observers (the points of an extended laboratory) this notion of
coordinate time emphasizes the conventional nature of the
synchronization of the clocks of the observers. }. A {\it
reference frame} (or {\it system of reference} or {\it platform})
is a congruence of time-like world-lines, namely a unit vector
field $u^{\mu}(x)$ having these world-lines as integral curves.
\bigskip

The basic problem of relativity is the {\it absence of an absolute
notion of simultaneity}, so that the synchronization of distant
clocks, the definition of an instantaneous 3-space, the spatial
distance between events at space-like separation and the one-way
velocity of light are {\it frame-dependent} concepts. Regarding
light there are {\it two independent postulates} in special
relativity, which state that the {\it round-trip} (or {\it
two-way} \footnote{Its definition implies only one observer and
therefore only one clock.}) velocity of light is the same, $c$, in
every inertial system (the round-trip postulate) and {\it
isotropic} (the light postulate). Instead the {\it one-way}
velocity of light between two events depends on the definition of
synchronization of the two clocks on the two world-lines
containing the given events and may be i) non-isotropic, ii)
lesser or higher than $c$ \cite{48}. The standard theory of
measurements in special relativity is defined in inertial systems,
where the Cartesian 4-coordinates select the simultaneity surfaces
$x^o = c\, t = const.$ as the instantaneous 3-spaces $R^3$ inside
which all the clocks are synchronized with Einstein's convention
\footnote{It is based on the choice of the rays of light as
preferred tools to measure time and length. In a given inertial
system the clock $A$, associated to the time-like world-line
$\gamma_A$, emits a light signal at its time $x^o_{Ai}$,
corresponding to an event $Q_i$ on $\gamma_A$, towards the
time-like world-line $\gamma_B$ carrying the clock $B$. When the
signal arrives at a point $P$ on $\gamma_B$, it is reflected
towards $\gamma_A$, where it is detected at time $x^o_{Af}$,
corresponding to an event $Q_f$ on $\gamma_A$. Then the clock $B$
at the event $P$ on $\gamma_B$ is synchronized to the time $x^o_A
= {1\over 2}\, (x^o_{Ai} + x^o_{Af})$, corresponding to an event
$Q$ in between $Q_i$ and $Q_f$. It can be checked that $Q$ and $P$
{\it lie on the same space-like hyper-plane orthogonal to the
world-line} $\gamma_A$, i.e. that they are simultaneous events for
the chosen inertial observer.} and spatial distances are defined
as Euclidean distances.

Modern metrology, with also its post-Newtonian extension to
general relativity, is reviewed in Refs.\cite{47}: after a
conventional definition of a standard of coordinate time, a
statement about the value of the round-trip velocity of light $c$
and the choice of Einstein's convention of simultaneity (valid on
$x^o = const.$ hyper-planes in inertial systems and replacing the
old slow transport of clocks) is made. Then the derived unit of
length (replacing the old {\it rods}) is defined either as $c\,
\triangle t / 2$, where $\triangle t$ is the round-trip time from
the location of the clock to another fixed location (where light
is reflected back) or as the wave-length of the radiation emitted
from the atomic clock.

\bigskip

After these preliminaries, let us remark that the notion of
inertial observer is an idealized limit concept: all actual
observers are accelerated. Since there is no relativity principle
concerning non-inertial observers, their interpretation of
experiments relies on the {\it hypothesis of locality} (see Refs
\cite{49,50,51}):  {\it an accelerated observer at each instant
along its world-line is physically equivalent to an otherwise
identical momentarily comoving inertial observer}, namely a
non-inertial observer passes through a continuous infinity of
hypothetical momentarily comoving inertial observers. While this
hypothesis is verified in Newtonian mechanics and in those
relativistic cases in which a phenomenon can be reduced to
point-like coincidences of classical point particles and light
rays (geometrical optic approximation), its validity is
questionable in presence of electro-magnetic waves (see
Refs.\cite{21,51}), when the wave-length $\lambda$ of the
radiation under scrutiny, emitted by an accelerated charge, is
comparable to the {\it acceleration length} ${\cal L}$ of the
observer \footnote{${\cal L} = {{c^2}\over a}$ for an observer
with translational acceleration $a$; ${\cal L} = {c\over
{\Omega}}$ for an observer rotating with frequency $\Omega$. The
hypothesis of locality is clearly valid in many Earth-based
experiments since $c^2 / g_{Earth} \approx 1\, lyr$, $c /
\Omega_{Earth} \approx 20\, AU$.}.

\bigskip

The fact that we can describe phenomena only locally near the
observer and that the actual observers are accelerated leads to
the {\it 1+3 point of view} (or threading splitting). Assuming we
know the world-line $\gamma$ of an accelerated observer, we must
try to define a notion of simultaneity and 4-coordinates (for
instance the Fermi normal ones) around the observer \footnote{The
observer is assumed endowed with a tetrad, whose time axis is the
unit 4-velocity and whose space axes are identified by three
orthogonal gyroscopes with a prescribed, but arbitrary,
prescription for their transport along the world-line (often the
Fermi-Walker transport is preferred due to the associated notion
of non-rotation) }. However, the knowledge of the observer
world-line only allows, in each point of $\gamma$, to split the
4-vectors on $\gamma$ in a part parallel to the observer
4-velocity (the tangent vector to $\gamma$) and in a part
orthogonal to it. The 3-dimensional orthogonal sub-spaces of the
tangent space $TM^4$ restricted to $\gamma$ are the {\it local
observer rest frames}: they are taken as a {\it substitute} of
instantaneous simultaneity 3-spaces, orthogonal to the world-line
$\gamma$, over which to define 3-coordinates. However, this is not
a good notion of simultaneous 3-spaces because these hyper-planes
intersect each other at a distance of the order of the
acceleration length ${\cal L}$ of the observer, invalidating the
global validity of the Fermi coordinates centered on accelerated
observers. Therefore, even if all the locally measured quantities
are coordinate-independent tetradic quantities referred to the
tetrads associated to the observer, it is not possible to write
equations of motion with a well defined Cauchy problem for these
tetradic quantities due to the lack of a good notion of
simultaneity. As a consequence, statements like the conservation
of energy cannot be demonstrated using only the 1+3 point of view.
\bigskip

While these problems are less serious in the case of linearly
accelerated observers, they become dramatic for rotating ones as
it is shown by the enormous number of papers dealing with the {\it
rotating disk} and the {\it Sagnac effect} (see the bibliography
of Refs.\cite{20,21}). Here we are concerned with congruences of
time-like observers [defined by a unit vector field $U^{\mu}(x)$]
which are not synchronizable due to the non-zero vorticity of the
congruence: this implies that simultaneity 3-spaces orthogonal to
all the world-lines of the observers do not exist and that it is
impossible to synchronize the clocks on the rotating disk with
Einstein's convention.
\bigskip

This state of affairs implies the necessity of considering the 1+3
point of view as embedded in the dual complementary {\it 3+1 point
of view}, in which the starting point is the preliminary
introduction of all the possible 3+1 splittings of Minkowski
space-time with foliations whose leaves are arbitrary space-like
hyper-surfaces and not only space-like hyper-planes. Each of these
hyper-surfaces is both a {\it simultaneity surface}, i.e. a
conventional notion of synchronization of distant clocks, and a
{\it Cauchy surface} for the equations of motion of the
relativistic system of interest. Each 3+1 splitting has well
defined notions of spatial length and of one-way velocity of
light.

\medskip

The 3+1 point of view is less physical (it is impossible to
control the initial data on a non-compact space-like Cauchy
surface), but it is the only known way to establish a well posed
Cauchy problem for the dynamics, so to be able to use the
mathematical theorems on the existence and uniqueness of the
solutions of field equations for identifying the predictability of
the theory. A posteriori, a non-inertial observer can try to
separate the part of the dynamics, implied by these solutions,
which is determined at each instant from the (assumed known)
information coming from its causal past from the part coming from
the rest of the universe.

\bigskip

To implement this program we have to come back to M$\o$ller's
formalization \cite{50} (Chapter VIII, Section 88) of the notion
of simultaneity. Given a relativistic  inertial system ${\cal K}$
with Cartesian 4-coordinates $x^{\mu}$ in Minkowski space-time and
with the $x^o = const.$ simultaneity hyper-planes, M$\o$ller
defines the {\it admissible coordinates transformations}
$x^{\mu}\, \mapsto\, y^{\mu} = f^{\mu}(x)$ [with inverse
transformation $y^{\mu}\, \mapsto\, x^{\mu} = h^{\mu}(y)$] as
those transformations whose associated metric tensor
$g_{\mu\nu}(y) = {{\partial h^{\alpha}(y)}\over {\partial
y^{\mu}}}\, {{\partial h^{\beta}(y)}\over {\partial y^{\nu}}}\,
\eta_{\alpha\beta}$ satisfies the following conditions

\bea
 && \sgn\, g_{oo}(y) > 0,\qquad
  \sgn\, g_{ii}(y) < 0,\qquad \begin{array}{|ll|} g_{ii}(y)
 & g_{ij}(y) \\ g_{ji}(y) & g_{jj}(y) \end{array}\, > 0, \nonumber
 \\
 &&\sgn\, det\, [g_{ij}(y)]\, < 0,\qquad
 \Rightarrow det\, [g_{\mu\nu}(y)]\, < 0.
 \label{II1}
 \eea

These are the necessary and sufficient conditions for having
${{\partial h^{\mu}(y)}\over {\partial y^o}}$ behaving as the
velocity field of a relativistic fluid, whose integral curves, the
fluid flux lines, are the world-lines of time-like observers.
Eqs.(\ref{II1}) say:

i) the observers are time-like because $\sgn g_{oo} > 0$;

ii) that the hyper-surfaces $y^o = f^{o}(x) = const.$ are good
space-like simultaneity surfaces.

\medskip

Moreover we must ask that $g_{\mu\nu}(y)$ tends to a finite limit
at spatial infinity on each of the hyper-surfaces $y^o = f^{o}(x)
= const.$ If, like in the ADM canonical formulation of metric
gravity \cite{52,53}, we write $g_{oo} = \sgn\, (N^2 - g_{ij}\,
N^i\, N^j)$, $g_{oi} = g_{ij}\, N^j$ introducing the lapse ($N$)
and shift ($N^i$) functions, this requirement says that the lapse
function (i.e. the proper time interval between two nearby
simultaneity surfaces) and the shift functions (i.e. the
information about which points on two nearby simultaneity surfaces
are connected by the so-called {\it evolution vector field}
${{\partial h^{\mu}(y)}\over {\partial y^o}}$) do not diverge at
spatial infinity. This implies that at spatial infinity on each
simultaneity surface there is {\it no asymptotic either
translational or rotational acceleration} and this implies that
{\it the simultaneity surfaces must tend to space-like
hyper-planes at spatial infinity}.

\bigskip

Let us remark that admissible coordinate transformations $x^{\mu}
\mapsto y^{\mu} = f^{\mu}(x)$ constitute the most general
extension of the Poincare' transformations $x^{\mu} \mapsto
y^{\mu} = a^{\mu} + \Lambda^{\mu}{}_{\nu}\, x^{\nu}$ compatible
with special relativity. A special family of admissible
transformations are the {\it frame-preserving} ones: $x^o \mapsto
y^o = f^o(x^o, \vec x)$, $\vec x \mapsto \vec y = \vec f(\vec x)$.

\bigskip

It is then convenient to describe \cite{54,22,55} the simultaneity
surfaces of an admissible foliation (3+1 splitting of Minkowski
space-time) with {\it naturally adapted Lorentz scalar admissible
holonomic coordinates} $x^{\mu}\, \mapsto \sigma^A = (\tau ,\vec
\sigma ) = f^A(x)$ [with inverse $\sigma^A\, \mapsto\, x^{\mu} =
z^{\mu}(\sigma ) = z^{\mu}(\tau ,\vec \sigma )$] such that:

i) the scalar time coordinate $\tau$ labels the leaves
$\Sigma_{\tau}$ of the foliation ($\Sigma_{\tau} \approx R^3$);

ii) the scalar curvilinear 3-coordinates $\vec \sigma = \{
\sigma^r \}$ on each $\Sigma_{\tau}$ are defined with respect to
an arbitrary time-like centroid $x^{\mu}(\tau )$ chosen as their
origin;

iii) if $y^{\mu} = f^{\mu}(x)$ is any admissible coordinate
transformation describing the same foliation, i.e. if the leaves
$\Sigma_{\tau}$ are also described by $y^o = f^o(x) = const.$,
then, modulo reparametrizations, we must have $y^{\mu} =
f^{\mu}(z(\tau ,\vec \sigma )) = {\tilde f}^{\mu}(\tau ,\vec
\sigma ) = A^{\mu}{}_A\, \sigma^A$ with $A^o{}_{\tau} = const.$,
$A^o{}_r = 0$, so that we get $y^o = const.\, \tau$, $y^i =
A^i{}_A(\tau ,\vec \sigma )\, \sigma^A$. The $\tau$ and $\vec
\sigma$ adapted admissible coordinates may be called {\it
radar-like 4-coordinates} with respect to the arbitrary
non-inertial observer, whose world-line $x^{\mu}(\tau ) =
z^{\mu}(\tau ,\vec 0)$ is chosen as origin of the 3-coordinates:
since the 3-surfaces $\Sigma_{\tau}$ are {\it  not orthogonal} to
this world-line, the pathologies of the Fermi coordinates are
avoided.  Therefore these foliations describe possible notions of
simultaneity for the non-inertial observer $x^{\mu}(\tau )$.

\medskip

The use of these Lorentz-scalar adapted coordinates allows to make
statements depending only on the foliation but not on the
4-coordinates $y^{\mu}$ used for Minkowski space-time.

\bigskip

The simultaneity hyper-surfaces $\Sigma_{\tau}$ are described by
their embedding $x^{\mu} = z^{\mu}(\tau ,\vec \sigma )$ in
Minkowski space-time [$R^3\, \mapsto \, \Sigma_{\tau} \subset M^4
\approx R \times R^3$, $(\tau ,\vec \sigma ) \mapsto z^{\mu}(\tau
,\vec \sigma )$] and the induced metric is $g_{AB}(\tau ,\vec
\sigma ) = z^{\mu}_A(\tau ,\vec \sigma )\, z^{\nu}_B(\tau ,\vec
\sigma )\, \eta_{\mu\nu}$ with $z^{\mu}_A =
\partial z^{\mu} / \partial \sigma^A$ (they are flat tetrad fields
over Minkowski space-time). Since the vector fields
$z^{\mu}_r(\tau ,\vec \sigma )$ are tangent to the surfaces
$\Sigma_{\tau}$, the time-like vector field of normals
$l^{\mu}(\tau ,\vec \sigma )$ is proportional to
$\epsilon^{\mu}{}_{\alpha\beta\gamma}\, z^{\alpha}_1(\tau ,\vec
\sigma )\, z^{\beta}_2(\tau ,\vec \sigma )\, z^{\gamma}_3(\tau
,\vec \sigma )$. Instead the time-like evolution vector field is
$z^{\mu}_{\tau}(\tau ,\vec \sigma ) = N(\tau ,\vec \sigma )\,
l^{\mu}(\tau ,\vec \sigma ) + N^r(\tau ,\vec \sigma )\,
z^{\mu}_r(\tau ,\vec \sigma )$, so that we have $dz^{\mu}(\tau
,\vec \sigma ) = z^{\mu}_{\tau}(\tau ,\vec \sigma )\, d\tau +
z^{\mu}_r(\tau ,\vec \sigma )\, d\sigma^r = N(\tau ,\vec \sigma
)\, d\tau\, l^{\mu}(\tau ,\vec \sigma ) + (N^r(\tau ,\vec \sigma
)\, d\tau + d\sigma^r)\, z^{\mu}_r(\tau ,\vec \sigma )$.

Since the 3-surfaces $\Sigma_{\tau}$ are {\it equal time} 3-spaces
with all the clocks synchronized, the spatial distance between two
equal-time events will be $dl_{12} = \int^2_1 dl\,
\sqrt{{}^3g_{rs}(\tau ,\vec \sigma (l))\, {{d\sigma^r(l)}\over
{dl}}\, {{d\sigma^s(l)}\over {dl}}}\,\,$ [$\vec \sigma (l)$ is a
parametrization of the 3-geodesic $\gamma_{12}$ joining the two
events on $\Sigma_{\tau}$]. Moreover, by using test rays of light
we can define the {\it one-way} velocity of light between events
on different $\Sigma_{\tau}$'s.

\medskip

The main property of {\it each foliation with simultaneity
surfaces} associated to an admissible 4-coordinate transformation
is that the embedding of the leaves of the foliation automatically
determine two time-like vector fields and therefore {\it two
congruences of (in general) non-inertial time-like observers}:

i) The time-like vector field $l^{\mu}(\tau ,\vec \sigma )$ of the
normals to the simultaneity surfaces $\Sigma_{\tau}$ (by
construction surface-forming, i.e. irrotational), whose flux lines
are the world-lines of the so-called (in general non-inertial)
Eulerian observers. The simultaneity surfaces $\Sigma_{\tau}$ are
(in general non-flat) Riemannian 3-spaces in which the physical
system is visualized and in each point the tangent space to
$\Sigma_{\tau}$ is the local observer rest frame $R_{\tilde
l(\tau_{\gamma})}$ of the Eulerian observer through that point.
This 3+1 viewpoint is called {\it hyper-surface 3+1 splitting}.

ii) The time-like evolution vector field $z^{\mu}_{\tau}(\tau
,\vec \sigma ) / \sqrt{\sgn\, g_{\tau\tau}(\tau ,\vec \sigma ) }$,
which in general is not surface-forming (i.e. it has non-zero
vorticity like in the case of the rotating disk). The observers
associated to its flux lines have the local observer rest frames
$R_{\tilde u(\tau_{\gamma})}$ not tangent to $\Sigma_{\tau}$:
there is no intrinsic notion of instantaneous 3-space for these
observers (1+3 point of view or {\it threading splitting}) and no
visualization of the physical system in large. However these
observers can use the notion of simultaneity associated to the
embedding $z^{\mu}(\tau ,\vec \sigma )$, which determines their
4-velocity. This 3+1 viewpoint is called {\it slicing 3+1
splitting}. In the case of the uniformly rotating disk all the
existing rotating 4-coordiinate systems have a {\it coordinate
singularity} ($g_{oo}(y^o, \vec y) = 0$) where $\omega\, r = c$:
there the time-like observers of the congruence would become null
observers like on the horizon of a Schartzschild black hole and
this is not acceptable in absence of a horizon.

\bigskip

As shown in Ref.\cite{21} the 3+1 point of view allows to get the
following results:

i) To define the special class of foliations implementing the idea
behind the locality hypothesis, that a non-inertial observer is
equivalent to a continuous family of comoving inertial observers,
but with restrictions coming from the admissibility conditions.
The main  byproduct of these restrictions will be that there exist
admissible 4-coordinate transformations interpretable as {\it
rigid systems of reference with arbitrary translational
acceleration}. However there is {\it no admissible 4-coordinate
transformation corresponding to a rigid system of reference with
rotational motion}. When rotations are present, the admissible
4-coordinate transformations give rise to a continuum of local
systems of reference like it happens in general relativity ({\it
differential rotations}).

ii) The simplest foliation of the previous class, whose
simultaneity surfaces are space-like hyper-planes with
differentially rotating 3-coordinates is given by the embedding

\bea
  &&z^{\mu}(\tau ,\vec \sigma ) = x^{\mu}(\tau ) + \epsilon^{\mu}_r\,
R^r{}_s(\tau , \sigma )\, \sigma^s\,
 {\buildrel {def}\over =}\, x^{\mu}(\tau ) + b^{\mu}_r(\tau
 ,\sigma )\, \sigma^r,\nonumber \\
 &&{}\nonumber \\
 &&R^r{}_s(\tau ,\sigma ) {\rightarrow}_{\sigma \rightarrow
 \infty} \delta^r_s,\qquad \partial_A\, R^r{}_s(\tau
 ,\sigma )\, {\rightarrow}_{\sigma \rightarrow
 \infty}\, 0,\nonumber \\
 &&{}\nonumber \\
 &&b^{\mu}_s(\tau ,\sigma ) = \epsilon^{\mu}_r\, R^r{}_s(\tau
 ,\sigma )\, {\rightarrow}_{\sigma \rightarrow
 \infty}\, \epsilon^{\mu}_s,\quad [b^{\mu}_r\, \eta_{\mu\nu}\, b^{\nu}_s](\tau ,\sigma )
 = - \sgn\, \delta_{rs},\nonumber \\
 &&{}\nonumber \\
 &&R = R(\alpha ,\beta ,\gamma ),\quad with\, Euler\, angles\,
 satisfying\nonumber \\
 &&{}\nonumber \\
 && \alpha (\tau ,\sigma) =F(\sigma )\, \tilde \alpha (\tau
 ),\qquad
 \beta (\tau ,\sigma ) = F(\sigma )\, \tilde \beta (\tau
 ),\qquad
 \gamma (\tau ,\sigma )=F(\sigma )\, \tilde \gamma (\tau
 ),\nonumber \\
 &&{}\nonumber \\
  &&0< F(\sigma ) <
 {m\over {2\, K\, M_1\, \sigma}}\,(K-1)=\frac{1}{M\,\sigma},\qquad
 {{d F(\sigma )}\over {d \sigma }} \not= 0,\nonumber \\
 &&{}\nonumber \\
 \mbox{ or }\qquad&&| \partial_{\tau} \alpha (\tau ,\sigma )|,
  | \partial_{\tau} \beta (\tau ,\sigma )|,
   | \partial_{\tau} \gamma (\tau ,\sigma )| <
   {{m}\over {2\, K\, \sigma}}\,(K-1).
 \label{II2}
 \eea

iii) To solve the following inverse problem: given a time-like
unit vector field, i.e. a (in general not irrotational) congruence
of non-inertial observers like that associated with a rotating
disk, find an admissible foliation with simultaneity surfaces such
that $z^{\mu}_{\tau}(\tau ,\vec \sigma )$ is proportional to the
given vector field.

iv) To define an operational method, generalizing Einstein's
convention from inertial hyper-planes to arbitrary admissible
simultaneity surfaces. It can be used to build a grid of radar
4-coordinates to be used by a set of satellites of the
Global-Positioning-System type.

v) To give the 3+1 point of view on the rotating disk and the
Sagnac effect by using the embedding (\ref{II2}) as an example.
Now there are a foliation-dependent instantaneous 3-space and a
foliation-dependent 3-geometry for the rotating disk.

vi) To use the foliation (\ref{II2}) to describe Earth rotation in
the determination of the one-way time transfer for the propagation
of light from an Earth station to a satellite, with the
consequence that the ESA ACES mission on the synchronization of
clocks \cite{56a} can be re-interpreted as a measure of the
deviation of this admissible simultaneity convention from
Einstein's one.

vii) To use the description of electro-magnetism as a parametrized
Minkowski theory to arrive at Maxwell equations in non-inertial
frames.

\bigskip

In conclusion, the absence of an absolute simultaneity and of an
absolute notion of instantaneous 3-space, replaced by the absolute
chrono-geometrical structure of Minkowski space-time, forces every
time-like observer to choose an admissible 3+1 splitting of it to
formulate a theory of measurement and a well posed Cauchy problem
for the dynamics. As we have seen this can be done in many ways,
which generalize Einstein's convention for inertial observers (the
traditional foliation with the hyper-planes $x^o = const.$) by
relaxing the condition that the observer's world-line is
orthogonal to the {\it equal time} 3-spaces $\Sigma_{\tau}$. Then
the new problem is whether all these possible notions of
simultaneity lead to an equivalent description of phenomena. In
the next Subsection we introduce parametrized Minkowski theories,
in which this equivalence is realized as a {\it gauge equivalence}
in the sense of Dirac theory of constraints.

\subsection{Parametrized Minkowski Theories.}

The previous two Subsections justify the attempt \cite{54} (see
also Refs.\cite{55} and Appendix A of Ref.\cite{53}) to
reformulate every isolated system on arbitrary space-like {\it
equal time} 3-surfaces $\Sigma_{\tau}$ leaves of an admissible 3+1
splitting of Minkowski space-time, so to get a parametrized field
theory already in a form suited to the transition to general
relativity in its ADM canonical formulation. The starting point
was given by Dirac \cite{10} with his reformulation of classical
field theory on space-like hyper-surfaces foliating Minkowski
space-time $M^4$.
\medskip

If $z^{\mu}(\tau ,\vec \sigma )$ is the embedding of the leaves
$\Sigma_{\tau}$, particle world-lines $x^{\mu}_i(\tau ) =
z^{\mu}(\tau , {\vec \eta}_i(\tau ))$ are identified by scalar
3-coordinates ${\vec \eta}_i(\tau )$ labeling the intersection of
the world-lines with $\Sigma_{\tau}$ with respect to the centroid
$x^{\mu}(\tau ) = z^{\mu}(\tau , \vec 0)$ chosen as origin. This
solves the problem of the relative times (all the particles are at
the {\it same} scalar time $\tau$), forcing us to solve the
mass-shell constraints $p^2_i - m^2_i \approx 0$ (in the free
case) and to choose the sign of the energy $p_i^o = \pm\,
\sqrt{m^2_i + {\vec p}_i^2}$ of each particle. In this way we get
a $3N$-dimensional configuration space for the $N$-particle system
without mass-shell constraints. For a scalar field we replace the
traditional $\tilde \phi (x^o, \vec x)$ with the new field $\phi
(\tau ,\vec \sigma ) = \tilde \phi (z(\tau ,\vec \sigma ))$ which
{\it knows} the non-local information of the chosen simultaneity,
being a function of the adapted radar 4-coordinates. For the
electro-magnetic field we use $A_A(\tau ,\vec \sigma ) =
{{\partial z^{\mu}(\tau ,\vec \sigma )}\over {\partial
\sigma^A}}\, A_{\mu}(z(\tau ,\vec \sigma ))$ and so on. All the
new fields are Lorentz scalar, having only surface indices,
satisfy (if any) Lorentz scalar constraints and, due to the
absence of the mass-shell constraints, the problem quoted at the
end of the Subsection on relativistic mechanics disappears. This
treatment of the fields is the classical basis of the
Tomonaga-Schwinger quantum field theory \cite{46}.

\medskip

Then one rewrites the Lagrangian density of the given isolated
system in the form required by the minimal coupling to an external
gravitational field. Instead of considering the 4-metric as
describing a gravitational field, here one replaces the 4-metric
with the the induced metric $g_{ AB}[z] =z^{\mu }_{A}\eta_{\mu\nu
}z^{\nu }_{B}$ on $\Sigma_{\tau}$, which is a functional of
$z^{\mu }$ and considers the embedding  $z^{\mu }(\tau ,\vec
\sigma )$ as {\it new configurational fields}. In this way the
Lagrangian density becomes a functional of the embedding and of
the isolated system.

The action of the system $S = \int d\tau d^3\sigma\, {\cal L}(\tau
,\vec \sigma )$ [${\cal L}(\tau ,\vec \sigma ) = \pm m\,
\delta^3(\vec \sigma - {\vec \eta}(\tau ))\,
\sqrt{g_{\tau\tau}(\tau ,\vec \sigma ) + 2 g_{\tau r}(\tau ,\vec
\sigma  )\, \eta^r(\tau ) + r_{rs}(\tau ,\vec \sigma )\,
\eta^r(\tau )\, \eta^s(\tau )}$ for a scalar particle of mass $\pm
m$] is invariant under frame-preserving reparametrizations: $\tau
\mapsto \tau^{'} (\tau ,\vec \sigma )$ and $\vec \sigma \mapsto
{\vec \sigma}^{'}(\vec \sigma )$. This is a {\it non-trivial
special relativistic type of general covariance} implying that the
embedding configuration variables $z^{\mu}(\tau ,\vec \sigma )$
are {\it gauge variables}, so that  {\it the physical results
about the system do not depend on the choice of the notion of
simultaneity}. Therefore, in parametrized Minkowski theories the
{\it conventionalism of simultaneity} is rephrased as a {\it gauge
problem}, i.e. as the {\it arbitrary choice of a gauge fixing
selecting a well defined notion of simultaneity among those
allowed by the gauge freedom}.
\bigskip

From this Lagrangian, besides a Lorentz-scalar form of the
constraints of the given system, we get four extra primary first
class constraints

\bea
 {\cal H} _{\mu }(\tau ,\vec \sigma )&=&\rho_{\mu }(\tau ,\vec \sigma
)-l_{\mu }(\tau ,\vec \sigma )T_{sys}^{\tau\tau}(\tau ,\vec \sigma
)-z_{r \mu }(\tau ,\vec \sigma )T_{sys}^{\tau r}(\tau ,\vec \sigma
) \approx 0,\nonumber \\
 &&\lbrace {\cal H}_{\mu }(\tau ,\vec \sigma
), {\cal H}_{\nu }(\tau ,{\vec \sigma}^{'}) \rbrace =0.
 \label{III6}
 \eea

\noindent Here $T_{sys}^{\tau\tau}(\tau ,\vec \sigma )$, $T_{sys}
^{\tau r}(\tau ,\vec \sigma )$, are the components of the
energy-momentum tensor in the holonomic coordinate system,
corresponding to the energy- and momentum-density of the isolated
system, $\rho_{\mu}(\tau ,\vec \sigma )$ is the canonical momentum
conjugate to $z^{\mu}(\tau ,\vec \sigma )$ and $l^{\mu}(\tau ,\vec
\sigma )$ is the unit normal to $\Sigma_{\tau}$. These constraints
are the generators of Hamiltonian gauge transformations implying
the independence of the description from the choice of the 3+1
splitting, i.e. from the choice of the foliation with space-like
hyper-sufaces. The evolution vector is given by $z^{\mu
}_{\tau}=N_{[z](flat)}l^{\mu }+ N^{r}_{[z](flat)}z^{\mu }_{r}$ and
$N_{[z](flat)}(\tau ,\vec \sigma )$, $N_{[z](flat)}^r(\tau ,\vec
\sigma )$ are the flat lapse and shift functions defined through
the metric like in general relativity: however, now they are not
independent variables but functionals of $z^{\mu }(\tau ,\vec
\sigma )$.

The Dirac Hamiltonian contains the piece $\int d^3\sigma
\lambda^{\mu }(\tau ,\vec \sigma ){\cal H}_{\mu } (\tau ,\vec
\sigma )$ with $\lambda^{\mu }(\tau ,\vec \sigma )$ Dirac
multipliers. It is possible to rewrite the integrand in the form
[${}^3g^{rs}$ is the inverse of $g_{rs}$]

\bea
 \lambda_{\mu }(\tau ,\vec \sigma ){\cal H}^{\mu }(\tau ,\vec
\sigma )&=&[(\lambda_{\mu }l^{\mu }) (l_{\nu }{\cal H}^{\nu
})-(\lambda_{\mu }z^{\mu }_{r})({}^3g ^{rs} z_{s \nu }{\cal
H}^{\nu })](\tau ,\vec \sigma ) =\nonumber \\
 &&{\buildrel {def}
\over =}\, N_{(flat)}(\tau ,\vec \sigma ) (l_{\mu }{\cal H}^{\mu
})(\tau ,\vec \sigma ) -N_{(flat) r}(\tau ,\vec \sigma )
({}^3g^{rs} z_{ s \nu } {\cal H}^{\nu })(\tau ,\vec \sigma ),
 \label{III7}
 \eea

\noindent with the (nonholonomic form of the) constraints $(l_{\mu
}{\cal H}^{\mu })(\tau ,\vec \sigma )\approx 0$, $({}^3g^{rs} z_{s
\mu } {\cal H}^{\mu })$\hfill\break $(\tau ,\vec \sigma )\approx
0$, satisfying the universal Dirac algebra of the ADM constraints
of canonical metric gravity. In this way we have defined new  flat
lapse and shift functions $N_{(flat)}(\tau ,\vec \sigma )=
\lambda_{\mu }(\tau ,\vec \sigma ) l^{\mu }(\tau ,\vec \sigma )$,
$N_{(flat) r}(\tau ,\vec \sigma )= \lambda_{\mu }(\tau ,\vec
\sigma ) z^{\mu }_{r}(\tau ,\vec \sigma )$, which have the same
content of the arbitrary Dirac multipliers $\lambda_{\mu }(\tau
,\vec \sigma )$, namely they multiply primary first class
constraints satisfying the Dirac algebra. In Minkowski space-time
they are quite distinct from the previous lapse and shift
functions $N_{[z](flat)}$, $N_{[z](flat) r}$, defined starting
from the metric.
\medskip

In special relativity, it is convenient to restrict ourselves to
arbitrary space-like hyper-planes $z^{\mu }(\tau ,\vec \sigma
)=x^{\mu }_s(\tau )+ b^{\mu }_{r}(\tau ) \sigma^{r}$. Since they
are described by only 10 variables, after this restriction we
remain only with 10 first class constraints determining the 10
variables conjugate to the hyperplane in terms of the variables of
the system:

\beq
 {\cal H}^{\mu }(\tau )=p^{\mu }_s-p^{\mu } _{(sys)}\approx
 0,\qquad
{\cal H}^{\mu\nu }(\tau )=S^{\mu\nu} _s-S^{\mu\nu }_{(sys)}\approx
0.
 \label{III8}
 \eeq

\noindent After the restriction to space-like hyper-planes the
previous piece of the Dirac Hamiltonian is reduced to ${\tilde
\lambda}^{\mu }(\tau ){\cal H}_{\mu }(\tau ) -{1\over 2}{\tilde
\lambda}^{\mu\nu }(\tau ){\cal H}_{\mu\nu }(\tau )$. Since at this
stage we have $z^{\mu }_{r}(\tau ,\vec \sigma )\approx b^{\mu
}_{r}(\tau )$, so that $z^{\mu } _{\tau}(\tau ,\vec \sigma
)\approx N_{[z](flat)}(\tau ,\vec \sigma )l ^{\mu }(\tau ,\vec
\sigma )+N^{r}_{[z](flat)}(\tau ,\vec \sigma )$ $b^{\mu }_{r}(\tau
,\vec \sigma )\approx {\dot x}^{\mu }_s(\tau )+{\dot b}^{\mu
}_{r}(\tau ) \sigma^{r}=-{\tilde \lambda}^{\mu }(\tau )-{\tilde
\lambda}^{\mu\nu }(\tau )b_{r \nu }(\tau )\sigma^{r}$, it is only
now that we get the coincidence of the two definitions of flat
lapse and shift functions:

\bea
 &&N_{[z](flat)}(\tau ,\vec \sigma
)\approx N_{(flat)}(\tau ,\vec \sigma )= -{\tilde \lambda} _{\mu
}(\tau )l^{\mu }-l^{\mu }{\tilde \lambda}_{\mu\nu }(\tau )b ^{\nu
}_{s}(\tau ) \sigma^{s},\nonumber \\
 &&N_{[z](flat)r}(\tau ,\vec
\sigma )\approx N_{(flat )}(\tau ,\vec \sigma )=-{\tilde \lambda}
_{\mu }(\tau )b^{\mu }_{r}(\tau )-b^{\mu }_{r}(\tau ){\tilde
\lambda}_{\mu\nu }(\tau ) b^{\nu }_{s}(\tau ) \sigma^{s}.
 \label{III9}
 \eea

\noindent The 20 variables for the phase space description of a
hyperplane are: \hfill\break

i) $x^{\mu }_s(\tau ), p^{\mu }_s$, [or $T_s = p_s \cdot {\tilde
x}_s/\epsilon_s$, $\epsilon_s = \sqrt{\sgn\, p^2_s}$, ${\vec
z}_s$, ${\vec k}_s$] parametrizing the origin of the coordinates
on the family of space-like hyper-planes. The four constraints
${\cal H}^{\mu }(\tau ) \approx 0$ say that $p_s^{\mu }$ is
determined by the 4-momentum of the isolated system.\hfill\break

ii) $b^{\mu }_A(\tau )$ (with the $b^{\mu }_r(\tau )$'s being
three orthogonal space-like unit vectors generating the time-like
unit normal $b^{\mu }_{\tau}(\tau )=l^{\mu }(\tau )$ to the
hyper-planes) and a spin tensor $S^{\mu\nu }_s = -S^{\nu\mu }_s$
with the orthonormality constraints $b^{\mu }_A\, {}^4\eta_{\mu\nu
} b^{\nu }_B={}^4\eta_{AB}$ \footnote{Enforced by assuming the
Dirac brackets $\{ S^{\mu\nu }_s,b^{\rho }_A \}={}^4\eta^{\rho\nu
} b^{\mu }_A-{}^4\eta^{\rho\mu } b^{\nu }_A$, $\{ S^{\mu\nu
}_s,S^{\alpha\beta }_s \} =C^{\mu\nu\alpha\beta }_{\gamma\delta }
S^{\gamma\delta }_s$ with $C^{\mu\nu\alpha\beta }_{\gamma\delta }$
the structure constants of the Lorentz algebra.}. In these
variables there are hidden six independent pairs of degrees of
freedom. The six constraints ${\cal H}^{\mu \nu } (\tau )\approx
0$ say that $S_s^{\mu\nu }$ coincides the spin tensor of the
isolated system. Then one has that $p^{\mu }_s$, $J^{\mu\nu }_s=x
^{\mu }_sp^{\nu }_s-x^{\nu }_sp^{\mu }_s+S^{\mu\nu }_s$, satisfy
the algebra of the Poincar\'e group. Finally the requirement of a
$\tau$-independent normal $l^{\mu}$  is equivalent to three more
gauge fixings,forbidding the action of Lorentz boosts, and
reducing to seven the surviving first class constraints. \medskip

Let us remark that after the restriction to these hyper-planes
with constant normal the only surviving congruence of time-like
observers (see Subsection B) is composed by inertial observers
having the unit normal as 4-velocity. Instead, with more general
admissible 3+1 splittings, we have two congruences of non-inertial
observers and in Ref.\cite{21} there  is a preliminary study of
their description of the dynamics using the electro-magnetic field
as an example.

\subsection{The Rest-Frame Wigner-Covariant Instant Form.}

Let us remark that, for each configuration of an isolated system
there is a privileged family of hyper-planes (the {\it Wigner
hyper-planes orthogonal to $p^{\mu }_s$}, existing when $\sgn\,
p^2_s > 0$), namely a {\it preferred notion of simultaneity},
corresponding to the {\it intrinsic rest-frame of the isolated
system} \cite{54}. If we choose these hyper-planes with suitable
gauge fixings, we remain with only  the four constraints ${\cal
H}^{\mu }(\tau )\approx 0$, which can be rewritten as

\bea
 \epsilon_s &\approx& [invariant\, mass\, of\, the\,
isolated\, system\, under\, investigation]= M_{sys}, \nonumber
\\
 {\vec p}_{sys} &=& [3-momentum\, of\, the\, isolated\, system\,
inside\, the\, Wigner\, hyperplane]\approx 0.\nonumber \\
 &&{}
 \label{III10}
 \eea

\noindent There is no more a restriction on $p_s^{\mu }$, because
$u^{\mu }_s(p_s)=p^{\mu }_s/\epsilon_s$ gives the orientation of
the Wigner hyper-planes containing the isolated system with
respect to an arbitrary given external inertial observer.

In this special gauge we have $b^{\mu }_A\equiv L^{\mu
}{}_A(p_s,{\buildrel \circ \over p}_s)$ (the standard Wigner boost
for time-like Poincar\'e orbits), $S_s^{\mu\nu }\equiv
S_{system}^{\mu\nu }$, and the only remaining canonical variables
are the non-covariant  canonical  coordinate ${\tilde x}^{\mu
}_s(\tau )$ (living on the Wigner hyper-planes) and $p^{\mu }_s$.
The embedding for the Wigner hyper-planes is $z^{\mu}(\tau ,\vec
\sigma ) = x^{\mu}(0) + b^{\mu}_A\, \sigma^A$. Now 3 degrees of
freedom of the isolated system (an {\it internal} center-of-mass
3-variable ${\vec \sigma}_{sys}$ defined inside the Wigner
hyperplane and conjugate to ${\vec p}_{sys}$) become gauge
variables \footnote{The natural gauge fixing is ${\vec
\sigma}_{sys}\approx 0$, so that it coincides with the centroid
$x^{\mu }_s(\tau )=z^{\mu }(\tau ,\vec \sigma =0)$ origin of the
3-coordinates on the Wigner hyper-plane.}, while the ${\tilde
x}^{\mu }$ is playing the role of a kinematical Newton-Wigner-like
{\it external} 4-center of mass for the isolated system and may be
interpreted as a decoupled observer with his parametrized clock
({\it point particle clock}). All the fields living on the Wigner
hyperplane are now either Lorentz scalar or with their 3-indices
transforming under Wigner rotations (induced by Lorentz
transformations in Minkowski space-time) as any Wigner spin 1
index.

One obtains in this way a new kind of instant form of the dynamics
(see Ref.\cite{37}), the  {\it Wigner-covariant 1-time rest-frame
instant form} \cite{54} with a universal breaking of Lorentz
covariance. It is the special relativistic generalization of the
non-relativistic separation of the center of mass from the
relative motion [$H={{ {\vec P}^2}\over {2M}}+H_{rel}$]. The role
of the center of mass is taken by the Wigner hyperplane,
identified by the point ${\tilde x}^{\mu } (\tau )$ and by its
normal $p^{\mu }_s$. The invariant mass $M_{sys}$ of the system
replaces the non-relativistic  Hamiltonian $H_{rel}$ for the
relative degrees of freedom, after the addition of the
gauge-fixing $T_s-\tau \approx 0$  ($T_s = p_s \cdot  {\tilde x}_s
/ \epsilon_s = p_s \cdot x_s / \epsilon_s$) and generates the
evolution in this rest-frame time.

\medskip

The Wigner hyperplane with its natural Euclidean metric structure
offers a natural solution to the problem of boost for lattice
gauge theories and realizes explicitly the machian aspect of
dynamics that only relative motions are relevant.

\medskip

The isolated systems till now analyzed to get their rest-frame
Wigner-covariant generalized Coulomb gauges, i.e. the subset of
global Shanmugadhasan canonical bases, which, for each Poincar\'e
stratum, are also adapted to the geometry of the corresponding
Poincar\'e orbits with their little groups, are:

a) The system of N scalar particles with Grassmann electric
charges plus the electromagnetic field \cite{54}. The final
Dirac's observables are: i) the transverse radiation field
variables ${\vec A}_{\perp}$, ${\vec E}_{\perp}$; ii) the particle
canonical variables ${\vec \eta}_i(\tau )$, ${{\vec \kappa}}
_i(\tau )$, dressed with a Coulomb cloud. The physical Hamiltonian
contains the mutual instantaneous Coulomb potentials extracted
from field theory and there is a regularization of the Coulomb
self-energies due to the Grassmann character of the electric
charges $Q_i$ [$Q^2_i=0$, $Q_i\, Q_j = Q_j \, Q_i \not= 0$ for
$i\not= j$]. In Ref.\cite{56} there is the study of the
Lienard-Wiechert potentials and of Abraham-Lorentz-Dirac equations
in this rest-frame Coulomb gauge. In the semi-classical
approximation of Ref.\cite{55}, the electro-magnetic degrees of
freedom are re-expressed in terms of the particle variables by
means of the Lienard-Wiechert solution in the framework of the
rest-frame instant form. In this way it has been possible to
derive the exact semi-classical relativistic form of the
action-at-a-distance Darwin potential (or the Salpeter one for
spinning particles) in the reduced phase space of the particles.
Note that these potentials, till now deduced only from quantum
field theory through the Bethe-Salpeter equation, are independent
of the choice of the Green function in the Lienard-Wiechert
solution due to the semi-classical regularization.

Also the rest-frame 1-time relativistic statistical mechanics has
been developed \cite{54}.

b) The system of N scalar particles with Grassmann-valued color
charges plus the color SU(3) Yang-Mills field\cite{57}: it gives
the pseudoclassical description of the relativistic scalar-quark
model, deduced from the classical QCD Lagrangian and with the
color field present. The physical invariant mass of the system is
given in terms of the Dirac observables. From the reduced Hamilton
equations the second order equations of motion both for the
reduced transverse color field and the particles are extracted.
Then, one studies  the N=2 (meson) case. A special form of the
requirement of having only color singlets, suited for a
field-independent quark model, produces a {\it pseudoclassical
asymptotic freedom} and a regularization of the quark self-energy.
With these results one can covariantize the bosonic part of the
standard model given in Ref.\cite{33}.

c) The system of N spinning particles of definite energy
[$({1\over 2},0)$ or $(0,{1\over 2})$ representation of SL(2,C)]
with Grassmann electric charges plus the electromagnetic
field\cite{58} and that of a Grassmann-valued Dirac field plus the
electromagnetic field (the pseudoclassical basis of QED)
\cite{59}.

d) Relativistic perfect fluids have been reformulated \cite{60} as
parametrized Minkowski theories and their rest-frame instant form
is known.

\medskip

In conclusion all the fields (Klein-Gordon, Yang-Mills and Dirac)
appearing in the $SU(3) \times SU(2) \times U(1)$ model of
elementary particles have been reformulated as parametrized
Minkowski theories.

\subsection{Relativistic Kinematics and the M$\o$ller Radius.}

The formulations of relativistic mechanics with first class
constraints in a $8N$-dimensional phase space $x^{\mu}_i(\tau )$,
$p^{\mu}_i(\tau )$ (see Subsection A) led to discover a
quasi-Shanmugadhasan canonical transformation for the $N$-body
problem adapted to the $N-1$ constraints ${\bar \phi}_a \approx 0$
and having $N-1$ Lorentz-scalar relative times (generalizing $T_R
= p \cdot r/\epsilon$ of the $N=2$ case) as conjugate gauge
variables. This adapted basis \cite{62,54} also contains: i) a
pair $T = p \cdot \tilde x/\epsilon$, $\epsilon = \sqrt{\sgn\,
p^2}$ (with the mass-spectrum $\epsilon$ to be determined from the
constraint ${\bar \phi}_{+} \approx 0$); ii) a decoupled
non-covariant 3-center of mass $\vec z$ and its conjugate momentum
$\vec k$; iii) $N-1$ pairs of relative variables ${\vec \rho}_a$,
${\vec \pi}_a$ (Dirac observables gauge  invariant with respect to
the gauge transformations generated by the ${\bar \phi}_a$'s).
Since $H_c \equiv 0$, on each branch of the mass spectrum ${\vec
\rho}_a$, ${\vec \pi}_a$ can be replaced with an equal number of
constant Jacobi data representing the true Dirac observables of
the frozen picture associated with the reduced phase space of
these reparametrization invariant theories, by  means of a
branch-dependent Shanmugadhasan canonical transformation
\footnote{In the free case it can be done. In general it can be
done every time the dynamics if Liouville integrable. }.
\medskip

The understanding of this type of $N$-body kinematics was of help
in developing a new relativistic kinematics \cite{62} adapted to
the framework of parametrized Minkowski theories and in particular
to the rest-frame instant form of dynamics. For a positive-energy
$N$-body system we start with a $6N$-dimensional phase space
${\vec \eta}_i(\tau )$, ${\vec \kappa}_i(\tau )$ [$x^{\mu}_i(\tau
) = z^{\mu}(\tau ,{\vec \eta}_i(\tau ))$ and $p^{\mu}_i(\tau )$
($p^2_i = m^2_i$, $sign\, p^o_i = +$) are derived dependent
quantities] and we can study the canonical transformations to a
new canonical basis containing a canonical {\it internal} 3-center
of mass ${\vec \sigma}_{sys}$ conjugate to ${\vec p}_{sys} \approx
0$. The rest-frame instant form leads to a {\it doubling of
viewpoints and concepts}:
\medskip

1) The {\it external} viewpoint, taken by an arbitrary inertial
Lorentz observer, who describes the Wigner hyper-planes determined
by the time-like configurations of the isolated system. A change
of inertial observer by means of a Lorentz transformation rotates
the Wigner hyper-planes and induces a Wigner rotation of the
3-vectors inside each Wigner hyperplane. Every such hyperplane
inherits an induced {\it internal Euclidean structure} while an
{\it external} realization of the Poincar\'e group induces an {\it
internal} Euclidean action.

2) The {\it internal} viewpoint, taken by an observer inside the
Wigner hyper-planes. This viewpoint is associated to a unfaithful
{\it internal} realization of the Poincar\'e algebra: the total
{\it internal} 3-momentum of the isolated system vanishes due to
the rest-frame conditions. The {\it internal} energy and angular
momentum are the invariant mass $M_{sys}$ and the spin
$S^{rs}_{sys}$ (the angular momentum with respect to ${\tilde
x}^{\mu}_s(\tau )$) of the isolated system, respectively.
\medskip

The determination of ${\vec \sigma}_{sys}$ may be done with the
group theoretical methods of Ref.\cite{63}: given a realization on
the phase space of a given system of the ten Poincar\'e generators
one can build three 3-position variables only in terms of them. In
our case of a system on the Wigner hyperplane with ${\vec
p}_{sys}\approx 0$ and using the internal Poincare' algebra they
are: i) a canonical center of mass (the {\it internal} center of
mass ${\vec \sigma}_{sys}$); ii) a non-canonical M\o ller center
of energy ${\vec \sigma}^{(E)}_{sys}$; iii) a non-canonical
Fokker-Pryce center of inertia ${\vec \sigma}^{(FP)}_{sys}$. Due
to ${\vec p}_{sys}\approx 0$, we have ${\vec \sigma}_{sys} \approx
{\vec \sigma}^{(E)}_{sys} \approx {\vec \sigma}^{(FP)}_{sys}$. By
adding the gauge fixings ${\vec \sigma}_{sys}\approx 0$ one can
show that the origin $x_s^{\mu }(\tau )$ becomes  simultaneously
the Dixon center of mass of an extended object and both the Pirani
and Tulczyjew centroids. With similar methods, starting from the
external Poincare' algebra, one can construct three {\it external}
collective positions (all located on the Wigner hyper-plane): i)
the {\it external} canonical non-covariant center of mass ${\tilde
x}_s^{\mu }$; ii) the {\it external} non-canonical and
non-covariant M\o ller center of energy $R^{\mu }_s$; iii) the
{\it external} covariant non-canonical Fokker-Pryce center of
inertia $Y^{\mu }_s$ (when there are the gauge fixings ${\vec
\sigma}_{sys}\approx 0$ it also coincides with the centroid
$x^{\mu }_s$ origin of the 3-coordinates).

\medskip

In the gauge where $\epsilon_s \equiv M_{sys}$, $T_s \equiv \tau
$, the canonical basis ${\vec z}_s$, ${\vec k}_s$, ${\vec
\eta}_i$, ${\vec \kappa}_i$ is restricted by the three pairs of
second class constraints ${\vec \kappa}_{+}=\sum_{i=1}^N{\vec
\kappa}_i \approx 0$ (the rest-frame condition), ${\vec
\sigma}_{sys} \approx 0$, so that 6N canonical variables describe
the N particles like in the non-relativistic case. We still need a
canonical transformation ${\vec \eta}_i$, ${\vec \kappa}_i$ $\,\,
\mapsto \,\,$ ${\vec \sigma}_{sys} [\approx 0]$, ${\vec
\kappa}_{+} [\approx 0]$, ${\vec \rho}_a$, ${\vec \pi}_a$
[$a=1,..,N-1$] identifying a set of {\it relative canonical
variables}. The final 6N-dimensional canonical basis is ${\vec
z}_s$, ${\vec k}_s$, ${\vec \rho}_a$, ${\vec \pi}_a$. To get this
result  we need a highly non-linear (but point in the momenta)
canonical transformation\cite{62}, which can be obtained by
exploiting the Gartenhaus-Schwartz singular transformation
\cite{64}.
\medskip

At the end we obtain {\it the Hamiltonian for the relative motions
as a sum of N square roots}, each one containing a squared mass
and a quadratic form in the relative momenta, which goes into the
non-relativistic Hamiltonian for relative motions in the limit
$c\, \rightarrow \infty$.

\medskip

The N quadratic forms in the relative momenta appearing in the
relative Hamiltonian {\it cannot be simultaneously diagonalized}
and it can be shown that concepts like {\it reduced masses}, {\it
Jacobi normal relative coordinates} and {\it tensor of inertia}
cannot be extended to special relativity. Instead in the
non-relativistic N-body problem the fact that the non-relativistic
kinetic energy of the relative motions is a quadratic form in the
relative velocities allows the introduction of special sets of
relative coordinates, the {\it Jacobi normal relative coordinates}
that diagonalize the quadratic form and correspond to different
patterns of clustering of the centers of mass of the particles.
Each set of Jacobi normal relative coordinates organizes the N
particles into a {\it hierarchy of clusters}, in which each
cluster of two or more particles has a mass given by an eigenvalue
({\it reduced masses}) of the quadratic form; Jacobi normal
coordinates join the centers of mass of pairs of clusters.\medskip

Morever, the non-Abelian nature of the rotation symmetry group
whose associated Noether constants of motion (the conserved total
angular momentum) are not in involution, prevents the possibility
of a global separation of absolute rotations from the relative
motions, so that there is no global definition of absolute {\it
vibrations}. Consequently, an {\it isolated} deformable body can
undergo rotations by changing its own shape (see the examples of
the {\it falling cat} and of the {\it diver}). It was just to deal
with these problems that the theory of the orientation-shape SO(3)
principal bundle approach\cite{65} has been developed. Its
essential content is that any {\it static} (i.e.
velocity-independent) definition of {\it body frame} for a
deformable body must be interpreted as a gauge fixing in the
context of a SO(3) {\it gauge} theory. Both the laboratory and the
body frame angular velocities, as well as the orientational
variables of the static body frame, become thereby {\it
unobservable gauge} variables. This approach is associated with a
set of {\it point} canonical transformations, which allow to
define the body frame components of relative motions in a
velocity-independent way. \bigskip

Since in many physical applications (e.g. nuclear physics,
rotating stars,...) angular velocities are viewed as {\it
measurable} quantities, one would like to have an alternative
formulation complying with this requirement and possibly
generalizable to special relativity. This has been done in
Ref.\cite{66} starting from the canonical basis ${\vec \rho}_a$,
${\vec \pi}_a$. First of all, for $N \geq 3$, we have constructed
a class of {\it non-point} canonical transformations which allow
to build the so called {\it canonical spin bases}: they are
connected to the patterns of the possible {\it clusterings of the
spins} associated with relative motions. The definition of these
{\it spin bases} is independent of Jacobi normal relative
coordinates, just as the patterns of spin clustering are
independent of the patterns of center-of-mass Jacobi clustering.
We have  found two basic frames associated to each spin basis: the
{\it spin frame} and the {\it dynamical body frame}. Their
construction is guaranteed by the fact that in the relative phase
space, besides the natural existence of a Hamiltonian symmetry
{\it left} action of SO(3), it is possible to define as many
Hamiltonian non-symmetry  {\it right} actions of SO(3) as the
possible patterns of spin clustering. While for N=3 the unique
canonical spin basis coincides with a special class of global
cross sections of the trivial orientation-shape SO(3) principal
bundle, for $N \geq 4$ the existing {\it spin bases} and {\it
dynamical body frames} turn out to be unrelated to the local cross
sections of the {\it static} non-trivial orientation-shape SO(3)
principal bundle, and {\it evolve} in a dynamical way dictated by
the equations of motion. In this new formulation {\it both} the
orientation variables and the angular velocities become, by
construction, {\it measurable} quantities in each canonical spin
basis.\medskip

For each N, every allowed spin basis provides a physically
well-defined separation between {\it rotational} and {\it
vibrational} degrees of freedom. The non-Abelian nature of the
rotational symmetry implies that there is no unique separation of
{\it absolute rotations} and {\it relative motions}. The unique
{\it body frame} of rigid bodies is replaced here by a discrete
number of {\it evolving dynamical body frames} and of {\it spin
canonical bases}, both of which are grounded in patterns of spin
couplings, direct analog of the coupling  of quantum angular
momenta.
\bigskip

This study of relativistic kinematics for the N-body system has
been completed \cite{67} by evaluating the rest-frame Dixon
multipoles \cite{68} and then by analyzing the role of Dixon's
multipoles for open subsystems. The basic technical tool is the
standard definition of the {\it energy momentum tensor} of the N
positive-energy {\it free} particles on the Wigner hyperplane. On
the whole, it turns out that the Wigner hyperplane is the natural
framework for reorganizing a lot of kinematics connected with
multipoles. Only in this way, moreover, a concept like the {\it
barycentric tensor of inertia} can be introduced in special
relativity, specifically by means of the quadrupole
moments.\medskip

\bigskip

Finally let us remark that, as shown in Refs.\cite{54,31}, the
rest-frame instant form of dynamics automatically gives a physical
ultraviolet cutoff in the spirit of Dirac and Yukawa: it is the
{\it M$\o$ller radius} \cite{50} $\rho =\sqrt{- \sgn\,
W^2}/p^2=|\vec S|/\sqrt{\sgn\, p^2}$ ($W^2=-p^2{\vec S}^2$ is the
Pauli-Lubanski Casimir when $\sgn\, p^2 > 0$), namely the
classical intrinsic radius of the world-tube, around the covariant
non-canonical Fokker-Pryce center of inertia $Y^{\mu }$, inside
which the non-covariance of the canonical center of mass ${\tilde
x}^{\mu}$ is concentrated. At the quantum level $\rho$ becomes the
Compton wavelength of the isolated system multiplied its spin
eigenvalue $\sqrt{s(s+1)}$ , $\rho \mapsto \hat \rho =
\sqrt{s(s+1)} \hbar /M=\sqrt{s(s+1)} \lambda_M$ with
$M=\sqrt{\sgn\, p^2}$ the invariant mass and $\lambda_M=\hbar /M$
its Compton wavelength. Therefore, the criticism to classical
relativistic physics, based on quantum pair production, concerns
the testing of distances where, due to the Lorentz signature of
space-time, one has intrinsic classical covariance problems: it is
impossible to localize the canonical center of mass ${\tilde
x}^{\mu}$ adapted to the first class constraints of the system
(also named Pryce center of mass and having the same covariance of
the Newton-Wigner position operator) in a frame independent way.

Let us remember \cite{54} that $\rho$ is also a remnant in flat
Minkowski space-time of the energy conditions of general
relativity: since the M$\o$ller non-canonical, non-covariant
center of energy $R^{\mu }$has its non-covariance localized inside
the same world-tube with radius $\rho$ (it was discovered in this
way) \cite{50}, it turns out that for an extended relativistic
system with the material radius smaller of its intrinsic radius
$\rho$ one has: i) its peripheral rotation velocity can exceed the
velocity of light; ii) its classical energy density cannot be
positive definite everywhere in every frame.

Now, the real relevant point is that this ultraviolet cutoff
determined by $\rho$ exists also in Einstein's general relativity
(which is not power counting renormalizable) in the case of
asymptotically flat space-times, taking into account the
Poincar\'e Casimirs of its asymptotic ADM Poincar\'e charges (when
supertranslations are eliminated with suitable boundary
conditions).

Moreover, the extended Heisenberg relations  of string theory
\cite{69}, i.e. $\triangle x ={{\hbar}\over {\triangle
p}}+{{\triangle p}\over {T_{cs}}}= {{\hbar}\over {\triangle
p}}+{{\hbar \triangle p}\over {L^2_{cs}}}$ implying the lower
bound $\triangle x > L_{cs}=\sqrt{\hbar /T_{cs}}$ due to the
$y+1/y$ structure, have a counterpart in the quantization of the
M$\o$ller radius \cite{54}: if we ask that, also at the quantum
level, one cannot test the inside of the world-tube, we must ask
$\triangle x > \hat \rho$ which is the lower bound implied by the
modified uncertainty relation $\triangle x ={{\hbar}\over
{\triangle p}}+{{\hbar \triangle p}\over {{\hat \rho}^2}}$. This
could imply that the center-of-mass canonical non-covariant
3-coordinate $\vec z=\sqrt{\sgn\, p^2}({\vec {\tilde x}}-{{\vec
p}\over {p^o}}{\tilde x}^o)$ \cite{54} cannot become a
self-adjoint operator. See Hegerfeldt's theorems (quoted in
Refs.\cite{31,54}) and his interpretation pointing at the
impossibility of a good localization of relativistic particles
(experimentally one determines only a world-tube in space-time
emerging from the interaction region). Since the eigenfunctions of
the canonical center-of-mass operator are playing the role of the
wave function of the universe, one could also say that the
center-of-mass variable has not to be quantized, because it lies
on the classical macroscopic side of Copenhagen's interpretation
and, moreover, because, in the spirit of Mach's principle that
only relative motions can be observed, no one can observe it (it
is only used to define a decoupled {\it point particle clock}). On
the other hand, if one rejects the canonical non-covariant center
of mass in favor of the covariant non-canonical Fokker-Pryce
center of inertia $Y^{\mu}$, $\{ Y^{\mu},Y^{\nu} \} \not= 0$, one
could invoke the philosophy of quantum groups to quantize
$Y^{\mu}$ to get some kind of quantum plane for the center-of-mass
description. Let us remark that the quantization of the square
root Hamiltonian done in Ref.\cite{70} is consistent with this
problematic.

In conclusion, the best set of canonical coordinates adapted to
the constraints and to the geometry of Poincar\'e orbits in
Minkowski spacetime and naturally predisposed to the coupling to
canonical tetrad gravity has emerged for the electromagnetic, weak
and strong interactions with matter described either by fermion
fields or by relativistic particles with a definite sign of the
energy.

\vfill\eject

\section{A Model of ADM Metric and Tetrad Gravity.}

Let us now look at general relativity taking into account what has
been learned about special relativity in the previous Section. We
started  an attempt \cite{53,71,72} to revisit classic metric
gravity \cite{53} and its ADM Hamiltonian formulation \cite{52} to
see whether it is possible to define a model of general relativity
able to incorporate fields and particles and oriented to a
background-independent quantization. First of all to include
fermions it is natural to resolve the metric tensor in terms of
cotetrad fields \cite{71,72} [$g_{\mu\nu}(x) = E^{(\alpha
)}_{\mu}(x)\, \eta_{(\alpha )(\beta )}\, E^{(\beta )}_{\nu}(x)$;
$\eta_{(\alpha )(\beta )}$ is the flat Minkowski metric in
Cartesian coordinates] and to reinterpret the gravitational field
as a {\it theory of time-like observers endowed with tetrads},
whose dynamics is controlled by the ADM action thought as a
function of the cotetrad fields. The model of general relativity
we are going to describe gives an idealized description of an
isolated system like the solar system. It can be extended to
describe astrophysical systems like our galaxy, but has no
relevance for cosmology at this stage.

\subsection{Selection of a Class of Non-Compact Space-Times where
the Time Evolution is Ruled by the Weak ADM Energy.}

Since the standard model of elementary particles and its
extensions are a chapter of the theory of representations of the
Poincare' group on the non-compact Minkowski space-time and we
look for a Hamiltonian description, the mathematical
pseudo-Riemannian 4-manifold $M^4$ introduced to describe
space-time is assumed to be {\it non-compact} and {\it globally
hyperbolic}. This means that it admits 3+1 splittings with
foliations whose leaves are space-like Cauchy 3-surfaces assumed
diffeomorphic to $R^3$ (so that any two points on them are joined
by a unique 3-geodesic). As in special relativity these 3-surfaces
are also {\it simultaneity surfaces}, namely a convention for the
synchronization of clocks. Therefore, if $\tau$ is the
mathematical time labeling these 3-surfaces, $\Sigma_{\tau}$, and
$\vec \sigma$ are 3-coordinates (with respect to an arbitrary
observer, a centroid $x^{\mu}(\tau )$, chosen as origin) on them,
then $\sigma^A = (\tau ,\vec \sigma )$ can be interpreted as
Lorentz-scalar radar 4-coordinates and the surfaces
$\Sigma_{\tau}$ are described by embedding functions $x^{\mu} =
z^{\mu}(\tau ,\vec \sigma )$. In these coordinates the metric is
$g_{AB}(\tau ,\vec \sigma ) = z^{\mu}_A(\tau ,\vec \sigma )\,
g_{\mu\nu}(z(\tau ,\vec \sigma ))\, z^{\nu}_B(\tau ,\vec \sigma )$
[$g_{AB}(\tau ,\vec \sigma ) = E^{(\alpha )}_A(\tau ,\vec \sigma
)\, \eta_{(\alpha )(\beta )}\, E^{(\beta )}_B(\tau ,\vec \sigma )$
in tetrad gravity]: differently from special relativity the
$z^{\mu}_A(\tau ,\vec \sigma )$ are not tetrad fields but only
transition coefficients to (radar) 4-coordinates $\sigma^A$
adapted to the 3+1 splitting. While in parametrized Minkowski
theories the embedding $z^{\mu}(\tau ,\vec \sigma )$ are the
Lagrangian configuration variables, now the commponents of the
4-metric tensor $g_{AB}(\tau ,\vec \sigma )$ [or the cotetrad
field $E^{(\alpha )}_A(\tau ,\vec \sigma )$] are the configuration
variables, while the allowed embeddings are  determined only a
posteriori after the solution of Einstein's equations. As in
special relativity, the Hamiltonian description has naturally
built in the tools (essentially the 3+1 splitting) to make contact
with experiments in a relativistic framework, where simultaneity
is a frame-dependent property. The manifestly covariant
description using Einstein's equations is the natural one for the
search of exact solutions, but is inadequate to describe
experiments.

\medskip

Other requirements \cite{53,72} on the Cauchy and simultaneity
3-surfaces $\Sigma_{\tau}$ induced by particle physics are:

i) Each $\Sigma_{\tau}$ must be a Lichnerowitz 3-manifold
\cite{73}, namely it must admit an involution so that a
generalized Fourier transform can be defined and the notion of
positive and negative frequencies can be introduced (otherwise the
notion of particle is missing like it happens in quantum field
theory in arbitrary curved space-times \cite{2}).

ii) Both the cotetrad fields (and the metric tensor) and the
fields of the standard model of elementary particles must belong
to the {\it same family of suitable weighted Sobolev spaces} so
that simultaneously there are no Killing vector fields on the
space-time (this avoids the cone-over-cone structure of
singularities in the space of metrics) and no Gribov ambiguity
(either gauge symmetries or gauge copies \cite{31}) in the
particle sectors; in both cases no well defined Hamiltonian
description is available.

iii) The space-time must be {\it asymptotically flat at spatial
infinity} and with boundary conditions there attained in a way
independent from the direction (like it is needed to define the
non-Abelian charges in Yang-Mills theory \cite{31}). This
eliminates the {\it supertranslations} (the obstruction to define
angular momentum in general relativity) and reduces the {\it spi
group} of asymptotic symmetries to the {\it ADM Poincare' group}
\footnote{This group has 10 generators (Noether constants) given
in the form of either 10 {\it strong} Poincare' charges (defined
as a flux through the surface at spatial infinity) or 10 {\it
weak} Poincare' charges (defined as volume integrals over
$\Sigma_{\tau}$) differing from the strong ones by integrals
 of the secondary  constraints.}. The constant ADM Poincare'
generators should become the standard conserved Poincare'
generators of the standard model of elementary particles when
gravity is turned off and the space-time (modulo a possible
renormalization of the ADM energy to subtract an infinite term
coming from its dependence on both $G$ and $1/G$) reduces to the
Minkowski one. As a consequence, as shown in Ref.\cite{53}, the
{\it admissible foliations} of the space-time must have the
simultaneity surfaces $\Sigma_{\tau}$ tending in a
direction-independent way to Minkowski space-like hyper-planes at
spatial infinity, where they must be orthogonal to the ADM
4-momentum \footnote{Incidentally, this is the first example of
consistent {\it deparametrization} of general relativity. In
presence of matter we get the description of matter in Minkowski
space-time foliated with the space-like hyper-planes orthogonal to
the total matter 4-momentum (Wigner hyper-planes intrinsically
defined by matter isolated system). Of course, in closed
space-times, the ADM Poincare' charges do not exist and the
special relativistic limit is lost.}. But, in absence of matter,
these are the conditions satisfied by the
Christodoulou-Klainermann space-times \cite{74} , which are near
Minkowski space-time in a norm sense and have a {\it rest-frame}
condition of zero ADM 3-momentum. Therefore the surfaces
$\Sigma_{\tau}$ define the {\it rest frame} of the $\tau$-slice of
the universe and in this model there are asymptotic inertial
observers to be identified with astronomers' {\it fixed stars}
(the standard origin of rotations to study the precession of
gyroscopes in space).

As a consequence in this class of space-times there is an {\it
asymptotic Minkowski metric} (asymptotic background), which allows
to define weak gravitational field configurations {\it without
splitting the metric} in a background one plus a perturbation and
without being a bimetric theory of gravity.

As shown in Ref.\cite{53} these properties are concretely enforced
by using a technique introduced by Dirac \cite{10} for the
selection of space-times admitting asymptotically flat
4-coordinates at spatial infinity. As a consequence the admissible
embeddings of the simultaneity leaves $\Sigma_{\tau}$ have the
following direction-independent limit at spatial infinity:
$z^{\mu}(\tau ,\vec \sigma ) = X^{\mu}(\tau ) + F^{\mu}(\tau ,\vec
\sigma ) \rightarrow_{|\vec \sigma | \rightarrow \infty}\,
X^{\mu}_{(\infty)}(0) + \epsilon^{\mu}_A\, \sigma^A =
X^{\mu}_{(\infty)}(\tau ) + \epsilon^{\mu}_r\, \sigma^r$. Here
$X^{\mu}_{(\infty)}(\tau ) = X^{\mu}_{(\infty)}(0) +
\epsilon^{\mu}_{\tau}\, \tau$ is just the world-line of an
asymptotic inertial observer having $\tau$ as proper time and
$\epsilon^{\mu}_A$ denotes an asymptotic constant tetrad with
$\epsilon^{\mu}_{\tau}$ parallel to the ADM 4-momentum (it is
orthogonal to the asymptotic space-like hyper-planes). Such
inertial observers corresponding to the {\it fixed stars} can be
endowed with a spatial triad ${}^3e^r_{(a)} = \delta^r_{(a)}$,
$a=1,2,3$. Then the asymptotic spatial triad ${}^3e^r_{(a)}$ can
be transported in a dynamical way (on-shell) by using the
Sen-Witten connection \cite{75} (it depends on the extrinsic
curvature of the $\Sigma_{\tau}$'s) in the Frauendiener
formulation \cite{76} in {\it every point} of $\Sigma_{\tau}$,
where it becomes a well defined triad ${}^3e_{(a)}^{(WSW) r}(\tau
,\vec \sigma )$. This defines a {\it local compass of inertia}, to
be compared with the local gyroscopes (whether Fermi-Walker
transported or not). The Wigner-Sen-Witten (WSW) local compass of
inertia consists in pointing to the fixed stars with a telescope.
It is needed in a satellite like Gravity Probe B to detect the
frame-dragging (or gravito-magnetic Lense-Thirring effect) of the
inertial frames by means of the rotation of a Fermi-Walker
transported gyroscope.

Finally from Eq.(12.8) of Ref.\cite{53} we get the set of partial
differential equations for the determination of the embedding
$x^{\mu} = z^{\mu}(\tau ,\vec \sigma )$ ($x^{\mu}$ is an arbitrary
4-coordinate system in which the asymptotic hyper-planes of the
$\Sigma_{\tau}$'s have $\epsilon^{\mu}_A$ as asymptotic tetrad):
$z^{\mu}(\tau ,\vec \sigma ) = X^{\mu}_{(\infty)}(0) + F^A(\tau
,\vec \sigma )\, {{\partial z^{\mu}(\tau ,\vec \sigma )}\over
{\partial \sigma^A}}$ with $F^{\tau}(\tau ,\vec \sigma ) =
{{-\epsilon\, \tau}\over {-\epsilon + n(\tau ,\vec \sigma )}}$ and
$F^r(\tau ,\vec \sigma ) = \sigma^r + [{}^3e_{(a)}^{(WSW) r}(\tau
,\vec \sigma )\, - \delta^r_{(a)} ]\, \delta_{(a)s}\, \sigma^s +
{{\epsilon\, n^r(\tau ,\vec \sigma )}\over {-\epsilon + n(\tau
,\vec \sigma )}}$.

\medskip

As shown in Ref.\cite{53}, a consistent treatment of the boundary
conditions at spatial infinity requires the explicit separation of
the {\it asymptotic} part of the lapse and shift functions from
their {\it bulk} part: $N(\tau ,\vec \sigma ) = N_{(as)}(\tau
,\vec \sigma ) + n(\tau , \vec \sigma )$, $N_r(\tau ,\vec \sigma )
= N_{(as)r}(\tau ,\vec \sigma ) + n_r(\tau , \vec \sigma )$, with
$n$ and $n_r$ tending to zero at spatial infinity in a
direction-independent way. On the contrary, $N_{(as)}(\tau ,\vec
\sigma ) = - \lambda_{\tau}(\tau ) - {1\over 2}\, \lambda_{\tau
u}(\tau )\, \sigma^u$ and $N_{(as)r}(\tau ,\vec \sigma ) = -
\lambda_{r}(\tau ) - {1\over 2}\, \lambda_{r u}(\tau )\,
\sigma^u$. In the Christodoulou-Klainermann space-times \cite{74}
we have $N_{(as)}(\tau ,\vec \sigma ) = \epsilon$, $N_{(as)
r}(\tau ,\vec \sigma ) = 0$.

\medskip

We start off with replacement of the ten components
${}^4g_{\mu\nu}$ of the 4-metric tensor by the configuration
variables of ADM canonical gravity: the {\it lapse} $N(\tau , \vec
\sigma ) = \sgn + n(\tau ,\vec \sigma )$ and {\it shift} $N_r(\tau
,\vec \sigma ) = n_r(\tau ,\vec \sigma )$ functions and the six
components of the {\it 3-metric tensor} on $\Sigma_{\tau}$,
${}^3g_{rs}(\tau , \vec \sigma )$. We have $ {}^4g_{AB}=\left(
\begin{array}{ll} {}^4g_{\tau\tau}= \epsilon
(N^2-{}^3g_{rs}N^rN^s)& {}^4g_{\tau s}=- \epsilon \, {}^3g_{su}N^u
\\ {}^4g_{\tau r}=- \epsilon \, {}^3g_{rv}N^v&
{}^4g_{rs}=-\epsilon \, {}^3g_{rs} \end{array} \right)$.
Einstein's equations are then recovered as the Euler-Lagrange
equations of the ADM action. Besides the ten configuration
variables listed above, the ADM functional phase space is {\it
coordinatized} by ten canonical momenta ${\tilde \pi}^n(\tau ,\vec
\sigma )$, ${\tilde \pi}^r_{\vec n}(\tau ,\vec \sigma )$,
${}^3{\tilde \Pi}^{rs}(\tau , \vec \sigma )$ and there are eight
{\it first class} constraints ${\tilde \pi}^n(\tau ,\vec \sigma )
\approx 0$ , ${\tilde \pi}^r_{\vec n}(\tau ,\vec \sigma ) \approx
0$, ${\tilde {\cal H}}(\tau ,\vec \sigma ) \approx 0$,
${}^3{\tilde {\cal H}}^r(\tau ,\vec \sigma  \approx 0$. While the
first four are {\it primary} constraints, the remaining four are
the super-hamiltonian and super-momentum  {\it secondary}
constraints. The behavior at spatial infinity, $r = |\vec \sigma |
\rightarrow \infty$, of the components of the 4-metric tensor and
of the cotriads is $n(\tau ,\vec \sigma )\, \rightarrow\,
O(r^{-(2+\epsilon )})$, $n_r(\tau ,\vec \sigma )\, \rightarrow\,
O(r^{-\epsilon })$, ${}^3g_{rs}(\tau ,\vec \sigma )\,
\rightarrow\, (1 + {M\over r})\, \delta_{rs} +O(r^{-3/2})$,
${}^3e_{(a)r}(\tau ,\vec \sigma )\, \rightarrow\, (1 + {M\over
{2r}})\, \delta_{(a)r} + O(r^{-3/2})$ ($\epsilon > 0$).
\bigskip

Instead in ADM tetrad gravity \cite{71,72} there are 16
configuration variables: the cotetrad fields can be parametrized
in terms of $n(\tau ,\vec \sigma )$, $n_r(\tau ,\vec \sigma )$, 3
boost parameters $\varphi_{(a)}(\tau ,\vec \sigma )$, 3 angles
$\alpha_{(a)}(\tau ,\vec \sigma )$ and cotriads $e_{(a)r}(\tau
,\vec \sigma )$ on $\Sigma_{\tau}$ [$a = 1,2,3$]. There are 14
first class constraints $\pi_n(\tau ,\vec \sigma ) \approx 0$,
$\pi_{n_r}(\tau ,\vec \sigma ) \approx 0$,
$\pi_{\varphi_{(a)}}(\tau ,\vec \sigma ) \approx 0$ (the
generators of local Lorentz boosts), $M_{(a)}(\tau ,\vec \sigma )
\approx 0$ (the generators of local rotations), $\Theta^r(\tau
,\vec \sigma ) \approx 0$ (the generators of the changes of
3-coordinates on $\Sigma_{\tau}$) and ${\cal H}(\tau ,\vec \sigma
) \approx 0$ (the super-hamiltonian constraint). Only the last
four are secondary constraints.
\bigskip

It can be shown \cite{53,72} that the addition of the DeWitt
surface term to the Dirac Hamiltonian (needed to make the
Hamiltonian theory well defined in the non-compact case) implies
that the Hamiltonian does not vanish on the constraint surface
({\it no frozen reduced phase space picture} in this model of
general relativity) but is proportional to the {\it weak ADM
energy}, which governs the $\tau$-evolution \cite{77}, so that
{\it an effective evolution takes place in mathematical time
$\tau$}. Moreover the Hamiltonian gauge transformations generated
by the super-hamiltonian constraint {\it do not have the
Wheeler-DeWitt interpretation} (evolution in local time), but
transform an admissible 3+1 splitting into another admissible one
(so that all the admissible notions of simultaneity are gauge
equivalent).

\bigskip

It follows, therefore, that the boundary conditions of this model
of general relativity imply that the real Dirac Hamiltonian is

\bea
 H_D &=& E_{ADM} + H_{(D)ADM} \approx E_{ADM},\nonumber \\
 &&{}\nonumber \\
 H_{(D)ADM}&=&\int d^3\sigma \, \Big[ n\, {\tilde {\cal H}}+n_r\,
{}^3{\tilde {\cal H}}^r + \lambda_n\, {\tilde \pi}^n + \lambda
^{\vec n}_r\, {\tilde \pi}^r_{\vec n}\Big](\tau ,\vec \sigma )
\approx 0,\nonumber \\
 &&(metric\, gravity),\nonumber \\
 &&{}\nonumber \\
 H_{(D)ADM}&=&\int d^3\sigma \, \Big[ n\, {\tilde {\cal H}}+n_r\,
{}^3{\tilde {\cal H}}^r + \lambda_n\, {\tilde \pi}^n + \lambda
^{\vec n}_r\, {\tilde \pi}^r_{\vec n} + \lambda^{\vec
\varphi}_{(a)}\, \pi_{\varphi_{(a)}} + \lambda^{\vec
\alpha}_{(a)}\, M_{(a)}\Big](\tau ,\vec \sigma )
\approx 0,\nonumber \\
 &&(tetrad\, gravity),
 \label{III3}
 \eea

\noindent where $\lambda_n(\tau , \vec \sigma )$ and
$\lambda^r_{\vec n}(\tau ,\vec \sigma )$ [and $\lambda^{\vec
\varphi}_{(a)}(\tau ,\vec \alpha )$, $\lambda^{\vec
\alpha}_{(a)}(\tau ,\vec \sigma )$] are {\it arbitrary Dirac
multipliers} in front of the primary constraints. The resulting
hyperbolic system of Hamilton-Dirac equations has the same
solutions of the non-hyperbolic system of (Lagrangian) Einstein's
equations with the same boundary conditions.

\bigskip

The weak ADM energy, and also the other nine asymptotic weak
Poincare' charges ${\vec P}_{ADM}$, $J^{AB}_{ADM}$, {\it are
Noether constants of the motion whose numerical value has to be
given as part of the boundary conditions}. The numerical value of
$E_{ADM}$ is the {\it mass} of the $\tau$-slice of the universe,
while $J^{rs}_{ADM}$ gives the value of the {\it spin} of the
universe. The weak ADM energy $E_{ADM} = \int d^3\sigma\, {\cal
E}_{ADM}(\tau ,\vec \sigma )$ has a density ${\cal E}_{ADM}(\tau
,\vec \sigma )$, which, like every type of energy density in
general relativity, is non-tensorial and gauge-dependent, because
it contains variables whose evolution depends on the arbitrary
Dirac multipliers.

Since, in our case, space-time is of the Christodoulou-Klainermann
type \cite{74}, the ADM 3-momentum has to vanish. This implies
three first class constraints

\beq
 {\vec P}_{ADM} \approx 0,
 \label{III5}
 \eeq

\noindent which identify the {\it rest frame of the universe}. As
shown in Ref.\cite{53}, the natural gauge fixing to these three
constraints is the requirement the the ADM boosts vanish: $J^{\tau
r}_{ADM} \approx 0$. In this way we decouple from the universe its
3-center of mass by making a choice of the centroid
$X^{\mu}_{(\infty)}(\tau )$, origin of the 3-coordinates on each
$\Sigma_{\tau}$, and only {\it relative motions} survive,
recovering a Machian flavour.

\subsection{Meaning of the Hamiltonian Gauge Transformations and
of the Gauge Fixings: Extended Space-Time Laboratories.}

The first class constraints are the generators of Hamiltonian
gauge transformations, under which the ADM action is
quasi-invariant (second Noether theorem).

\medskip

The eight infinitesimal off-shell Hamiltonian gauge
transformations generating the Hamiltonian gauge orbits, have the
following interpretation \cite{53}:

i) those generated by the four primary constraints modify the
lapse and shift functions: these in turn determine how densely the
space-like hyper-surfaces $\Sigma_{\tau}$ are distributed in
space-time and which points have the same 3-coordinates $\vec
\sigma$ on each $\Sigma_{\tau}$ (this is also a convention about
gravito-magnetism);

ii) those generated by the three super-momentum constraints induce
a transition on $\Sigma_{\tau}$ from a given 3-coordinate system
to another one;

iii) that generated by the super-hamiltonian constraint induces a
transition from a given 3+1 splitting of $M^4$ to another, by
operating normal deformations \cite{78} of the space-like
hyper-surfaces, and shows that, like in special relativity, all
the admissible notions of simultaneity are gauge-equivalent;

iv) those generated by the three rest-frame constraints
(\ref{III5}) can be interpreted as a change of centroid to be used
as origin of the 3-coordinates.

v) in tetrad gravity those generated by $\pi_{\varphi_{(a)}}(\tau
,\vec \sigma ) \approx 0$ and $M_{(a)}(\tau ,\vec \sigma ) \approx
0$ change the cotetrads with local Lorentz transformations.
\medskip

As usual, to get a completely fixed Hamiltonian gauge we add four
gauge fixings to the super-hamiltonian and super-momentum
constraints, which fix the form of $\Sigma_{\tau}$ (i.e. the
simultaneity) and a 3-coordinate system on it \footnote{Since the
diffeomorphism group has no canonical identity, this gauge fixing
has to be done in the following way. We choose a 3-coordinate
system by choosing a parametrization of the six components
${}^3g_{rs}(\tau ,\vec \sigma )$ of the 3-metric in terms of {\it
only three} independent functions. This amounts to fix the three
functional degrees of freedom associated with the diffeomorphism
parameters $\xi^r(\tau ,\vec \sigma )$. For instance, a
3-orthogonal coordinate system is identified by ${}^3g_{rs}(\tau
,\vec \sigma ) = 0$ for $r \not= s$ and ${}^3g_{rr} = \phi^2\,
exp(\sum_{\bar a = 1}^2 \gamma_{r\bar a} r_{\bar a})$. Then, we
impose the gauge fixing constraints $\xi^r(\tau ,\vec \sigma ) -
\sigma^r \approx 0$ as a way of identifying this system of
3-coordinates with a conventional origin of the diffeomorphism
group manifold. }, respectively. Their $\tau$-constancy generates
the gauge fixings determining the lapse and shift functions, so
that the 3+1 splitting is fixed. Further $\tau$-constancy
determines the Dirac multipliers. In tetrad gravity we have also
to fix a cotetrad field. Therefore, a complete Hamiltonian gauge
corresponds to a {\it extended non-inertial space-time
laboratory}, which can be shown to correspond to a 4-coordinate
system of the Einstein space-time (the $\sigma^A$ adapted to the
3+1 splitting) on the solutions of Einstein's equations, with a
fixed dynamical chrono-geometrical structure: i) a well defined
simultaneity convention for the synchronization of distant clocks
(the 3-spaces $\Sigma_{\tau}$); ii) a unit of proper time in each
point of $\Sigma_{\tau}$ (the lapse function); iii) a convention
for gravito-magnetism (the shift functions); iv) a 3-metric in
$\Sigma_{\tau}$ for measuring spatial distances; v) a 4-metric for
determining the local light-cone in each point of $\Sigma_{\tau}$
and, then, the one-way velocity of light in the geometrical optic
approximations (light rays along null geodesic).

\bigskip

\subsection{Quasi-Shanmugadhasan Canonical Transformation and the Generalized
Inertial and Tidal Effects.}

The discussion of the previous Subsection shows that there are 8
(14) {\it arbitrary gauge variables} in metric (tetrad) gravity.
In both cases we have to identify a canonical basis $r_{\bar
a}(\tau ,\vec \sigma )$, $\pi_{\bar a}(\tau ,\vec \sigma )$, $\bar
a = 1,2$, of {\it Dirac observables} (DO) as the {\it physical
degrees of freedom of the gravitational field}, i.e. a canonical
basis of predictable gauge-invariant quantities satisfying
deterministic Hamilton equations governed by the weak ADM energy.
This can be achieved by means of a Shanmugadhasan canonical
transformation adapted to 7 (13) of the 8 (14) first class
constraints (not to the super-hamiltonian one), which turns out to
be a {\it point} canonical transformation as a consequence of the
form of the finite gauge transformations. As a consequence,  the
old momenta are linear functionals of the new ones, with the
kernels determined by a set of elliptic partial differential
equations. In the new canonical basis 7 (13) new momenta vanish
due to the 7 (13) constraints and their 13 conjugate configuration
variables are {\it Abelianized gauge variables}.
\medskip

Since it can be shown \cite{53} that Lichnerowicz's identification
of the conformal factor of the 3-metric on $\Sigma_{\tau}$ ($\phi
= [det\, {}^3g]^{1/12}$) as the unknown in the super-hamiltonian
constraint is the correct one, as a consequence the gauge variable
describing the normal deformations \cite{78} of the simultaneity
surfaces $\Sigma_{\tau}$ is the momentum $\pi_{\phi}(\tau ,\vec
\sigma )$ canonically conjugate to $\phi (\tau ,\vec \sigma )$
(and not the trace of the extrinsic curvature of $\Sigma_{\tau}$,
the so called intrinsic York time).  In this way for the first
time we can identify a canonical basis of non-local and
non-tensorial DO $r_{\bar a}(\tau ,\vec \sigma )$, $\pi_{\bar
a}(\tau ,\vec \sigma )$ , which remains canonical in the class of
gauges where $\pi_{\phi}(\tau ,\vec \sigma ) \approx 0$, even if
no one knows how to solve the super-hamiltonian constraint, i.e.
the Lichnerowicz equation for the conformal factor.

\bigskip

In order to visualize the meaning of the various types of degrees
of freedom we need the construction of a {\it Shanmugadhasan
canonical basis}  of metric gravity having the following structure
with (a similar basis exists for tetrad gravity)

 \bea
\begin{minipage}[t]{3cm}
\begin{tabular}{|l|l|l|} \hline
$n$ & $n_r$ & ${}^3g_{rs}$ \\ \hline ${\tilde \pi}^n \approx 0$ &
${\tilde \pi}_{\vec n}^r \approx 0$ & ${}^3{\tilde \Pi}^{rs}$ \\
\hline
\end{tabular}
\end{minipage} &&\hspace{2cm} {\longrightarrow \hspace{.2cm}} \
\begin{minipage}[t]{4 cm}
\begin{tabular}{|ll|l|l|l|} \hline
$n$ & $n_r$ & $\xi^{r}$ & $\phi$ & $r_{\bar a}$\\ \hline
 ${\tilde \pi}^n \approx 0$ & ${\tilde \pi}_{\vec n}^r \approx 0$
& ${\tilde \pi}^{{\vec {\cal H}}}_r \approx 0$ &
 $\pi_{\phi}$ & $\pi_{\bar a}$ \\ \hline
\end{tabular}
\end{minipage} \nonumber \\
 &&{}\nonumber \\
&& {\longrightarrow \hspace{.2cm}} \
\begin{minipage}[t]{4 cm}
\begin{tabular}{|ll|l|l|l|} \hline
$n$ & $n_r$ & $\xi^{r}$ & $Q_{\cal H} \approx 0$ & $r^{'}_{\bar
a}$\\ \hline
 ${\tilde \pi}^n \approx 0$ & ${\tilde \pi}^r_{\vec n} \approx 0$
& ${\tilde \pi}^{{\vec {\cal H}}}_r \approx 0$ &
 $\Pi_{\cal H}$ & $\pi^{'}_{\bar a}$ \\ \hline
\end{tabular}
\end{minipage}.
 \label{III7}
 \eea

\noindent It is seen that we need a sequence of two canonical
transformations.\bigskip

a) The first transformation replaces seven first-class constraints
with as many Abelian momenta ($\xi^r$ are the gauge parameters,
namely coordinates on the group manifold,  of the passive
3-diffeomorphisms generated by the super-momentum constraints) and
introduces the conformal factor $\phi$ of the 3-metric as the
configuration variable to be determined by the super-hamiltonian
constraint. Note that the final gauge variable, namely the
momentum $\pi_{\phi}$ conjugate to the conformal factor, is the
only gauge variable of momentum type: it plays the role of a {\it
time} variable, so that the Lorentz signature of space-time is
made manifest by the Shanmugadhasan transformation in the set of
gauge variables $(\pi_{\phi}; \xi^r)$. More precisely, the first
canonical transformation should be called a {\it
quasi-Shanmugadhasan } transformation, because nobody has
succeeded so far in Abelianizing the super-hamiltonian constraint.

\medskip

Since it is not known how to build a global atlas of coordinate
charts for the group manifold of diffeomorphism groups, it is not
known either how to express the $\xi^r$'s, $\pi_{\phi}$ and the DO
$r_{\bar a}$, $\pi_{\bar a}$ in terms of the original ADM
canonical variables.However, since the  transformation
(\ref{III7}) is a {\it point} canonical transformation, we know
the inverse point canonical transformation from the form of finite
gauge transformations (see Ref.\cite{71,72} for the case of tetrad
gravity)

\bea
 {}^3g_{rs}(\tau ,\vec \sigma ) &=& {{\partial \xi^u(\tau ,\vec \sigma )}
 \over {\partial \sigma^r}}\, {{\partial \xi^v(\tau ,\vec \sigma )}
 \over {\partial \sigma^s}}\, \phi^4(\tau ,\vec \sigma )\,
 {}^3{\hat g}_{uv}[r_{\bar a}(\tau , \vec \xi (\tau ,\vec \sigma
 ))], \nonumber \\
  &&{}\nonumber \\
 {}^3\Pi^{rs}(\tau ,\vec \sigma ) &\approx& \int d^3\sigma_1\,
 {\cal K}^{rs}_{(a)}(\tau ,\vec \sigma ,{\vec \sigma}_1|\vec \xi
 ,\phi , r_{\bar a})\, \pi_{(a)}(\tau ,{\vec \sigma}_1),
 \label{a}
 \eea

\noindent where i) ${}^3{\hat g}_{rs}$ is a 3-metric with unit
determinant depending only on the two independent functions
$r_{\bar a}(\tau , \vec \xi (\tau ,\vec \sigma ))$; ii) ${\cal
K}^{rs}_{(a)}$ is a kernel determined by the requirement of
canonicity of the transformation.

\medskip

In absence of explicit solutions of the Lichnerowicz equation, the
best we can do is to construct the {\it quasi-Shanmugadhasan}
transformation.

\bigskip

b) The second canonical transformation would be instead a {\it
complete Shanmugadhasan} transformation, where $Q_{{\cal H}}(\tau
,\vec \sigma ) \approx 0$ would denote the Abelianization of the
super-hamiltonian constraint\footnote{If $\tilde \phi [r_{\bar a},
\pi_{\bar a}, \xi^r, \pi_{\phi}]$ is the solution of the
Lichnerowicz equation, then $Q_{{\cal H}}=\phi - \tilde \phi
\approx 0$. Other forms of this canonical transformation should
correspond to the extension of the York map \cite{79} to
asymptotically flat space-times: in this case the momentum
conjugate to the conformal factor would be just York time and one
could add the maximal slicing condition as a gauge fixing. Again,
however, nobody has been able so far to build a York map
explicitly.}. The variables $n$, $n_r$, $\xi^r$, $\Pi_{\cal H}$
are the final {\it Abelianized Hamiltonian gauge variables}, while
$r^{'}_{\bar a}$, $\pi^{'}_{\bar a}$ are the final DO.
\bigskip

\bigskip

Let us stress the important fact that the Shanmugadhasan canonical
transformation is a {\it highly non-local} (it involves the whole
3-space) transformation: this feature has a Machian flavor,
although in a non-Machian context.

\bigskip

\subsection{The Hamilton-Dirac Equations, Time Evolution and the
Dynamical Determination of the Allowed Notions of Simultaneity.}

In a completely fixed Hamiltonian gauge {\it all the gauge
variables} $\xi^r$, $\pi_{\phi}$, $n$, $n_r$ {\it become uniquely
determined functions of the DO} $r_{\bar a}(\tau ,\vec \sigma )$,
$\pi_{\bar a}(\tau ,\vec \sigma )$, which at this stage are four
arbitrary fields. Moreover, also the (unknown) solution $\phi
(\tau ,\vec \sigma )$ of the Lichnerowicz equation becomes a
uniquely determined functional of the DO, and this implies that
all the geometrical tensors like the 3-metric ${}^3g_{rs}(\tau
,\vec \sigma )$, the extrinsic curvature ${}^3K_{rs}(\tau ,\vec
\sigma )$ of the simultaneity surfaces $\Sigma_{\tau}$, and the
4-metric ${}^4g_{AB}(\tau ,\vec \sigma )$ {\it become uniquely
determined functionals of the DO only}.
\medskip

This is true in particular for the {\it weak ADM energy} $E_{ADM}
= \int d^3\sigma\, {\cal E}_{ADM}(\tau ,\vec \sigma )$, since the
energy density ${\cal E}_{ADM}(\tau ,\vec \sigma )$ depends not
only on the DO but also on $\phi$ and on the gauge variables
$\xi^r$ and $\pi_{\phi}$. In a fixed gauge we get $E_{ADM} = \int
d^3\sigma\, {\cal E}^G_{ADM}(\tau ,\vec \sigma )$ and this becomes
the functional that rules the Hamilton equations \cite{77} for the
DO in the completely fixed gauge

 \beq
 {{\partial r_{\bar a}(\tau ,\vec \sigma )}\over {\partial \tau }}
  = \{ r_{\bar a}(\tau ,\vec \sigma ), E_{\mathrm{ADM}}\}^*,
\quad {{\partial \pi_{\bar a}(\tau ,\vec \sigma )}\over {\partial
\tau}} = \{\pi_{\bar a}(\tau ,\vec \sigma ), E_{\mathrm{ADM}}\}^*,
 \label{III8}
  \eeq

\noindent  where the $\{\cdot,\cdot\}^*$ are Dirac Brackets. By
using the inversion of the first set of Eqs.(\ref{III8}) to get
$\pi_{\bar a} = \pi_{\bar a}[r_{\bar b}, {{\partial r_{\bar
b}}\over {\partial \tau}}]$, we arrive at the second order in time
equations ${{\partial2 r_{\bar a}(\tau ,\vec \sigma )}\over
{\partial \tau^2}} = F_{\bar a}[r_{\bar b}(\tau ,\vec \sigma ),
{{\partial r_{\bar b}(\tau ,\vec \sigma )}\over {\partial \tau}},
spatial\, gradients\, of\, r_{\bar b}(\tau ,\vec \sigma )]$, where
the $F_{\bar a}$'s are {\it effective forces} whose functional
form depends on the gauge.\bigskip

These Hamilton-Dirac equations have a well posed Cauchy problem as
a consequence of the use of Dirac constraint theory in the
Hamiltonian framework. Instead at the configurational level the
Cauchy problem for Einstein's equations and the identification of
the predictable quantities (DO) are so complicated that their
modern treatment \cite{80} simulates the Hamiltonian strategy.

Thus, once we have chosen any surface of the foliation as initial
Cauchy surface $\Sigma_{\tau_o}$ and assigned the initial data
$r_{\bar a}(\tau_o ,\vec \sigma )$, $\pi_{\bar a}(\tau_o ,\vec
\sigma )$ of the DO, we can calculate the solution of the
Einstein-Hamilton equations corresponding to these initial data.

This identifies an Einstein space-time including its dynamical
chrono-geometrical structure {\it including the associated
admissible dynamical definitions of simultaneity, distant clocks
synchronization and gravito-magnetism}.
\medskip

The admissible dynamical simultaneity notions in our class of
space-times are much less in number than the non-dynamical
admissible simultaneity notions in special relativity: as shown in
Section VIII of Ref.\cite{72}, if Minkowski space-time is thought
of as a special solution (with vanishing DO) of Einstein-Hamilton
equations, then its allowed 3+1 splittings must have 3-conformally
flat simultaneity 3-surfaces (due to the vanishing of the DO the
Cotton-York tensor of $\Sigma_{\tau}$ vanishes), a restriction
absent in special relativity considered as an autonomous  theory.

\bigskip

Let us stress that the subdivision of canonical variables in two
sets (gauge variables and DO) is a peculiar outcome of the
quasi-Shanmugadhasan canonical transformation which has no simple
counterpart within the {\it Lagrangian viewpoint} at the level of
the Hilbert action and/or of Einstein's equations. This
subdivision amounts to an extra piece of (non-local) information
which should be added to the traditional wisdom of the {\it
equivalence principle} asserting the local impossibility of
distinguishing gravitational from inertial effects. Indeed, it
allows to distinguish and visualize which aspects of the local
physical effects on test matter contain a {\it genuine
gravitational} component (think to the geodesic  deviation
equation) and which aspects depend solely upon the choice of the
{\it global non-inertial space-time laboratory} with the
associated atlas of 4-coordinate systems in a topologically
trivial space-time: these latter effects could then be named {\it
inertial}, in analogy with what happens in the non-relativistic
Newtonian case in {\it global rigid} non-inertial reference
frames. This interpretation is possible because the Hamiltonian
point of view leads naturally to a re-reading of geometrical
features in terms of the traditional concept of {\it force}. As a
consequence, we can say that in a completely fixed Hamiltonian
gauge  the 8 (14) gauge variables describe {\it generalized
inertial effects} and the DO describe {\it generalized tidal
effects} seen by the associated extended non-inertial space-time
laboratory.

Let us also remark that the {\it reference standards} of time and
length correspond to units of {\it coordinate time and length} and
not to proper times and proper lengths \cite{47}: this is not in
contradiction with general covariance, because an extended {\it
laboratory}, in which one defines the reference standards,
corresponds to a particular {\it completely fixed on-shell
Hamiltonian gauge} plus a local congruence of time-like observers.

\medskip

The picture we have presented is not altered by the presence of
matter. The only new phenomenon besides the above purely
gravitational, {\it inertial and tidal} effects, is that from the
solution of the super-hamiltonian and super-momentum constraints
emerge {\it  action-at-a-distance  Newtonian-like and
gravito-magnetic} effects among matter elements.\medskip

\subsection{The Hole Argument and the Physical Identification of
Point-Events.}

In Ref.\cite{81} there is a review of the various implications of
Einstein's Hole Argument, a consequence of the {\it general
covariance} of the theory in its various forms: i) invariance of
the Hilbert action under passive diffeomorphisms (general
coordinate transformations); ii) quasi-invariance of the ADM
action under passive Hamiltonian gauge transformations; iii)
invariance of Einstein's equations under active diffeomorphisms
(see Ref.\cite{82} for their passive re-interpretation). Its
consequences are: i) the absence of determinism (only two of
Einstein's equations contain dynamical information: four are
restrictions on initial data and four are void due to Bianchi
identities), i.e. the presence of arbitrary gauge variables; ii)
absence of a physical individuation of the mathematical points of
the mathematical pseudo-Riemannian 4-manifold $M^4$ as physical
point-events of space-time. In Refs.\cite{81,83}, final
re-elaboration of Refs.\cite{84}, there is a complete study and a
solution of the interpretational problems connected with the Hole
Argument, which, even if obsolete in physics, is still source of
an open debate on the ontology of space-time in philosophy of
science (see for instance Ref.\cite{85}).

\medskip

In absence of space-time symmetries, Stachel \cite{4} suggested to
identify the point-events by means of the Bergmann-Komar  {\it
intrinsic pseudo-coordinates} \cite{86} (used as individuating
fields), i.e. as four suitable functions ${\bar \sigma}^{\bar
A}(\sigma ) = F^{\bar A}[\Lambda^{(k)}_W [{}^4g(\sigma),
\partial\, {}^4g(\sigma)]], \,(\bar A = 1,2,...,4)$ of the four
invariant scalar eigenvalues $\Lambda^{(k)}_W(\tau ,\vec \sigma
)$, $k=1,..,4$, of the Weyl tensor.

Since the Weyl eigenvalues do not depend on the lapse and shift
functions, in a completely fixed gauge $G$ they are functions only
of the DO: $\Lambda^{(k)}_W(\tau ,\vec \sigma ){|}_G =  {\tilde
\Lambda}^{(k)}_W[{}^3g(\tau ,\vec \sigma), {}^3\Pi(\tau ,\vec
\sigma)]{|}_G = \Lambda_G^{(k)}[r_{\bar a}(\tau ,\vec \sigma ),
\pi_{\bar a}(\tau ,\vec \sigma )]$

\bigskip

The space-time points, {\it mathematically individuated} by the
quadruples of real numbers $\sigma^A$, corresponding to a {\it
completely arbitrary mathematical} radar coordinate system
$\sigma^A \equiv [\tau,\sigma^a]$ adapted to the $\Sigma_\tau$
surfaces, become now {\it physically individuated point-events}
through the imposition of the following gauge fixings to the four
secondary constraints

\beq
 {\bar \chi}^A(\tau ,\vec \sigma )\- {\buildrel {def} \over =}\-  \sigma^A -
 {\bar \sigma}^{\bar A}(\tau ,\vec \sigma ) = \sigma^A -
 F^{\bar A}\Big[{\tilde \Lambda}^{(k)}_W[{}^3g(\tau ,\vec \sigma), {}^3\Pi(\tau ,\vec
\sigma)]\Big] \approx 0.
 \label{IV5}
 \eeq

\noindent where the four functions $F^{\bar A}[{
\Lambda}^{(k)}_W(\tau ,\vec \sigma )]$ ({\it  physical
individuating fields} ) are chosen so that the ${\bar \chi}^A(\tau
,\vec \sigma )$'s satisfy the {\it orbit conditions} $det\, | \{
{\bar \chi}^A(\tau ,\vec \sigma ), {\tilde {\cal H}}^B(\tau ,{\vec
\sigma}^{'}) \} | \not= 0$ with ${\tilde {\cal H}}^B(\tau ,\vec
\sigma ) = \Big( {\tilde {\cal H}}(\tau ,\vec \sigma );
{}^3{\tilde {\cal H}}^r(\tau ,\vec \sigma ) \Big) \approx 0$.
These conditions enforce the Lorentz signature, namely the
requirement that $F^{\bar \tau}$ be a {\it time} variable, and
imply that {\it the $F^{\bar A}$'s are not DO}.
\medskip

The above gauge fixings allow in turn the determination of the
four Hamiltonian gauge variables $\xi^r(\tau ,\vec \sigma )$,
$\pi_{\phi}(\tau ,\vec \sigma )$. Then, their time constancy
induces the further gauge fixings ${\bar \psi}^A(\tau ,\vec \sigma
) \approx 0$ for the determination of the remaining gauge
variables, i.e., the lapse and shift functions in terms of the DO
and then of the Dirac multipliers.

\bigskip

If, after this complete breaking of general covariance, we go to
Dirac brackets, we enforce the point-events individuation in the
form of the {\it identity} $\sigma^A \equiv {\bar \sigma}^{\bar A}
= {\tilde F}^{\bar A}_{G}[ r_{\bar a}(\tau ,\vec \sigma ),
\pi_{\bar a}(\tau , \vec \sigma)] = F^{\bar
A}[\Lambda^{(k)}_W(\tau , \vec \sigma )]{|}_G$, and on-shell this
is a coordinate chart of the atlas of $M^4$.

\bigskip

Summarizing, the effect of the whole procedure is that {\it the
values of the {\it DO}, whose dependence on space (and on {\it
parameter} time) is indexed by the chosen radar coordinates $(\tau
,\vec \sigma )$, reproduces precisely such $(\tau ,\vec \sigma )$
as the Bergmann-Komar {\it intrinsic coordinates} in the chosen
gauge $G$}. In this way {\it mathematical points have become
physical individuated point-events} by means of the highly
non-local structure of the DO and each coordinate system
$\sigma^A$ is determined on-shell by the values of the 4 canonical
degrees of freedom of the gravitational field in that gauge. This
is tantamount to claiming that {\it the physical role and content
of the gravitational field in absence of matter is just the very
identification of the points of Einstein space-times into physical
point-events by means of its four independent phase space degrees
of freedom}. The existence of physical point-events in general
relativity appears here as a synonym of the existence of the DO,
i.e. of the true physical degrees of freedom of the gravitational
field.
\medskip

The addition of matter does not change this conclusion, because we
can continue to use the gauge fixing (\ref{IV5}). However, matter
changes the Weyl tensor through Einstein's equations and
contributes to the separation of gauge variables from DO in the
quasi-Shanmugadhasan canonical transformation through the presence
of its own DO. In this case we have DO both for the gravitational
field and for the matter fields, which satisfy coupled Hamilton
equations. Therefore, since the gravitational DO will still
provide the individuating fields for point-events according to our
procedure, {\it matter will come to influence the very physical
individuation of points}.

\bigskip

Let us conclude by noting that the above gauge fixings induce a
{\it coordinate-dependent non-commutative Poisson bracket
structure} upon the {\it physical point-events} of space-time by
means of the associated Dirac brackets implying $\{ {\tilde
F}^{\bar A}_G(r_{\bar a}(\tau , \vec \sigma ), \pi_{\bar a}(\tau
,\vec \sigma )), {\tilde F}^{\bar B}_G(r_{\bar a}(\tau ,{\vec
\sigma}_1), \pi_{\bar a}(\tau ,{\vec \sigma}_1)) \}^* \not= 0$.
The meaning of this structure in view of quantization is worth
investigating.

\subsection{Bergmann Observables versus Dirac Observables: a
Conjecture.}

Let us now consider the problem of the {\it observables} of the
gravitational field. Two fundamental definitions of {\it
observable} have been proposed in the literature.
\bigskip

1) The {\it  Hamiltonian non-local Dirac observables} (DO) which,
by construction, satisfy hyperbolic Hamilton equations of motion
and are, therefore, deterministically {\it predictable}. In
general, as already said, they are neither tensorial quantities
nor invariant under the passive diffeomorphisms of  $M^4$ (PDIQ).
\bigskip

2) The {\it configurational Bergmann observables} ({\it BO})
\cite{87}: they are quantities defined on $M^4$ {\it which not
only are independent of the choice of the coordinates}, i.e. they
are quantities invariant under passive diffeomorphisms of $M^4$
(PDIQ), but are also {\it uniquely predictable from the initial
data}, namely they  are also DO.

\medskip

In order to give consistency to Bergmann's  multiple definition of
BO and, in particular, to his (strictly speaking unproven) claim
\cite{87} about the {\it existence} of DO that are simultaneously
BO, the following {\it conjecture} should be true:

\bigskip
\noindent {\bf A Main Conjecture}: "The Darboux basis in the
quasi-Shanmugadhasan canonical basis (\ref{III7}) of I {\it can be
replaced} by a Darboux basis whose 16  variables are all PDIQ (or
tetradic variables), such that four of them are simultaneous DO
and BO, eight vanish because of the first class constraints, and
the other 8 are coordinate-independent gauge variables."
\bigskip

If this conjecture is sound, it would be possible to construct an
{\it intrinsic tensorial} Darboux basis  of the Shanmugadhasan
type. More precisely, we would have a family of
quasi-Shanmugadhasan canonical bases in which all the variables
are PDIQ and include 7 PDIQ first class constraints that play the
role of momenta. It would be interesting, in particular, to check
the form of the extra constraint replacing the standard
super-hamiltonian constraint. In this way an {\it intrinsic
characterization of inertial and tidal effects} would emerge. The
same strategy applies to tetrad gravity.

\medskip

Further strong support to the conjecture comes from Newman-Penrose
formalism \cite{88} where the basic tetradic fields (evaluated by
using null tetrads suggested by the Hamiltonian formalism) are the
20 Weyl and Ricci scalars which are PDIQ by construction . While
the vanishing of the Ricci scalars is equivalent to Einstein's
equations (and therefore to a scalar form of the super-hamiltonian
and super-momentum constraints), the 10 Weyl scalars plus 10
scalars describing the ADM momenta (restricted by the four primary
constraints) should lead to the construction of a Darboux basis
spanned only by PDIQ restricted by eight PDIQ first class
constraints. Again, a quasi-Shanmugadhasan transformation should
produce the Darboux basis of the conjecture. The problem of the
phase space re-formulation of Newman-Penrose formalism is now
under investigation.
\bigskip

Such an intrinsic basis would allow to start a new program of
quantization of gravity, based on the idea of quantizing only the
tidal effects (the BO) and not the inertial ones (the gauge
variables), since the latter describe only the {\it appearances}
of the phenomena. A prototype of this quantization is under study
in special relativity to arrive at a formulation of atomic physics
in non-inertial systems: while for relativistic particles (and
their non-relativistic limit) there are already preliminary
results \cite{89}, for the inclusion of the electro-magnetic field
we have to find a way out from the Torre-Varadarajan no-go theorem
\cite{19}, the obstruction to the Tomonaga-Schwinger formalism
\cite{46}.
\medskip

Moreover, if the weak ADM energy in a completely fixed Hamiltonian
gauge can be expressed in terms of BO, this would help to clarify
the problem of the coordinate-dependence of the energy density in
general relativity, which we think is a preliminary step for a
correct understanding of the cosmological constant and dark energy
problems.

\subsection{An Operational Determination of Space-Time.}

Lacking solutions to Einstein's equations with matter
corresponding to simple systems to be used as idealizations for a
measuring apparatuses described by matter DO (hopefully also BO),
a generally covariant theory of measurement as yet does not exist.
\bigskip

In the meanwhile let us sketch here a scheme for implementing - at
least in principle - the physical individuation of points as an
experimental setup and protocol for positioning and orientation.
\medskip

a) A {\it radar-gauge} system of coordinates can be defined in a
finite four-dimensional volume by means of a network of artificial
spacecrafts similar to the Global Position System (GPS) \cite{90}.
Let us consider a family of spacecrafts, whose navigation is
controlled from the Earth by means of the standard GPS. Note that
the GPS receivers are able to determine their actual position and
velocity because the GPS system is based on the advanced knowledge
of the gravitational field of the Earth and of the satellites'
trajectories, which in turn allows the {\it coordinate}
synchronization of the satellite clocks . During the navigation
the spacecrafts are test objects. Since the geometry of space-time
and the motion of the spacecrafts are not known in advance in our
case, we must think of the receivers as obtaining four, so to
speak, {\it conventional} coordinates by operating a full-ranging
protocol involving bi-directional communication to four {\it
super-GPS} that broadcast the time of their standard
a-synchronized clocks. This first step parallels the axiomatic
construction of Ehlers, Pirani and Schild \cite{5} of the {\it
conformal structure} of space-time.

Once the spacecrafts have arrived in regions with non weak fields,
like near the Sun or Jupiter, they become the (non test but with
world-lines assumed known from GPS space navigation) elements of
an experimental setup and protocol for the determination of a
local 4-coordinate system and of the associated 4-metric. Each
spacecraft, endowed with an atomic clock and a system of
gyroscopes, may be thought as a time-like observer (the spacecraft
world-line assumed known) with a tetrad (the time-like vector is
the spacecraft 4-velocity (assumed known) and the spatial triad is
built with gyroscopes) and one of them is chosen as the origin of
the radar-4-coordinates we want to define. This means that the
natural framework should be tetrad gravity instead of metric
gravity.

\medskip

b) At this point we have to synchronize the atomic clocks by means
of radar signals \cite{91}. Since the geometry and the admissible
simultaneity conventions of the solar system   Einstein space-time
are not known in advance, we could only lay down the lines of an
approximation procedure starting from an arbitrary simultaneity
convention. The spacecraft $A$ chosen as origin (and using the
proper time $\tau$ along the assumed known world-line) sends radar
signals to the other spacecrafts, where they are reflected back to
$A$. For each radar signal sent to a spacecraft $B$, the
spacecraft $A$ records four data: the emission time $\tau_o$, the
emission angles $\theta_o$, $\phi_o$ and the absorption time
$\tau_f$. Given four admissible (see Ref.\cite{21}) functions
${\cal E}(\tau_o, \theta_o, \phi_o, \tau_f)$, ${\vec {\cal
G}}(\tau_o, \theta_o, \phi_o, \tau_f)$ the point $P_B$ of the
world-line of the spacecraft $B$, where the signal is reflected,
is given radar coordinates $\tau_{(R)}(P_B) = \tau_o + {\cal
E}(\tau_o, \theta_o, \phi_o, \tau_f)\, (\tau_f - \tau_o)$, ${\vec
\sigma}_{(R)}(P_B) = {\vec {\cal G}}(\tau_o, \theta_o, \phi_o,
\tau_f)$ and will be simultaneous (according to this convention)
to the point $Q$ on the world-line of the spacecraft $A$
identified by $\tau {|}_Q = \tau_{(R)}(P_B)$ \footnote{Einstein's
simultaneity convention would correspond to ${\cal E} = {1\over
2}$ and to space-like hyper-planes as simultaneity surfaces.}.

This allows establishing a {\it radar-gauge system of
4-coordinates} (more exactly a coordinate grid) $\sigma^A_{(R)} =
( \tau_{(R)}; \sigma^r_{(R)})$ in a finite region, with
$\tau_{(R)} = const$ defining the radar simultaneity surfaces of
this convention. By varying the functions ${\cal E}$, ${\vec {\cal
G}}$ we change the simultaneity convention among the admissible
ones .

\medskip

Then the navigation system provides determination of the
4-velocities (time-like tetrads) of the satellites, namely of the
${}^4g_{(R)\tau\tau}(\sigma_{(R)}^A)$ component of the 4-metric in
these coordinates. Then, employing test gyroscopes and light
signals (i.e. only the {\it conformal structure}),  by means of
exchanges (two-ways signals) of {\it polarized} light it should be
possible to determine how the spatial triads of the satellites are
rotated with respect to the triad of the satellite chosen as
origin. Once we have the tetrads ${}^4E^A_{(r)(\alpha
)}(\tau_{(R)}, {\vec \sigma}_{(R)})$ in radar coordinates, we can
build from them the inverse 4-metric ${}^4g_{(R)}^{AB}(\tau_{(R)},
{\vec \sigma}_{(R)}) = {}^4E^A_{(r)(\alpha )}(\tau_{(R)}, {\vec
\sigma}_{(R)})\, {}^4\eta^{(\alpha )(\beta )}\, {}^4E^B_{(r)(\beta
)}(\tau_{(R)}, {\vec \sigma}_{(R)})$, and the the 4-metric, in
radar coordinates.

\medskip

c) By measuring the spatial and temporal variation of
${}^4{g}_{(R)AB}(\sigma^C_{(R)})$, the components of the Weyl
tensor and the Weyl eigenvalues can in principle be determined.
\bigskip

d) Points a), b) and c) furnish {\it operationally} a slicing of
space-time into surfaces $\tau_{(R)} = const$, a system of
coordinates $\sigma^r_{(R)}$ on the surfaces, as well as a
determination of the components of the metric
${}^4{g}_{(R)AB}(\sigma^C_{(R)})$. The components of the Weyl
tensor (= Riemann in void) and the local value of the Weyl
eigenvalues, with respect to the radar-gauge coordinates
$(\tau_{(R)}, \sigma^r_{(R)})$ are also thereby determined. By
assuming the validity of Eintein's theory, it is then a matter of
computation:
\bigskip

i) To check whether Einstein's equations in these radar
coordinates are satisfied. If not, this means that the chosen
simultaneity $\tau_{(R)} = const.$ is not in the class of the
allowed dynamical notions of simultaneity of the Einstein solution
describing the solar system. By changing the functions ${\cal E}$,
${\vec {\cal G}}$, we can put up an approximation procedure
converging towards an admissible dynamical notion of
simultaneity.\bigskip

ii) If $(\tau_R, {\vec \sigma}_R)$ are the radar coordinates
corresponding to a dynamical synchronization of clocks, we can get
a numerical determination of the intrinsic coordinate functions
${\bar \sigma}^{\bar A}_R$ defining the radar gauge by the gauge
fixings $\sigma^A_R - F^{\bar A}[{\tilde
\Lambda}^{(k)}_W[{}^3g(\tau_R ,{\vec \sigma}_R), {}^3\Pi(\tau_R
,{\vec \sigma}_R)]] = \sigma^A_R - {\tilde F}^{\bar A}_R[ r_{\bar
a}(\tau_R ,{\vec \sigma}_R ), \pi_{\bar a}(\tau_R , {\vec
\sigma}_R)] \approx 0$ built as intrinsic coordinates functions of
the known eigenvalues of the Weyl tensor in the radar gauge.

\bigskip

This procedure of principle would close the {\it coordinative
circuit} of general relativity, linking individuation to
operational procedures.

\subsection{Hamiltonian Linearization and Background-Independent
Post-Minkowskian Gravitational Waves.}

As a first application of the previous Hamiltonian formalism, in
Ref.\cite{92} we made a background-independent Hamiltonian
linearization of vacuum tetrad gravity in a completely fixed {\it
3-orthogonal} gauge obtained by adding 14 suitable gauge fixings,
one of which is $\pi_{\phi}(\tau ,\vec \sigma ) \approx 0$. This
allows to express all the geometrical quantities in terms of two
pairs of canonically conjugated DO. In this gauge, which turns out
to be {\it non-harmonic} in the weak field regime, the 3-metric on
$\Sigma_{\tau}$ is {\it diagonal} and it corresponds to a unique
3-orthogonal 4-coordinate system on space-time (with an associated
admissible 3+1 splitting with well defined simultaneity leaves) on
the solutions of Hamilton-Dirac equations.

In this gauge it is possible to give a {\it
background-independent} definition of a weak gravitational field:
the DO $r_{\bar a}(\tau ,\vec \sigma )$, $\pi_{\bar a}(\tau ,\vec
\sigma )$ should be slowly varying on a wavelength of the
resulting post-Minkowskian gravitational wave, with the
configurational DO $r_{\bar a}$ replacing the two polarizations of
the harmonic gauges. A {\it Hamiltonian linearization} is defined
in the following way:

i) Assuming $ln\, \phi (\tau ,\vec \sigma ) = O(r_{\bar a})$, the
Lichnerowitz equation can be linearized and for the first time a
non-trivial solution for $\phi$ can be found. Using this solution
all the other constraints and the elliptic canonicity conditions
can be linearized and solved. By putting these solutions in the
integrand of the weak ADM energy, we get a well defined form for
the energy density in this gauge in terms of the DO, i.e. the
physical degrees of freedom of the gravitational field.

ii) The resulting ADM energy is approximated with the terms {\it
quadratic} in the DO and the resulting linearized Hamilton
equations are studied and solved.  It is explicitly checked that
the linearized Einstein's equations are satisfied by this
solution. Even if the gauge is not harmonic, the wave equation
$\Box r_{\bar a}(\tau ,\vec \sigma) = 0$ is implied by the
Hamilton equations and solutions satisfying the universe
rest-frame condition are found (they cannot be transverse waves in
the rest frame). These are the {\it post-Minkowskian
background-independent gravitational waves}. The deformation
patterns of a sphere of test particles induced by $r_{\bar 1}$ and
$r_{\bar 2}$ are determined by studying the geodesic deviation
equation. An explicit solution $r_{\bar a}(\tau ,\vec \sigma )$,
$\pi_{\bar a}(\tau ,\vec \sigma )$ of the Hamilton equations for
the DO is obtained and this allows to get the linearized 3-metric,
the linearized lapse and shift functions and  the linearized
cotetrads in this 3-orthogonal gauge.
\medskip

Therefore the dynamical chrono-geometrical structure of this
Einstein space-time is completely determined and the embeddding of
the associated dynamical simultaneity convention can be rebuilt
from the 3+1 splitting with the 3-spaces $\Sigma_{\tau}$ and the
boundary conditions:

\bea
 z^{\mu}(\tau ,\vec \sigma ) &=& x^{\mu}(0) + \epsilon^{\mu}_A\,
 \sigma^A -\nonumber \\
 &-& \epsilon^{\mu}_r\, \int^1_{-\infty}\, {{d\lambda}\over
 {\lambda^2}}\, \Big[{{\sqrt{3}}\over 2}\, (\lambda\, \sigma^r)\,
 \sum_{\bar au}\, \gamma_{\bar au}\, \int d^3\sigma_1\, {{\partial^2_{1 u}\,
 r_{\bar a}(\lambda\, \tau , {\vec \sigma}_1)}\over {4\pi\,
 |\lambda\, \vec \sigma - {\vec \sigma}_1|}} - n_r(\lambda\, \tau ,\lambda\, \vec \sigma
 )\Big],\nonumber \\
 &&{}\nonumber \\
 z^{\mu}_A(\tau ,\vec \sigma ) &=& {{\partial z^{\mu}(\tau ,\vec
 \sigma )}\over {\partial \sigma^A}}.
 \label{I1}
 \eea

\bigskip

We are now studying tetrad gravity coupled to a perfect fluid
described by a suitable singular Lagrangian \cite{60}. The
Hamiltonian linearization in the special 3-orthogonal gauge,
together with an adaptation to our formalism of the theory of
Dixon's multipoles \cite{68,67}, will allow to find the
post-Minkowskian (without any post-Newtonian approximation!)
generalization of the quadrupole emission formula and the explicit
form in this gauge of the action-at-a-distance Newton and
gravito-magnetic potentials inside the fluid together with its
tidal interactions. The resulting formalism should help to find a
description of binary systems in a post-Minkowskian regime, where
the post-Newtonian approximations fail. Moreover, the two-body
problem of general relativity in the post-Minkowskian weak field
regime will be studied by using a new semi-classical
regularization of the self-energies, implying the $i \not= j$ rule
like it happens in the electro-magnetic case \cite{55}. Also
tetrad gravity coupled to Klein-Gordon, electro-magnetic and Dirac
fields is under investigation.

It will be explored the possibility of defining a scheme of
Hamiltonian numerical relativity, based on expansions in the
Newton constant G (the so called post-Minkowskian approximations),
to study the strong field regime of tetrad gravity.

\bigskip

\vfill\eject

\section{Concluding Remarks and Future Developments.}

In conclusion a unified scenario for special and general
relativity (and their non-relativistic limit) taking into account
their non-dynamical and dynamical, respectively,
chrono-geometrical structures has emerged. It unifies many, often
unrelated, points of views and allows the incorporation of a great
body of phenomenology from experimental gravitation, space physics
till atom and particle physics. In particular it allows to extend
the description of physics from inertial frames to global
non-rigid non-inertial frames, the only ones existing in general
relativity, with a new insight on relativistic inertial forces and
with the hope to arrive at a better understanding of the
equivalence principle, especially after quantization. The
establishment of this classical scenario was possible due the
strength of the Hamiltonian formalism when a systematic use is
made of Dirac-Bergmann theory of constraints. In particular we
have identified a class of space-times in which it is possible to
arrive at a {\it background-independent} description of both the
gravitational field and elementary particles. It is then possible
to go towards Newtonian physics either by a direct post-Newtonian
approximation or first to make a post-Minkowskian approximation to
special relativity then followed by the non-relativistic limit.
Let us note that in the rest-frame instant form of relativistic
mechanics it is possible to show that there are interacting models
which are inequivalent at the special relativistic level but which
admit the same Newtonian level: this shows that it is impossible
to re-sum the series in $1/c^n$ of the post-Newtonian expansions.
\medskip

At the classical level the main unsolved problem is to find either
exact or approximate solutions of the Lichnerowicz  equation (the
super-hamiltonian constraint) beyond the post-Newtonian
approximation. Only in this way we can have an idea of the
coordinate-dependent modifications of Newton law (not to speak of
the action-at-a-distance gravito-magnetic potentials) between
matter elements implied by Einstein general relativity, before
looking for its extensions or modifications.

\medskip

The real challenge now is to see whether it is possible to define
a new background-independent quantization scheme allowing to
extend these classical results to the quantum regime in a way
consistent with relativistic causality.
\bigskip

Let us delineate the lines of the going on researches.
\medskip

A) Special relativity.
\medskip

a) Since there is no accepted formulation of quantum mechanics in
non-inertial frames, we are studying \cite{89} a new quantization
scheme for relativistic scalar particles on space-like
hyper-planes with the differentially rotating coordinates
(\ref{II2}), which correspond to an admissible family of non-rigid
non-inertial frames, in the framework of parametrized Minkowski
theories. The idea is to quantize only the physical particle
degrees of freedom and not the embedding, which only describes the
{\it appearances} of the phenomena by means of the inertial
forces: the degrees of freedom of the embedding are treated as
{\it generalized times} in a multi-temporal scheme. This framework
allows to make a non-relativistic limit to Newton mechanics in
both non-rigid and rigid non-inertial frames. In particular we
want to show that, after the separation of the center of mass, the
relative motions can be described in a way allowing to show that,
after quantization, the spectral lines of atoms are the same both
in inertial and non-inertial frames. In other words the inertial
forces should produce only a noise over-imposed to the continuum
spectrum of the center-of-mass free motion of the atom.

b) The next step is to try to quantize the electro-magnetic field
on the arbitrary admissible 3+1  splittings allowed by
parametrized Minkowski theories, but with the simultaneity
3-spaces $\Sigma_{\tau}$ restricted to admit a Fourier transform
(Lichnerowicz 3-manifolds \cite{73}). As already said, to arrive
to its Tomonaga-Schwinger description we have to overcome the
Torre-Varadarajan no-go theorem. This seems to require an
ultraviolet regularization already for {\it free} fields. We hope
to be able to use the M$\o$ller radius as a ultraviolet cutoff
allowing to define a Fock space on each $\Sigma_{\tau}$.

In particular this framework should allow to arrive to a
formulation of relativistic atomic physics, whose
semi-relativistic limit should provide a justification for the
existing formalism \cite{93}.

c) This framework should allow a relativistic extension of the
foundational problems of quantum mechanics like the entanglement
of macroscopic bodies \cite{3}, where till now it is impossible to
take into account Maxwell equations for the electro-magnetic
field, either considering relativistic quantum mechanics of
isolated systems with the preferred simultaneity of the Wigner
hyper-planes of the rest-frame instant form or by considering
relativistic atomic physics with the hope to arrive at
relativistic Bell's inequalities. Let us note that already at the
classical level every admissible notion of simultaneity, namely
the definition of instantaneous 3-spaces $\Sigma_{\tau}$,
introduces an unavoidable non-locality. Also an attempt to define
a {\it measuring apparatus} as those special wave functions in the
Hilbert space of a macroscopic body which do not spread in time
and which behave as macroscopic Newtonian bodies (Ehrenfest
theorem) could have a relativistic extension taking into account
the classical relativistic delocalization of the center of mass
connected with the M$\o$ller world-tube.

\bigskip

B) General Relativity
\medskip

a) The study of perfect fluids plus tetrad gravity will allow to
find in which regime it is still possible to linearize and solve
the Lichnerowicz equation and then to get a background-independent
post-Minkowskian quadrupole emission formula. It will help also to
understand the Cauchy problem for a ball of fluid (a star), which
till now has no formulation \cite{80} either in Einstein or
Newtonian gravity, since the surface of the ball is a free
boundary. It will also allow a calculation of the post-Minkowskian
rotation curves of galaxies, simulated by a ball of dust, to see
the modification from the Keplerian evaluation.

b) It is now possible to study the coordinate-dependence of the
gravito-magnetic effects \cite{94} and of the time-delay of radar
signals from satellites to Earth stations (effects of  order
$1/c^3$) \cite{95} . The main open problem is to try to understand
whether the empirical 4-coordinate grid used by NASA to describe
the sourroundigs of the Earth in the solar system is a harmonic or
a 3-orthogonal 4-coordinate system.

c) The study of the Hamiltonian 2-body problem by using a
semi-classical regularization of the self-energies to get a
$i\not= j$ rule in analogy to the electro-magnetic case with
Grasmann-valued electric charges \cite{55}. It should open the way
to define a general relativistic harmonic oscillator as a
prototype of a {\it dynamical clock} and to start to define a
theory of measurement in terms of dynamical and not of test
objects.

d) Then we will study the coupling of the electro-magnetic field
to tetrad gravity, we shall try to find a regime where the
Lichnerowicz equation can be solved and both the propagation of
light and gravitational lensing can be studied at the Hamiltonian
level. After an analogous study for the Klein-Gordon and Dirac
fields, we will have all the ingredients for studying the coupling
of the $SU(3) \times SU(2) \times U(1)$ standard model of
elementary particles to tetrad gravity. In particular the
Foldy-Wouthuysen transformation for the Dirac field coupled to
tetrad gravity is under investigation.

e) The search of Bergmann observables suggests, as already said,
to look at the Hamiltonian reformulation of the Newman-Penrose
formalism by using a set of null tetrads natural from the point of
view of canonical gravity. This would lead to an intrinsic
definition of inertial and tidal effects and would open the way to
an attempt to make a background- and coordinate-independent
multi-temporal quantization of gravity. In it only the DO=BO would
be quantized, while the coordinate independent gauge variables
(the inertial effects determining the appearances of the
phenomena) would be treated as c-number generalized times. If
ordering problems can be overcome (they should be less troublesome
than in the standard attempts the canonical quantization of
gravity), this quantization would respect relativistic causality
due the presence of the 3+1 splittings of space-time and, like in
the approach of Ref.\cite{89}, there would be a naturally defined
physical scalar product.

f) Due to the importance of black holes, a Hamiltonian formulation
of space-times with symmetries would be welcome but the cone over
cone structure of singularities in the space of 4-metrics is an
obstruction to formulate it. We are planning to start from
canonical gravity without symmetries and to approach the cone of
4-metrics with one Killing vector, by adding by hand a set of
Killing equations rewritten as Hamiltonian constraints.
Preliminary calculations seems to indicate that the effect of
these extra constraints is to forces the DO to become functions of
the gauge variables. If this is confirmed, it would mean that in
space-times with symmetries there are no independent degrees of
freedom of the gravitational field but only generalized inertial
effects besides eventual singularities.

g) Due to the dominance of dark energy in the present picture of
an accelerating universe, it is worthwhile to study better the
weak ADM energy of the gravitational field with its
coordinate-dependent energy density containing terms proportional
to both $G$ and $1/G$: could it contribute in a
coordinate-dependent way to the dark energy?

h) Finally Hamiltonian numerical gravity has to be developed by
considering an iterative post-Minkowskian scheme based on
developments in powers of Newton constant $G$.

\vfill\eject


\begin{thebibliography}{}

\bibitem{1}S.Weinberg, {\it The Theory of Fields}, 3 volumes
(Cambridge Univ. Press, Cambridge, 1995, 1996 and 2000).

\bibitem{2}N.D.Birrell and P.C.W.Davies, {\it Quantum Fields
in Curved Space} (Cambridge University Press, Cambridge, 1982).
\hfill\break
 P.C.W.Davis, {\it Particles do not Exist}, in {\it Quantum Theory
 of Gravity. Essays in Honor of the 60th Birthday of Bryce DeWitt.}, ed.
 S.Christensen (Hilger, Bristol, 1984).

\bibitem{2a}S.Carlip, {\it Quantum Gravity: A Status Reportt},
Rep.Prog.Phys.  {\bf 64}, 855 (2002).

\bibitem{3}D.Home, {\it Conceptual Foundations of Quantum Physics}
(Plenum Press, New York, 1997). \hfill\break
 M.Czachor, Phys.Rev. {\bf A55}, 72 (1997).\hfill\break
 A.Peres, P.F.Scudo and D.R.Terno, Phys.Rev.Lett. {\bf 88}, 230402
 (2002).\hfill\break
 R.M.Gingrich and C.Adami, Phys.Rev.Lett. {\bf 89}, 270402
 (2002).\hfill\break
 C.Soo and C.C.Y.Lin, {\it Wigner Rotations, Bell States and Lorentz
 Invariance of Entanglement and Von Neumann Entropy} (quant-ph/0307107).
     \hfill\break
 C.G.Timpson and H.R.Brown, {\it Entanglement and Relativity},
 Pittsburgh Univ. preprint 2002
 (http://philsci-archive.pitt.edu/archive/00001618/).
\hfill\break
 W.C.Myrvold, {\it Relativistic Quantum Becoming},Pittsburgh Univ. preprint
 2002
 (http://philsci-archive.pitt.edu/archive/00000569).\hfill\break
 S.Haroche, {\it Introduction to Quantum Optics and Decoherence},
 LXXIX Les Houches Summer School 2003
 (http://www.lkb.ens.fr/recherche/qedcav/french/frenchframes.html).

\bibitem{4}A.Einstein, {\it Die Grundlage der allgemeinin Relativit\"atstheorie},
Annalen der Physik {\bf 49}, 769 (1916); translated by W.Perrett
and G.B.Jeffrey, {\it The Foundations of of the General Theory of
Relativity}, in {\it The Principle of Relativity} (Dover, New
York, 1952), pp.117-118.\hfill\break
 J.Stachel, {\it Einstein's
Search for General Covariance, 1912-1915}, paper read at the Ninth
International Conference on General Relativity and Gravitation,
Jena 1980; published in {\it Einstein and the History of General
Relativity}, Einstein Studies, Vol.1, eds. D.Howard and J.Stachel
(Birkh\"auser, Boston, 1985), pp.63-100.

\bibitem{5}J.Ehlers, F.A.E.Pirani and A.Schild, {\it The Geometry of
Free-Fall and Light Propagation} in {\it General Relativity,
Papers in Honor of J.L.Synge}, ed. L.O'Raifeartaigh (Oxford
Univ.Press, London, 1972).



\bibitem{6}S.Weinstein, {\it Naive Quantum Gravity}, in {\it Physics
Meets Philosophy at the Planck Scale}, C.Callender and N.Huggett
eds. (Cambridge University Press, Cambridge 2001).


\bibitem{7}Zvi Bern, {\it Perturbative Quantum Gravity and Its
Relation to Gauge Theory}, Living Rev.Rel. {\bf 5}, 5 (2002)
(gr-qc/0206071).\hfill\break
 J.Polchinski, {\it String Theory}, 2 volumes (Cambridge Univ.
Press, Cambridge, 1998).


\bibitem{8}A.Ashtekar and J.Lewandowski, {\it Background
Independent Quantum Gravity: a Status Report}
(gr-qc/0404018).\hfill\break
 L.Smolin, {\it How far are we from the Quantum Theory of
 Gravity?}, 2003 (hep-th/0303185).\hfill\break
 C.Rovelli, {\it Loop Quantum
Gravity}, Living Rev.Rel. {\bf 1}, 1 (1998) (gr-qc/9710008).

\bibitem{9}J.L.Anderson and P.G.Bergmann, {\it Constraints in
Covariant Field Theories}, Phys.Rev. {\bf 83}, 1018 (1951).
\hfill\break
 P.G.Bergmann and J.Goldberg, {\it Dirac Bracket
Transformations in Phase Space}, Phys.Rev. {\bf 98}, 531 (1955).

\bibitem{10}P.A.M.Dirac, {\it Lectures on
Quantum Mechanics}, Belfer Graduate School of Science, Monographs
Series (Yeshiva University, New York, N.Y., 1964).


\bibitem{11}M.Henneaux, {\it Hamiltonian Form of the Path Integral
for Theories with Gauge Freedom}, Phys.Rep. {\bf 126}, 1
(1985).\hfill\break
 M.Henneaux and C.Teitelboim, {\it Quantization of Gauge Systems} (Princeton
University Press, Princeton, 1992).

\bibitem{12}L.Lusanna,  {\it An Enlarged Phase Space for Finite-Dimensional Constrained
Systems, Unifying their Lagrangian, Phase- and Velocity-Space
Descriptions}, Phys.Rep. {\bf 185}, 1 (1990).\hfill\break
 {\it The Second Noether Theorem as the Basis of the Theory of Singular
 Lagrangians and Hamiltonian Constraints}, Riv. Nuovo Cimento {\bf 14}, n.3, 1 (1991).
 \hfill\break
  M.Chaichian, D.Louis Martinez and
L.Lusanna, {\it Dirac's Constrained Systems: the Classification of
Second Class Constraints},  Ann.Phys.(N.Y.) {\bf 232}, 40 (1994).

\bibitem{13}H.Tetrode, Z.Phys. {\bf 10}, 317 (1922).\hfill\break
 A.D.Fokker, Z.Phys. {\bf 58}, 386 (1929).\hfill\break
 J.A.Wheeler and R.P.Feynman, Rev.Mod.Phys. {\bf 17}, 157 (1945);
 {\bf 21}, 425 (1949).\hfill\break
  H.VanDam and E.P.Wigner, Phys.Rev. {\bf B138}, 1576
  (1965).\hfill\break
  A.Staruskiewicz, Ann.Inst.H.Poincare' {\bf 14}, 69 (1971).

\bibitem{14}J.Droste, Versl.K.Akad.Wet.Amsterdam {\bf 25}, 460
(1916).\hfill\break
 H.A.Lorentz and J.Droste, Versl.K.Akad.Wet.Amsterdam {\bf 26},
 392 and 649 (1917).

\bibitem{15}O.Hara, S.Ishida and S.Naka eds., {\it Extended
Objects  and Bound Systems}, Karuizawa 1992 (World Scientific,
Singapore, 1992), see the talks of H.Crater, L.Lusanna and
H.Sazdjian.

\bibitem{16}C.J.Isham, {\it Canonical Quantum Gravity and the Problem of Time},
in {\it Integrable Systems, Quantum Groups and Quantum Field
Theories}, eds.L.A.Ibort and M.A.Rodriguez, Salamanca 1993
(Kluwer, London, 1993).
 {\it Conceptual and
Geometrical Problems in Quantum Gravity}, in {\it Recent Aspects
of Quantum Fields}, Schladming 1991, eds. H.Mitter and H.Gausterer
(Springer, Berlin, 1991).
 {\it Prima Facie Questions
in Quantum Gravity} and {\it Canonical Quantum Gravity and the
Question of Time}, in {\it Canonical Gravity: {}From Classical to
Quantum}, eds. J.Ehlers and H.Friedrich (Springer, Berlin, 1994).


\bibitem{17}K.Kuchar, {\it Time and Interpretations of Quantum Gravity}, in
Proc.4th Canadian Conf. on {\it General Relativity and
Relativistic Astrophysics}, eds. G.Kunstatter, D.Vincent and
J.Williams (World Scientific, Singapore, 1992).


\bibitem{18}I.Ciufolini and J.A.Wheeler, {\it Gravitation and
Inertia} (Princeton Univ.Press, Princeton, 1995).

\bibitem{19}C. G. Torre, M. Varadarajan, {\it Functional Evolution
of Free Quantum Fields}, Class. Quantum Grav. {\bf 16}, 2651
(1999); {\it Quantum Fields at Any Time}, Phys.Rev. {\bf D58},
064007 (1998).

\bibitem{20}G.Rizzi and M.L.Ruggiero eds., {\it Relativity in
Rotatiing Frames} (Kluwer, Dordrecht, 2004).

\bibitem{21}D.Alba and L.Lusanna, {\it Simultaneity, Radar 4-Coordinates
and the 3+1 Point of View about Accelerated Observers in Special
Relativity} (gr-qc/0311058).

\bibitem{22}L.Lusanna, {\it Towards a Unified Description of the Four
Interactions in Terms of Dirac-Bergmann Observables}, invited
contribution to the book {\it Quantum Field Theory: a 20th Century
Profile}, of the Indian National Science Academy, ed.A.N.Mitra,
forewards by F.J.Dyson (Hindustan Book Agency, New Delhi, 2000)
(hep-th/9907081).

\bibitem{23}S.Shanmugadhasan, {\it Canonical Formalism for Degenerate
Lagrangians}, J.Math.Phys. {\bf 14}, 677 (1973).\hfill\break
 L.Lusanna, {\it The Shanmugadhasan Canonical Transformation, Function Groups
 and the Second Noether Theorem}, Int.J.Mod.Phys. {\bf A8}, 4193 (1993).

\bibitem{24}L.Lusanna, {\it On the BRS's}, J.Math.Phys. {\bf 31}, 428 and {\it Lagrangian
 and Hamiltonian Many-Time Equations}, {\bf 31}, 2126
 (1990). {\it Classical Observables of Gauge Theories from the Multi-Temporal Approach},
 Comtemp.Math. {\bf 132}, 531
(1992).

\bibitem{25}I.A.Batalin and G.A.Vilkoviski, Nucl.Phys. {\bf B234}, 106 (1984).

\bibitem{26}J.A.Schouten and W.V.D.Kulk, {\it Pfaff's Problem and Its
Generalizations} (Clarendon, Oxford, 1949).


\bibitem{27}S.Lie, {\it Theorie der Transformation Gruppe}, Vol. II (B.G.Teubner,
Leipzig, 1890).\hfill\break
 A.R.Forsyth, {\it Theory of Differential
Equations}, Vol. V, Ch. IX (Dover, New York, 1959).\hfill\break
 L.P.Eisenhart, {\it Continuous Groups of Transformations} (Dover, New
York, 1961). \hfill\break
R.O.Fulp and J.A.Marlin, Pacific J. Math.
{\bf 67}, 373 (1976); Rep.Math.Phys. {\bf 18}, 295 (1980).


\bibitem{28}L.D.Faddeev and Popov, Phys.Lett. {\bf B25}, 30 (1967).


\bibitem{29}G.Longhi and L.Lusanna, Phys.Rev. {\bf D34}, 3707 (1986).

\bibitem{30}J.M.Arms, J.E.Marsden and V.Moncrief, Commun.Math.Phys. {\bf 78},
455 (1981).\hfill\break
 J.M.Arms, Acta Phys.Pol. {\bf B17}, 499 (1986).\hfill\break
 L.Bos and M.J.Gotay, J.Diff.Geom. {\bf 24}, 181 (1986).

\bibitem{31}L.Lusanna, {\it Classical Yang-Mills Theory with
Fermions}, {\it I) General Properties of a System with
Constraints}, Int.J.Mod.Phys. {\bf A10}, 3531 (1995); {\it II)
Dirac's Observables}, Int.J.Mod.Phys. {\bf A10}, 3675 (1995).


\bibitem{32}J.Sniatycki, in {\it Non-Linear Partial Differential Operators and
Quantization Procedures}, Clausthal 1981, Lecture Notes Math. 1037
(Springer, Berlin, 1983).\hfill\break
 J.Snyaticki and A.Weinstein, Lett.Math.Phys. {\bf 7}, 155 (1983).


\bibitem{33}L.Lusanna and P.Valtancoli, Int.J.Mod.Phys. {\bf A13}, 4605 (1998)
(hep-th/9707072).

\bibitem{34}L.Bel, Ann.Inst.Henri Poincare' {\bf A12}, 307 (1970);
     {\bf 14}, 189 (1971); {\bf 18}, 57 (1973); {\it Mecanica
     Relativista Predictiva}, report UABFT-34 de la Universidad Autonoma de
     Barcelona 1977 (unpublished).\hfill\break
     R.Arens, Arch.Rat.Mech.Anal. {\bf 47}, 255
     (1972).\hfill\break
     L.Bel and J.Martin, Ann.Inst.H.Poincare\ {\bf A22}, 173
     (1975); {\bf 33}, 409 (1980); {\bf 34}, 231
     (1981).\hfill\break
     D.Hirondel, J.Math.Phys. {\bf 15}, 1689 (1979).\hfill\break
     L.Bel and X.Fustero, Ann.Inst.H.Poincare' {\bf A25}, 411
     (1976).





\bibitem{35}D.G.Currie, Phys.Rev. {\bf 142}, 817
(1966).\hfill\break
 R.N.Hill, J.Math.Phys. {\bf 8}, 201 (1967).

\bibitem{36}l.Bel, Ann.Inst.Henri Poincare' {\bf A12}, 307 (1970)

\bibitem{37}P.A.M.Dirac, Rev.Mod.Phys. {\bf 21}, 392
(1949).

\bibitem{38}L.H.Thomas, Phys.Rev. {\bf 85}, 868 (1952).\hfill\break
     B.Bakamjian and L.H.Thomas, Phys.Rev. {\bf 92}, 1300 (1953).
     \hfill\break
     L.L.Foldy, Phys.Rev. {\bf 122}, 275 (1961); {\bf D15},
     3044 (1977).


\bibitem{39}D.G.Currie, T.F.Jordan and E.C.G.Sudarshan, Rev.Mod.Phys.
{\bf 35}, 350 (1965).\hfill\break
     H.Leutwyler, Nuovo Cimento {\bf 37}, 556 (1965).

\bibitem{40}E.C.G.Sudarshan and N.Mukunda, {\it Classical Mechanics: a Modern
Perspective} (Wiley, New York, N.Y. 1974).\hfill\break
 S.Chelkowski, J.Nietendel and R.Suchanek, Acta Phys.Pol. {\bf
B11}, 809 (1980).

\bibitem{41}Ph.Droz Vincent, Phys.Scripta, {\bf 2}, 129
(1970).\hfill\break
 J.G.Wray, Phys.Rev. {\bf D1}, 2212 (1970).\hfill\break
 L.Bel, Ann.Inst.H.Poincare' {\bf 14}, 189 (1971).\hfill\break
 R.Arens, Arch.Rat.Mech.Anal. {\bf 47}, 255 (1972).\hfill\break
 L.Bel and J.Martin, Ann.Inst.H.Poincare\ {\bf 22}, 173 (1975).


\bibitem{42}Ph.Droz Vincent, Lett.Nuovo Cim. {\bf 1}, 839 (1969);
     {\bf 7}, 206 (1973); Phys.Scripta {\bf 2}, 129 (1970); Rep.
     Math.Phys. {\bf 8}, 79 (1975); Ann.Inst.H.Poincar\'e {\bf
     27}, 407 (1977); Phys.Rev. {\bf D19}, 702 (1979).

\bibitem{43}I.T.Todorov, {\it Dynamics of Relativistic Point Particles as a
Problem with Constraints}, Comm.JINR E2-10125, Dubna 1976
(unpublished); Ann.Inst.Henri Poincare' {\bf 28A}, 207 (1978).{\it
Constraint Hamiltonian Approach to Relativistic
     Point Particle Dynamics}, SISSA report, Trieste 1980 (unpublished);
     {\it Constraint Hamiltonian Dynamics of Directly Interacting Relativistic
     Point Particle} in {\it Quantum Theory, Groups, Fields and Particles},
     (A.O.Barut, Ed.), Reidel, Dordrecht, 1983.\hfill\break
      V.A.Rizov, H.Sazdjian and I.T.Todorov, Ann.Phys. (N.Y.) {\bf 165}, 59 (1985).

\bibitem{44}A.Komar,  Phys.Rev. {\bf D18} 1881, 1887 and 3017 (1978);
     {\bf D19}, 2908 (1979).

\bibitem{45}L.Lusanna, Nuovo Cimento {\bf 65B}, 135
(1981); {\it From Relativistic Mechanics towards Green's
Functions:  Multi-Temporal Dynamics} in Proc.
     of the VII Seminar on Problems of High Energy Physics and Quantum
     Field Theory, Protvino USSR 1984, vol.I, p.123.





\bibitem{46}S. Tomonaga, Prog. Theor. Phys. {\bf 1}, 27
(1946).\hfill\break
 J. Schwinger, Phys. Rev. {\bf 74}, 1439 (1948).

\bibitem{47}M.H.Soffel, {\it Relativity in Astrometry, Celestial Mechanics
and Geodesy} (Springer, Berlin, 1989).\hfill\break
 M.Soffel, S.A.Klioner, G.Petit, P.Wolf, S.M.Kopeikin,
P.Bretagnon, V.A.Brumberg, N.Capitaine, T.Damour, T.Fukushima,
B.Guinot, T.Huang, L.Lindegren, C.Ma, K.Nordtvedt, J.Ries,
P.K.Seidelmann, D.Vokroulicky', C.Will and Ch.Xu, {\it The IAU
2000 Resolutions for Astrometry, Celestial Mechanics and Metrology
in the Relativistic Framework: Explanatory Supplement}
(astro-ph/0303376).\hfill\break
 B.Guinot, {\it Application of General Relativity to Metrology},
 Metrologia {\bf 34}, 261 (1997).


\bibitem{48}M.M.Capria, {\it On the Conventionality of Simultaneity in
Special Relativity}, Found.Phys. {\bf 31}, 775 (2001).

\bibitem{49}J.Stachel, {\it Einstein and the Rigidly Rotating
Disk}, in {\it General Relativity and Gravitation}, ed. H.Held
(Plenum, New York, 1980).

\bibitem{50}C.M. M$\o$ller, {\it The Theory of Relativity} (Oxford
Univ.Press, Oxford, 1957).

\bibitem{51}B.Mashhoon, {\it The Hypothesis of Locality and its
Limitations}, (gr-qc/0303029). {\it Limitations of Spacetime
Measurements}, Phys.Lett. {\bf A143}, 176 (1990). {\it The
Hypothesis of Locality in Relativistic Physics}, Phys.Lett. {\bf
A145}, 147 (1990). {\it Measurement Theory and General
Relativity}, in {\it Black Holes: Theory and Observation}, Lecture
Notes in Physics 514, ed. F.W.Hehl, C.Kiefer and R.J.K.Metzler
(Springer, Heidelberg, 1998), p.269.  {\it Acceleration-Induced
Nonlocality}, in {Advances in General Relativity and Cosmology},
ed. G.Ferrarese (Pitagora, Bologna, 2003) (gr-qc/0301065).
\hfill\break
 B.Mashhoon and U.Muench, {\it Length Measurement in
Accelerated Systems}, Ann.Phys. (Leipzig) {\bf 11}, 532 (2002).

\bibitem{52}R.Arnowitt, S.Deser and C.W.Misner, {\it Canonical Variables
for General Relativity}, Phys.Rev. {\bf 117}, 1595 (1960).
 \hfill\break
 {\it The Dynamics of General Relativity}, in {\it Gravitation: an
Introduction to Current Research}, ch. 7, ed.L.Witten (Wiley, New
York, 1962).

\bibitem{53}L.Lusanna, {\it The Rest-Frame Instant Form of Metric Gravity},
Gen.Rel.Grav. {\bf 33}, 1579 (2001) (gr-qc/0101048).

\bibitem{54}L.Lusanna, {\it The N- and 1-Time Classical
Descriptions of N-Body Relativistic Kinematics and the
Electromagnetic Interaction}, Int.J.Mod.Phys. {\bf A12}, 645
(1997).

\bibitem{55} H.Crater and L.Lusanna, Ann.Phys. (NY) {\bf 289}, 87 (2001)
(hep-th/0001046).\hfill\break
 D. Alba, H. Crater and L. Lusanna, Int. J. Mod.Phys. {\bf A16}, 3365 (2001)
(hep-th/0103109).

\bibitem{56}D.Alba and L.Lusanna,
Int.J.Mod.Phys. {\bf A13}, 2791 (1998) (HEP-TH/9705155).


\bibitem{56a}A.Spallicci, A.Brillet, G.Busca, G.Catastini, I.Pinto,
I.Roxburgh, C.Salomon, M.Soffel and C.Veillet, {\it Experiments on
Fundamental Physics on the Space Station}, Class.Quant.Grav. {\bf
14}, 2971 (1997).\hfill\break
 P.Lemonde, P.Laurent, G.Santarelli, M.Abgrall, Y.Sortais, S.Bize,
 C.Nicolas, S.Zhang, A.Clairon, N.Dimarcq, P.Petit, A.Mann, A.Luiten,
 S.Chang and C.Salomon, {\it Cold Atom Clocks on Earth and Space},
 in {\it Frequency Measurement and Control, Advanced Techniques
 and Future Trends}, ed.A.N.Luiten (Springer, Berlin, 2001).


\bibitem{57}D.Alba and L.Lusanna,
Int.J.Mod.Phys. {\bf A13}, 3275 (1998) (HEP-TH/9705156).

\bibitem{58}F.Bigazzi and L.Lusanna, Int.J.Mod.Phys. {\bf A14}, 1429 (1999)
(HEP-TH/9807052).

\bibitem{59}F.Bigazzi and L.Lusanna, Int.J.Mod.Phys. {\bf A14}, 1877 (1999)
(HEP-TH/9807054).

\bibitem{60}J. D. Brown, Class. Quantum Grav. {\bf 10}, 1579
(1993).\hfill\break
 L. Lusanna and D. Nowak-Szczepaniak, Int. J.
Mod. Phys. {\bf A15}, 4943 (2000).\hfill\break
 D.Alba and L.Lusanna, {\it Generalized Eulerian Coordinates for Relativistic
Fluids: Hamiltonian Rest-Frame Instant Form, Relative Variables,
Rotational Kinematics} (hep-th/0209032), to appear in
Int.J.Mod.Phys. A.


\bibitem{6l}L.Lusanna, Nuovo Cim. {\bf 64A}, 65 (1981).

\bibitem{62}D. Alba, L. Lusanna and M. Pauri, {\it Centers of Mass and
Rotational Kinematics for the Relativistic N-Body Problem in the
Rest-Frame Instant Form}, J. Math. Phys. {\bf 43}, 1677 (2002)
(hep-th/0102087).

\bibitem{63}M.Pauri and M.Prosperi, J.Math.Phys. {\bf 16}, 1503 (1975).

\bibitem{64}S.Gartenhaus and C.Schwartz, Phys.Rev. {\bf 108}, 842
(1957).

\bibitem{65}R.G.Littlejohn and M.Reinsch, Rev.Mod.Phys. {\bf 69},
213 (1997).

\bibitem{66}D. Alba, L. Lusanna and M. Pauri, J. Math. Phys. {\bf 43},
373 (2002) (hep-th/0011014).

\bibitem{67} D.Alba, L.Lusanna and M.Pauri, {\it Multipolar Expansions for
Closed and Open Systems of Relativistic Particles},
(hep-th/0402181).



\bibitem{68}W.G.Dixon, J.Math.Phys. {\bf 8}, 1591 (1967).

\bibitem{69}G.Veneziano, {\it Quantum Strings and the Constants of Nature}, in
{\it The Challenging Questions}, ed.A.Zichichi, the Subnuclear
Series n.27 (Plenum Press, New York, 1990).


\bibitem{70}C.L\"ammerzahl, J.Math.Phys. {\bf 34}, 3918 (1993).

\bibitem{71}L.Lusanna and S.Russo, {\it A New Parametrization for Tetrad Gravity},
Gen.Rel.Grav. {\bf 34}, 189 (2002)(gr-qc/0102074).

\bibitem{72}R.De Pietri, L.Lusanna, L.Martucci and S.Russo, {\it Dirac's
Observables for the Rest-Frame Instant Form of Tetrad Gravity in a
Completely Fixed 3-Orthogonal Gauge}, Gen.Rel.Grav. {\bf 34}, 877
(2002) (gr-qc/0105084).

\bibitem{73}A.Lichnerowicz, {\it Propagateurs, Commutateurs et
Anticommutateurs en Relativite Generale}, in Les Houches 1963,
{\it Relativity, Groups and Topology}, eds. C.DeWitt and B.DeWitt
(Gordon and Breach, New York, 1964).\hfill\break
 C.Moreno, {\it On the Spaces of Positive and Negative Frequency
 Solutions of the Klein-Gordon Equation in Curved Space-Times},
 Rep.Math.Phys. {\bf 17}, 333 (1980).

\bibitem{74}D.Christodoulou and S.Klainerman, {\it The Global Nonlinear
Stability of the Minkowski Space} (Princeton, Princeton, 1993).

\bibitem{75}A.Sen, {\it On the Existence of Neutrino "Zero-Modes" in
Vacuum Space-Times}, J.Math.Phys. {\bf 22}, 1781 (1981); {\it
Gravity as a Spin System}, Phys.Lett. {\bf 119B}, 89
(1982).\hfill\break
 E.Witten, {\it A New Proof of the Positive Energy Theorem},
 Commun.Math.Phys. {\bf 80}, 381 (1981).

\bibitem{76}J.Frauendiener, {\it Triads and the Witten Equation},
Class.Quantum Grav. {\bf 8}, 1881 (1991).

\bibitem{77}Y.Choquet-Bruhat, A.Fischer and J.E.Marsden, {\it Maximal
    Hypersurfaces and Positivity of Mass}, LXVII E.Fermi Summer School
  of Physics {\it Isolated Gravitating Systems in General
    Relativity}, ed. J.Ehlers (North-Holland, Amsterdam, 1979).

\bibitem{78}C.Teitelboim, {\it The Hamiltonian Structure of Space-Time}, in
{\it General Relativity and Gravitation}, ed.A.Held, Vol.I
(Plenum, New York, 1980).

\bibitem{79}J.Isenberg and J.E.Marsden, {\it The York Map is a Canonical
Transformation}, J.Geom.Phys. {\bf 1}, 85 (1984).

\bibitem{80}H.Friedrich and A.Rendall, {\it The Cauchy Problem for
Einstein Equations}, in {\it Einstein's Field Equations and their
Physical Interpretation}, ed. B.G.Schmidt (Springer, Berlin, 2000)
(gr-qc/0002074).\hfill\break
 A.Rendall, {\it Local and Global Existence Theorems for the
 Einstein Equations}, Online journal Living Reviews in Relativity
 {\bf 1}, n. 4 (1998) and {\bf 3}, n. 1 (2000) (gr-qc/0001008).

\bibitem{81}L.Lusanna and M.Pauri, {\it The Physical Role of Gravitational
and Gauge Degrees of Freedom in General Relativity - I: Dynamical
Synchronization and Generalized Inertial Effects},
(gr-qc/0403081).

\bibitem{82}P.G.Bergmann and A.Komar, {\it The Coordinate Group
Symmetries of General Relativity}, Int.J.Theor.Phys. {\bf 5}, 15
(1972).

\bibitem{83}L.Lusanna and M.Pauri, {\it The Physical Role of Gravitational
and Gauge Degrees of Freedom in General Relativity - II : Dirac
versus Bergmann Observables and the Objectivity of Space-Time}, in
preparation.



\bibitem{84}L.Lusanna and M.Pauri, {\it General Covariance and the
Objectivity of Space-Time Point-Events: The Physical Role of
Gravitational and Gauge Degrees of Freedom in General Relativity}
(gr-qc/0301040).\hfill\break
 M.Pauri and M.Vallisneri, {\it Ephemeral Point-Events: is there
a Last Remnant of Physical Objectivity?}, essay for the 70th
birthday of R.Torretti, Dialogos {\bf 79}, 263 (2002)
(gr-qc/0203014).\hfill\break
 L.Lusanna, {\it Space-Time, General Covariance, Dirac-Bergmann
Observables and Non-Inertial Frames}, talk at the 25th Johns
Hopkins Workshop {\it 2001: A Relativistic Space-Time Odyssey},
Firenze September 3-5, 2001 (gr-qc/0205039).




\bibitem{85}M.Dorato and M.Pauri, {\it Holism and Structuralism in Classical
and Quantum General Relativity}, Pittsburgh-Archive 2004
(http://philsci-archive.pitt.edu/archive/00001606/).

\bibitem{86}P.G.Bergmann and A.Komar, {\it Poisson Brackets between
Locally Defined Observables in General Relativity},
Phys.Rev.Letters {\bf 4}, 432 (1960).

\bibitem{87}P.G.Bergmann, {\it Observables in General Relativity},
Rev.Mod.Phys. {\bf 33}, 510 (1961).

\bibitem{88}J.Stewart, {\it Advanced General Relativity} (Cambridge Univ.
Press, Cambridge, 1993).

\bibitem{89}D.Alba and L.Lusanna, {\it Multi-Temporal Quantization
for Relativistic and Non-Relativistic Particles in Non-Inertial
Frames in Absence of Gravity}, in preparation.


\bibitem{90}N.Ashby and J.J.Spilker, {\it Introduction to
Relativistic Effects on the Global Positioning System}, in {\it
Global Positioning System: Theory and Applications}, Vol.1, eds.
B.W.Parkinson and J.J.Spilker (American Institute of Aeronautics
and Astronautics, 1995).

\bibitem{91}R.F.Marzke and J.A.Wheeler, {\it Gravitation as Geometry-
I: the Geometry of the Space-Time and the Geometrodynamical
Standard Meter}, in {\it Gravitation and Relativity}, eds.
H.Y.Chiu and W.F.Hoffman (Benjamin, New York, 1964).


\bibitem{92}J.Agresti, R.DePietri, L.Lusanna and L.Martucci, {\it
Hamiltonian Linearization of the Rest-Frame Instant Form of Tetrad
Gravity in a Completely Fixed 3-Orthogonal Gauge: a Radiation
Gauge for Background-Independent Gravitational Waves in a
Post-Minkowskian Einstein Space-Time} Gen.Rel.Grav. {\bf 36}, 1055
(2004) (gr-qc/0302084).


\bibitem{93}Ch.J.Borde', A.Karasiewicz and Ph.Tourrenc, {\it
General Relativistic Framework for Atomic Interferometry},
Int.J.Mod.Phys. {\bf D3}, 157 (1994).\hfill\break
 Ch.J.Borde', J.C.Houard and A.Karasiewicz, {\it Relativistic
 Phase Shifts for Dirac Particles Interacting with Weak
 Gravitational Fields in Matter-Wave Interferometers}, in
{\it Gyros, Clocks, and Interferometers: Testing Relativistic
Gravity in Space}, eds. C. Lammerzahl, C.W.F. Everitt, F.W. Hehl,
Lect.Notes Phys. {\bf 562}, p.403-438 (Springer, Berlin, 2001)
(gr-qc/0008033).

\bibitem{94}B.Mashhoon, {\it Gravitoelectromagnetism: a Brief
Review}, 2003 (gr-qc/0311030).


\bibitem{95}L.Blanchet, C.Salomon, P.Teyssandier and P.Wolf, {\it
Relativistic Theory for Time and Frequency Transfer to Order
$1/c^3$}, Astron.Astrophys. {\bf 370}, 320 (2000).



\bibitem{117a}R.Rynasiewicz, {\it Kretschmann's Analysis of
Covariance and Relativity Principles}, in {\it The Expanding
Worlds of General Relativity} (Einstein Studies, volume 7), eds.
H.Goenner, J.Renn, J.Ritter and T.Sauer (Birkh\"auser, Boston,
1999).


\end{thebibliography}
\end{document}